\documentclass[onecolumn,aps,superscriptaddress,prd,10pt,showpacs,preprintnumbers,nofootinbib,eqsecnum]{revtex4-2}
\usepackage{graphicx}
\graphicspath{{./}}
\usepackage{amsmath,amssymb,amsthm,dsfont,color}
\usepackage{subfigure}
\usepackage{bbold}
\usepackage{ulem}
\usepackage{natbib,ulem} 
\usepackage[dvipsnames]{xcolor}
\usepackage{newunicodechar}
\newunicodechar{−}{\ensuremath{-}}
\usepackage{multirow}
\usepackage{tikz}
\usepackage[colorlinks=true,citecolor=blue,linkcolor=blue,urlcolor=blue]{hyperref}
\setcitestyle{square, numbers, comma, sort&compress}

\newcommand\lsout{\bgroup\markoverwith{\textcolor{ForestGreen}{\rule[0.5ex]{2pt}{0.4pt}}}\ULon}
\newcommand\jlsout{\bgroup\markoverwith{\textcolor{blue}{\rule[0.5ex]{2pt}{0.4pt}}}\ULon}
\newcommand\msout{\bgroup\markoverwith{\textcolor{red}{\rule[0.5ex]{2pt}{0.4pt}}}\ULon}
\newcommand\tsout{\bgroup\markoverwith{\textcolor{purple}{\rule[0.5ex]{2pt}{0.4pt}}}\ULon}
\newcommand\cosout{\bgroup\markoverwith{\textcolor{olive}{\rule[0.5ex]{2pt}{0.4pt}}}\ULon}
\newcommand\clsout{\bgroup\markoverwith{\textcolor{magenta}{\rule[0.5ex]{2pt}{0.4pt}}}\ULon}

\newcommand{\fr}[2]{{\frac{#1}{#2}\,}}

\newcommand{\be}{\begin{equation}}
\newcommand{\ee}{\end{equation}}
\newcommand{\msbar}{{\overline{\mbox{\rm MS}}}}
\newcommand{\ba}{\begin{eqnarray}}
\newcommand{\ea}{\end{eqnarray}}
\newcommand{\nn}{\nonumber \\}
\renewcommand{\(}{\left(}
\renewcommand{\)}{\right)}
\renewcommand{\[}{\left[}

\newcommand{\as}{\ensuremath{\alpha_s}}

\newcommand{\mcm}{\ensuremath{m_s}}
\newcommand{\mub}{\ensuremath{\mu_{B}}}
\allowdisplaybreaks

\begin{document}
\title{Quark and hybrid stars with renormalization group improvement of NNLO perturbative QCD }

\author{Loïc Fernandez}
\email{loic.fernandez@helsinki.fi, loicfernandez00@gmail.com}
\affiliation{Helsinki Institute of Physics, P.O. box 64, FI-00014 University of Helsinki, Finland.}
\author{Jean-Loïc Kneur}
\email{jean-loic.kneur@umontpellier.fr}
\affiliation{Laboratoire Charles Coulomb (L2C), UMR 5221 Universit\'e de Montpellier, France}
\author{Marcus Benghi Pinto}
 \email{ marcus.benghi@ufsc.br}
\affiliation{Departamento de F\'{\i}sica, Universidade Federal de Santa Catarina, Florian\'{o}polis, SC, Brazil}
\author{Constança Providência}
\email{cp@uc.pt}
\affiliation{CFisUC, Department of Physics, University of Coimbra, 3004-516 Coimbra, Portugal }
\author{Claudia Ratti}
\email{cratti@central.uh.edu}
\affiliation{Department of Physics, University of Houston, Houston, TX 77204, USA}
\author{Tulio E. Restrepo}
\email{trestre2@central.uh.edu}
\affiliation{Department of Physics, University of Houston, Houston, TX 77204, USA}

\begin{abstract}
Recently, the NNLO perturbative QCD pressure  of cold and dense symmetric matter,  with arbitrary quark masses, has been resummed within the renormalization-group-optimized perturbation theory (RGOPT) framework. By being imbued with renormalization group properties, the  resulting pressure is less sensitive to renormalization scale ($\Lambda\equiv X \mu_B/3$) variations than the NNLO perturbative QCD pressure. Here, we extend  this by considering $\beta$-equilibrium and charge neutrality to evaluate the corresponding equation of state (EoS). 
We provide a compact ``pocket" fitting formula for the  EoS for $N_f=2+1$ massive quarks at different renormalization scale parameter ($X$) values. We describe pure quark stars as well as hybrid stars with quark-cores. 
Pure quark stars compatible with astrophysical observations were obtained
with $X=2.71-3.30$, whereas a larger value (3.84) is needed if
the secondary component of the observation GW190814 represents a
neutron star.
Hybrid stars were built considering three representative hadron models based on a relativistic mean-field description, and chosen to produce soft and stiff EoSs.
Stable hybrid stars with masses compatible with the massive pulsar PSR
J0740+6620 were obtained considering $X$ of the order of 2 to
$2.71$, the largest scale giving rise to hybrid stars with a large
quark core with a radius of 2 to 8 km , and the smallest to a small
quark core at the center of the star.
\end{abstract}

\maketitle

\section{Introduction}

The theoretical description of strongly interacting matter at high baryonic densities plays a central role in investigations aiming to describe the structure of neutron stars (NS). In  general, hadrons and/or quarks are considered to be the most relevant degrees of freedom representing the matter that makes up these  dense stellar objects. In particular, the possibility that hadron and quark matter form a hybrid NS became more appealing due to the recent observational data of two solar mass pulsars \cite{Demorest:2010bx,Fonseca:2021wxt,Antoniadis:2013pzd}  which  suggest  that quark matter may indeed be present within their core \cite{Annala:2023cwx,Albino:2025puc}. In this case, an appropriate equation of state (EoS), preferably obtained from the fundamental quantum chromodynamics (QCD) theory, should be used to describe the ultra-dense core region. 

An {\it ab-initio} determination of the
QCD EoS for cold and compressed quark matter
represents a challenging task, since lattice QCD (LQCD) calculations 
are limited to low baryon chemical potential values ($\mu_B$) due to the well-documented sign problem \cite{deForcrand:2009zkb}, which obstructs extensions to intermediate and large $\mu_B$ values. 
In the low baryon-density regime, chiral perturbation theory (chEFT) provides a precise description \cite{Drischler:2017wtt,Hebeler:2020ocj}.
At intermediate and higher $\mu_B$ values, one can consider effective models \cite{Oertel:2016bki,Dutra:2014qga,Cartaxo:2025jpi} as well as perturbative QCD (pQCD) \cite{Freedman:1976ub,Kurkela:2009gj,Fraga:2013qra,Gorda:2018gpy,Gorda:2021kme}, but unfortunately these alternatives also present some shortcomings. For instance, most effective quark models (such as the Nambu-Jona-Lasinio model \cite{Klevansky:1992qe,Hatsuda:1994pi,Rehberg:1995kh,Buballa:2003qv} and the quark-meson model \cite{Lenaghan:2000ey}) do not take asymptotic freedom and gluonic degrees of freedom into account. 
At the same time, pQCD applications are reliable only at high baryon densities of the order of $\rho_B \sim 25-40 \, \rho_0$  ($\rho_0 = 0.16 \, {\rm fm}^{-3}$) \cite{Gorda:2023mkk}, where asymptotic freedom allows for weak coupling expansions. However, for NS with masses $\sim 1.4 − 2M_\odot$ the typical central densities are expected to be around the
$5 − 10\, \rho_0$ range \cite{Baym:2017whm} where quark matter is still strongly coupled.
A bridge between the low- and high-density regions was elaborated via 
model-agnostic
equations of state (EoS) of compact stars  applying different statistical methods (see, e.g., \cite{Kurkela:2014vha,Annala:2019puf,Annala:2021gom,Altiparmak:2022bke, Gorda:2023usm,Annala:2023cwx,Komoltsev:2023zor}), 
via thermodynamical constrains\cite{Komoltsev:2021jzg},
and via microscopic models within a Bayesian inference approach (see, e.g., \cite{Lim:2019som,Traversi:2020aaa,Zhu:2022ibs,Malik:2022jqc,Malik:2023mnx,Takatsy:2023xzf,Zhou:2023zrm}), imposing NS observational constraints in addition to those from {\it ab-initio} pQCD and chEFT calculations. Such approaches are remarkable in that they produce EoSs over the largest possible $\mu_B$ range. However, they are limited by the present accuracy of theoretical chEFT and pQCD predictions at both ends and by the accuracy of available data. This is also complicated by the phase transition \cite{Gupta:2011wh} at intermediate $\mu_B$ values, which generally requires further model-dependent assumptions.

Concerning higher order pQCD evaluations,  
it is important to recall that these are 
notoriously complicated by thermal or medium properties \cite{Blaizot:2003tw,Kraemmer:2003gd,Ghiglieri:2020dpq}. 
 In particular, infrared divergences proliferate starting at order $\mathcal{O}(\alpha_s^2)$, which cancel only through a specific resummation of a specific class of Feynman diagrams, but producing 
 non-analytical contributions to the naive weak-coupling perturbative expansion.  
 For cold quark matter,
 these contributions are classified as ``soft'', associated to the scale $m_E\sim \sqrt{\as}\mu$, in contrast to the ``hard'' contributions provided by quarks at the Fermi surface given by the quark chemical potential scale, $\mu$.
  The groundbreaking calculation of Freedman and McLerran \cite{Freedman:1976ub} revealed the emergence of an $\as^2\ln\as$ dependence (in the massless quark approximation) in the next-to-next-to-leading order (NNLO)
pressure. These soft logarithm contributions emanate from the plasmon (ring) resummation of soft infrared divergences, nowadays well-understood also within the hard thermal loop (HTL) \cite{Braaten:1991gm} effective field theory framework \cite{Gorda:2021znl,Gorda:2021kme}, which has also led to the all-order resummation of the leading soft logarithms \cite{Fernandez:2021jfr}.
The cold quark matter pressure calculations have been generalized to include quark masses \cite{Fraga:2004gz,Laine:2006cp,Kurkela:2009gj,Graf:2015tda}; and thermal effects \cite{Ipp:2006ij,Kurkela:2016was}. 
In the recent few years, even the full NNNLO is in sight \cite{Gorda:2021znl,Gorda:2021kme,Gorda:2023mkk,Karkkainen:2025nkz}
in the massless quark approximation, expecting significantly improved pQCD 
accuracy. 
Despite this remarkable recent progress, at lower $\mu_B$ beyond the perturbative regime, the pQCD 
pressure exhibits a loss of accuracy, mainly reflected in a rapidly growing sensitivity on the arbitrary renormalization scale $\Lambda$ in the (${\overline{\rm MS}}$) scheme\cite{Kurkela:2009gj}.
Although the pQCD NNLO pressure is perturbatively renormalization group (RG) invariant, this implies that while formally of higher order ${\cal O}(\alpha_s^3)$, the residual scale dependence has sizable impact, even for not that small 
$\mu_B \sim 2 \, {\rm GeV}$, which corresponds to
a baryon density $\sim 40\, \rho_0$. 
It indicates that possible non-perturbative effects remain quite important in the lower/intermediate density range relevant to NSs. Moreover, it appears that this large residual scale dependence
originates principally \cite{Gorda:2021znl,Fernandez:2021jfr} from the hard sector, contributing predominantly  to the total pressure, while in comparison the soft sector is subdominant and under better control \cite{Gorda:2021znl,Gorda:2023mkk,Fernandez:2021jfr}.
This is further worsened if accounting for the strange quark mass effects, not always negligible at moderate and low $\mu_B$ values. For massive quarks, pQCD calculations  are even more involved and the state-of-the-art is
limited to NNLO \cite{Kurkela:2009gj}, thus impinging on scale uncertainty
reductions. \\
In the present work, we consider a resummation approach which incorporates 
higher order RG properties in order to mitigate the above mentioned pQCD problems in evaluations aiming to describe the quark sector of NSs. 
The  renormalization group optimized perturbation theory (RGOPT) \cite{Kneur:2013coa,Kneur:2015dda,Kneur:2015uha} 
is based on a modified perturbative expansion, 
around massive quasiparticle states regularizing infrared divergences.
Basically, this is similar to other approaches at finite temperature and densities based on interpolated Lagrangians augmented by prescriptions aiming to go beyond strictly perturbative expansions.
Such screened perturbation theory \cite{Parwani:1991gq,Karsch1997,Andersen:2000yj} (SPT) for the thermal $\phi^4$ model has found many applications, especially when used within the HTL Lagrangian, 
hence known as Hard Thermal Loop perturbation theory (HTLpt) \cite{Andersen:1999fw,Andersen:1999va}.
A specific feature of the RGOPT prescription is that it generates a ``RG-dressed'' screening mass, entailing an all-order RG-driven $\alpha_s$ dependence.
At finite temperature, RGOPT has been applied to $\phi^{4}$ theory \cite{Kneur:2015moa} up to NNLO \cite{Fernandez:2021jfr}, and to NLO for the hot QCD pressure \cite{Kneur:2021dfo,Kneur:2021feo}, where in both cases it drastically reduces the residual renormalization scale dependence
with respect to the standard weak-coupling expansion, SPT or HTLpt. 
For cold quark matter, the NLO RGOPT \cite{Kneur:2019tao} also reduces the residual scale dependence, although more moderately than in the $T\ne 0$ case.
Recently, it has been also used to describe non-strange and strange quark stars (QS) at NLO \cite{Restrepo:2022wqn,Restrepo:2025qgp}, yielding results which are less sensitive to scale variations than those provided by pQCD. 

Meanwhile, the  method has been considered at NNLO to describe the $N_f=2+1$ QCD EoS in the case of symmetric matter \cite{Fernandez:2024ilg}, $\mu_u=\mu_s=\mu_d\equiv \mu$. 
In the present work, the NNLO results of Ref \cite{Fernandez:2024ilg} will be extended to treat non-symmetric quark matter in $\beta$-equilibrium, so that an EoS tailored to describe pure quarks stars as well as the core of hybrid NSs can be made available. For practical purposes, we will also provide an approximate but accurate compact formula for our obtained (thermodynamically consistent) pressure as a function of the renormalization scale and baryon chemical potential, so that the EoS can be readily computed. 
When applying pQCD to cold and dense matter, one usually sets $\Lambda = X \, \mu_B/3$ where, for 
$N_f = 2+1$ flavors, $\mu_B$ relates to the individual quark chemical potentials through $\mu_B = (\mu_u+\mu_s+\mu_d)$. In general, $X$ is probed for $1 \le X \le 4$ with $X=2$ representing a 
``central" fiducial scale. 

In the present study, we will  compare state-of-the-art NNLO pQCD and RGOPT predictions
including massive quarks, also aiming to constrain the renormalization scale,  by imposing agreement with NS observational data as in Refs. \cite{Restrepo:2022wqn,Restrepo:2025qgp}. 
In particular, we will consider the mass and radius measurements from NASA’s
Neutron Star Interior Composition Explorer (NICER) mission for the pulsars PSR
J0030+0451 \cite{Riley:2019yda,Miller:2019cac}, PSR J0740+6620
\cite{Riley:2021pdl,Raaijmakers:2021uju,Miller:2021qha,Salmi:2024aum}, PSR
J0437+4715 \cite{Choudhury:2024xbk}, PSR J0614-3329 \cite{Mauviard:2025dmd}, and
 the light compact object in the supernova remnant HESS J1731-347
\cite{Doroshenko:2022nwp}.
A quite strong constraint is imposed by the  pulsar J0952−0607, a ``black widow'' pulsar with a gravitational mass above 2.27$M_\odot$ (2.12$M_\odot$) at 1$\sigma$ (3$\sigma$) confidence interval \cite{Romani:2022jhd,Romani:2025ytn}. 
However, the mass measurements of pulsars as PSR J0952−0607 depend on the optical modeling, which introduces larger uncertainties than  estimates from pulsar timing, as done by NICER observations. Additional constraints are imposed by the detection of gravitational waves by the LIGO Virgo Collaboration, in particular GW170817 \cite{LIGOScientific:2017vwq,LIGOScientific:2018cki} from a binary NS merger and GW190814 from the  coalescence of a  black hole with a 2.5-2.67 solar mass compact object, possibly a NS. 
Apart from discussing pure QSs, we will also build some hybrid EoSs by considering a Maxwell construction to describe a first order phase transition between the hadronic and quark phases, obtained within the NNLO RGOPT.  
In order to cover the present existing uncertainty in the hadronic phase, we select a few nucleonic models representative of soft and stiff equations of state to build the hybrid EoS, all based on a relativistic mean-field description of nuclear matter: SFHo  \cite{Steiner:2012rk} and DIDY \cite{Frohaug:2025okz} which represent a soft EoS and DD2, which can produce a reasonably hard EoS \cite{Typel:2009sy}. 
These equations of state satisfy the well-known nuclear matter and NS properties. 

The work is organized as follows. In Section \ref{Pr}, the quark pressure $P(X,\mu_B)$ is determined within the RGOPT resummation framework (this section may
be skipped by readers mainly interested in applications to compact stars).
In Sec. \ref{sec:thermo_consis} we present a pocket formula for the thermodynamically consistent pressure, for the case of $\beta$-equilibrium and charge neutral matter.
The RGOPT and pQCD results for quark and hybrid stars  are presented and compared  in Sec.~\ref{sec:results}.
Finally, in Sec.~\ref{sec:conclusions} we present our conclusions.

\section{RGOPT pressure }
\label{Pr}

\begin{figure}[h!]
      \begin{subfigure}
     \centering
     \includegraphics[width=.3\linewidth]{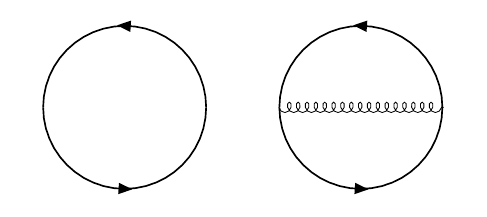}
    \end{subfigure}
    \raisebox{0.21cm}{\begin{subfigure}
    \centering
    \includegraphics[width=.36\linewidth]{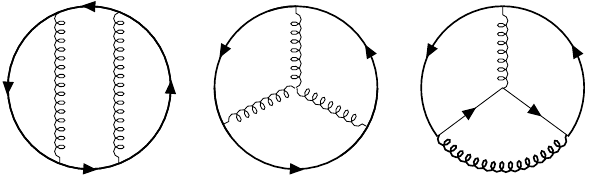}
    \end{subfigure}}
\caption{Feynman graphs contributing to the (infrared safe) NNLO weak coupling expansion.
In our case, both matter and vacuum contributions are considered.}
\label{fig:hard_graphs}
\end{figure}

\begin{figure}
      \begin{subfigure}
     \centering
     \includegraphics[width=.5\linewidth]{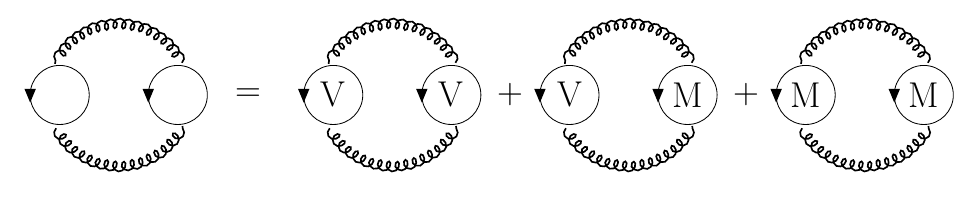}
    \end{subfigure}
    \raisebox{-0.25cm}{\begin{subfigure}
    \centering
    \includegraphics[width=.2\linewidth]{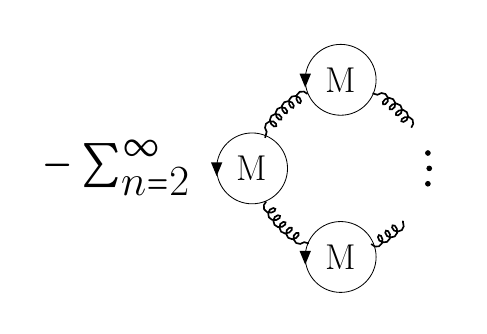}
    \end{subfigure} }
\caption{NNLO Feynman graphs involving infrared divergences requiring resummations, as indicated
in the rightmost graph.}
\label{fig:soft_graphs}
\end{figure}

In this Section, we review the derivation of the RGOPT pressure for cold dense quark 
matter at NNLO by summarizing the derivation contained in Ref. \cite{Fernandez:2024ilg}. 
As mentioned in the Introduction, 
by construction the RGOPT aims at reducing the residual renormalization scale dependence as compared to the standard weak coupling expansion.
Concerning specifically 
the cold and dense pQCD pressure, 
originally derived at NNLO with full quark mass dependence  \cite{Kurkela:2009gj},
as emphasized above, its sizable residual renormalization scale dependence (thus originating from unknown higher perturbative orders)
originates predominantly  from the hard sector contributions,  
which is further enhanced for massive quarks \cite{Fernandez:2024ilg}.
Accordingly, we consider the RGOPT
only in the massive quark
sector for simplicity\footnote{In principle, one could also apply the RGOPT to the effective HTL gluon mass, like it is done within HTLpt \cite{Andersen:1999va}, but it would require higher-order HTL contributions that are currently unknown.}. \\
Since this approach is basically a modification of the standard weak coupling expansion of a massive
theory, we start from the NNLO weak coupling expansion in $\alpha_s=g_s^2/(4\pi)$ of the cold and dense quark matter pressure,
whose full NNLO quark mass dependence was originally calculated in Ref. \cite{Kurkela:2009gj}.
The relevant Feynman graphs up to NNLO in the weak coupling expansion
are given in Figs. \ref{fig:hard_graphs} and \ref{fig:soft_graphs}.
Importantly, instead of considering massless $u,d$ and one massive strange quark as usual in pQCD, we anticipate that the RGOPT prescription induces an (all-order) dressed screening quark mass $m ={\cal O}(g_s\, \mu)$ common to the three flavors, as explained below (while we still approximate vanishing current $u,d$ quark masses). Thus, all graphs in Figs. \ref{fig:hard_graphs}, \ref{fig:soft_graphs} are evaluated with massive quarks. Moreover,
non-diagonal contributions occur from ultimately distinct quark masses $(m+m_u,m+m_d,m +m_s)$, where $m_s$ is the strange quark current mass and $m_u=m_d\equiv 0$. The graphs in Fig. \ref{fig:soft_graphs} produce such non-diagonal terms at NNLO, derived in Ref. \cite{Fernandez:2024ilg}, 
to which we refer for more details\footnote{Since in our application to beta-equilibrated matter below, quarks have different chemical potentials $\mu_i$,
keeping track of all mixing terms at NNLO would require cumbersome numerical fitting which cannot be achieved with good accuracy. However, 
we anticipate that the beta-equilibrium for $N_f= 2+1$ leads to very small values of the electron chemical potential $\mu_e\ll \mu$. Therefore, we systematically neglect $\mathcal{O}(\frac{\mu_e}{\mu})$ corrections within $\mathcal{O}(\as^2)$ contributions. The mixing of masses is one order of magnitude larger, thus we account for all such contributions.}. 

The resulting NNLO pressure for generic quark masses $m +m_i$ (that
we refer to in the sequel
as $P^{N_f=2^*+1^*}$ to distinguish from standard $N_f=2+1$ pQCD pressure),
with quark chemical potential $\mu_i$,
may be formally written as
\ba\label{Ptotal}
&&P(m,m_i,\mu_i)= \nn
&& \sum_{i=1}^{N_f}\left( P_{\rm LO}^{m}(m+m_i,\mu_i)
+P_{\rm NLO}^{m}(m+m_i,\mu_i) +P_{\rm 2GI}^{N_f=2^*+1^*}(m+m_i,\mu_i) 
+P_{\rm VM}^{N_f=2^*+1^*}(m+m_i,\mu_i)\right) +P_{\rm Ring}^{N_f=3^*}\!(m) \nn
&& +\sum_{i=1}^{N_f}\left(P_{\rm LO}^{v}(m+m_i)
+P_{\rm NLO}^{v}(m+m_i) +\sum_{i=1}^{N_f-1}
\left(P^{v}_{\rm NNLO,1}(m)+P_{\rm sub,1}(m)\right) 
+P^{v}_{\rm NNLO,2}(m+m_s) +P^{v,nd}_{\rm NNLO}(m,m+m_s)\right)\nn
&& +P_{\rm sub,2}(m+m_s)  +P_{\rm sub}^{nd}(m,m+m_s)\,,
\ea
where the different expressions may be found in Ref. \cite{Fernandez:2024ilg} (see also Appendix \ref{Appendix:A} for a summary) and we briefly comment on these here. 
Note that up to NLO, 
the contributions factorize {\it per flavor}, while at NNLO 
non-diagonal terms occur, as indicated by the ``nd'' quantities in Eq. (\ref{Ptotal}),
due to the occurrence of
independent quark loops in Fig. \ref{fig:soft_graphs}.

Accordingly, the NNLO contributions are conveniently decomposed 
into four classes: 
\begin{enumerate}
\item The NNLO weak coupling expansion of the matter ($m$) contributions, 
$P^{Nf=2^*+1^*}_{\rm 2GI}(m_i,\mu_i)$ in Eq. (\ref{Ptotal}) corresponds to all two-gluon irreducible contributions in Fig. \ref{fig:hard_graphs}. 
\item $P_{\rm VM}^{N_f=2^*+1^*}(m_i,\mu_i)$ corresponds to mixed vacuum-matter (VM) contributions
as indicated in Fig. \ref{fig:soft_graphs}.
\item The plasmon sum ``ring'' contributions, that result after the all-order resummation
sketched in the rightmost graph of Fig. \ref{fig:soft_graphs}: 
\be
P_{\rm Ring}(m) = -\Omega_{\rm Ring}^{N_f=3^\star}(m) \,.
\ee
Note that the latter contribution
could not be handled analytically while maintaining an exact dependence on arbitrary masses and chemical potentials. However, this ${\cal O}(\alpha_s^2)$ contribution is numerically an order of magnitude smaller than the NNLO $P_{\rm 2GI+VM}^{N_f=2^*+1^*}(m,\mu)$. Therefore, we neglect any mixing altogether within this contribution, i.e. we take a universal quark mass
$m$ for the three flavors \cite{Fernandez:2024ilg} (the actual $m$ value being subsequently determined from the RGOPT prescription, as explained below).
Contributions 1)--3) were originally derived in Ref. \cite{Kurkela:2009gj} with exact quark mass dependence, and extended to quark flavors with different masses in Ref. \cite{Fernandez:2024ilg}.
\item
Finally, a fourth class of contributions, not present within the standard weak coupling expansion\cite{Kurkela:2009gj} and crucially involved in the RGOPT prescription, is to (re)introduce vacuum contributions to the pressure, $P^{v}_{\rm LO}-P^{v}_{\rm NNLO}$
in Eq.(\ref{Ptotal}), given by the $\mu=0$ contributions in Fig. \ref{fig:hard_graphs}
and ``VV'' contribution in Fig. \ref{fig:soft_graphs}.
Importantly, such vacuum contributions should be supplemented with zero-point ``subtraction'' contributions \cite{Kneur:2015uha,Kneur:2019tao} (represented in Eq. (\ref{Ptotal}) by $P_{\rm sub,i}$ terms) directly related to the vacuum energy anomalous dimension \cite{Chetyrkin:1994ex,Baikov:2018nzi}, $\propto m^4\,\mathbb{1}$, which mixes with the $m \bar q q$ operator (see also Appendix \ref{Appendix:B}). Often disregarded in the thermal literature on account of being medium-independent,  those vacuum contributions play a crucial role for RG properties of a massive theory, as they are necessary to restore perturbative RG invariance of a {\it massive} pressure (equivalently vacuum energy)\footnote{As indicated in Eq. (\ref{Ptotal}), both $P^{v}_{\rm NNLO}$ and
$P_{\rm sub,i}$ involve non-diagonal contributions at NNLO
 from the VV contributions in Fig. \ref{fig:soft_graphs}.}.
\end{enumerate}
Next, the actual RGOPT prescription is implemented by the following steps (see, e.g., Refs. \cite{Kneur:2013coa,Kneur:2019tao} for a review):
\begin{itemize}
\item First,
the NNLO pressure in Eq. (\ref{Ptotal}), 
being formally a function $P(m,m_i,g_s^2,\mu_i)$, is modified by the following
substitution (performed after standard coupling and quark mass renormalization):
\begin{equation}
\{m_i, m \} \to \{ m_i, \eta\left(1-\delta\right)^a \}, \;\;  g_s^2\to \delta g_s^2\,,
 \label{delta}
\end{equation}
where the exponent $a$ is specified just below and, as already emphasized, the actual physical masses considered here are 
$m_u=m_d=0, m_s \ne 0$.
Initially, $\eta$ is an arbitrary trial mass, thus distinct from the physical quark mass $m_i$. Eq. (\ref{delta}) is such that $0\le \delta \le 1$ interpolates between the massive (free) theory and the original theory respectively, 
with $\delta\,\eta$ treated as an interaction.
The pressure is then re-expanded in powers of $\delta$ at the relevant 
perturbative order in $g_s^2$ (here NNLO), and taking the limit $\delta\to 1$ afterward to recover
the original theory. 
Truncating the series at finite $\delta$-order in this way leaves a remnant $\eta$-dependence, fixed by RG properties as specified below. The latter prescription applied to Eq.(\ref{Ptotal}) gives formally an expression denoted 
$P^{N_f=2^*+1^*}_{\rm RGOPT}(\eta,g_s^2,m_i,\mu_i)$ below.
\item A crucial feature of RGOPT is that
the exponent $a$ in Eq. (\ref{delta}) is uniquely fixed by the (massless) RG equation at LO to the critical value \cite{Kneur:2013coa,Kneur:2019tao}
\begin{equation}\label{acrit}
a= \frac{\gamma_0}{2b_0}\;\; = \frac{4}{9} \Big|_{N_c=N_f=3} \,,
\end{equation}
where $b_0$ and $\gamma_0$ are the leading coefficients for, respectively, the beta function and the quark mass anomalous dimension (see Appendix B for notation).
This prescription remains consistent with standard renormalization and maintains perturbative RG invariance\footnote{Note that for $a=1$ Eq. (\ref{delta}) reproduces the more 
familiar ``added and subtracted'' mass term prescription typically adopted e.g. in SPT \cite{Andersen:2000yj} or HTLpt \cite{Andersen:2002ey,Haque:2012my}.} even {\it after} the $\delta$-expansion is performed.
\item Lastly, in related optimized perturbation (OPT) approaches (for instance, in SPT\cite{Karsch1997} and HTLpt at NLO\cite{Andersen:2002ey,Haque:2012my}) a standard prescription to fix the arbitrary mass parameter $\eta$ is a stationarity principle, which
in our present context takes the form:
\begin{equation}\label{eq:OPTequation}
   {\rm f}_{\rm OPT}(\eta, g_s,\cdots) \equiv \frac{\partial P^{N_f=2^*+1^*}_{\rm RGOPT}}{\partial \eta}\Big|_{\eta=\tilde{\eta}}=0\,.
\end{equation}
However, the number of solutions of Eq. \eqref{eq:OPTequation} increases with perturbative order
and the solutions are not necessarily real-valued due to nonlinear $\eta$-dependence. In contrast, the RGOPT prescription in Eq. \eqref{acrit} crucially guarantees that, at successive orders, a single solution $\eta$ satisfies the leading behavior of the parameters RG flow in the original (massless) theory \cite{Kneur:2013coa} (i.e. for QCD the solution satisfies asymptotic freedom). 
At the same time, to fix the trial mass parameter $\eta$ beyond LO it appears more compelling 
to use the RG equation \begin{equation}\label{eq:RGred}
   {\rm f}_{\rm RG}(\eta, g_s,\cdots) \equiv \left(\Lambda\frac{\partial}{\partial\Lambda}+\beta(g_s^2)\frac{\partial}{\partial g_s^2} \right) P^{N_f=2^*+1^*}_{\rm RGOPT}(\eta,g_s^2,\cdots)\Big|_{\eta=\tilde \eta}=0,
\end{equation}
with the beta-function $\beta(g_s^2)$ (see Appendix B),
instead of Eq. \eqref{eq:OPTequation} (here for the simplified situation of vanishing physical mass $m_i$). Eq. (\ref{eq:RGred}) produces a ``RG-dressed'' screening quark mass, $\eta(g_s^2,\mu)$, incorporating an all-order summation of certain topologies \cite{Kneur:2019tao,Fernandez:2024ilg}.
\end{itemize}
Incidentally, for the cold quark matter pressure at NNLO, only Eq. \eqref{eq:RGred} gives a solution which is perturbatively consistent
with a screening behavior \cite{Fernandez:2024ilg}, $\tilde \eta_{RG} = {\cal O}(g_s \mu)$ for $g_s\to 0$, in contrast to Eq. \eqref{eq:OPTequation}. Accordingly, our prescription is essentially uniquely defined at NNLO.
Yet, such a RG solution is not guaranteed to be real-valued, as mentioned previously.
Indeed, in the $\msbar$ scheme the above prescription for the quark matter pressure leads to a complex-valued $\tilde \eta$ already at NLO \cite{Kneur:2019tao}.
We stress that this is a nonphysical artifact of the nonlinearity in $\eta$ arising from insisting on an ``exact'' solution $\tilde\eta(g_s,\mu)$.
Alternatively, $\tilde \eta(g_s,\mu)$ may be re-expanded perturbatively,
which makes it always real-valued, but this generally loses \cite{Kneur:2019tao,Kneur:2020bph} 
part of the sought RG improvement from higher orders.
To preserve the RG resummation benefits while circumventing the nonreal-solution issue,
we rather exploit the renormalization scheme freedom to slightly shift perturbative coefficients from the original $\msbar$ ones, so as to restore a real-valued $\tilde \eta$ solution. Our 
renormalization scheme change (RSC) is defined perturbatively \cite{Kneur:2019tao} consistently {\it prior
to} the modifications from Eq. (\ref{delta}), using
\begin{equation}\label{eq:RSC}
    \left(m, m_s\right) \to \left[m (1+B_2 g_s^4), m_s(1+B_2 g_s^4)\right]\,.
\end{equation}
After the modification induced from Eq. (\ref{delta}) and NNLO $\delta$-expansion, 
the arbitrary RSC parameter $B_2$ is uniquely fixed upon requiring the minimal departure from $\msbar$: it can be shown that this amounts to solve for $B_2$ 
and the trial mass $\eta$ simultaneously the RG Eq. (\ref{eq:RGred})
and the constraint \cite{Kneur:2013coa,Kneur:2019tao} 
\begin{equation}\label{eq:determinant}
     \rm det(f_{RG},f_{OPT})_{(g_s^2,\eta)} = \left(\frac{\partial f_{\rm RG}}{\partial g_s^2}\frac{\partial f_{\rm OPT}}{\partial \eta}- \frac{\partial f_{\rm OPT}}{\partial g_s^2} \frac{\partial f_{\rm RG}}{\partial \eta}\right)\Big|_{\eta=\tilde \eta,\ B_2=\tilde{B}_2}=0 \,.
\end{equation}
Actually, when accounting for non-zero physical strange quark mass, we modify
the RG operator, Eq. (\ref{eq:RGred}), as
\be\label{eq:RGredms}
\Lambda\frac{d}{d\Lambda}=\Lambda\frac{\partial}{\partial\Lambda}+\beta(g_s^2)\frac{\partial}{\partial g_s^2}-\mcm \gamma_{\mcm}(g_s^2) \frac{\partial}{\partial\mcm} \,,
\ee
with $\gamma_m(g_s^2)$ the anomalous quark mass dimension (see Appendix B). Note that the genuine physical mass $m_s$ thus remains 
unaffected by the $\delta$-expansion. \\
Applying the above RGOPT prescription 
to the cold dense $N_f=2^*+1^*$ NNLO pressure Eq. (\ref{Ptotal}) with respective flavor masses 
$(m,m,m+m_s)$, we obtain the doublet of real solutions ($\tilde\eta$,$\tilde{B}_2$)
and the RG-dressed mass spectrum $(\tilde\eta, \tilde\eta, \tilde\eta+m_s)$ respectively for $u,d,s$ quarks. Accordingly, 
the resulting NNLO RGOPT pressure reads formally 
\be\label{Prgopt_nnlo}
P_{\rm RGOPT} \equiv P^{N_f=2^*+1^*}_{\rm RGOPT}(\tilde \eta, g_s^2,
\tilde B_2,\mu_i,m_s) \,,
\ee
where we simply indicate its final dependence on relevant parameters since its explicit expression is quite involved \cite{Fernandez:2024ilg}. 
With the aim of providing a simpler procedure for further applications, it is more convenient to construct a sufficiently accurate approximation through an appropriate fitting procedure, as described in the Section below.
\section{Thermodynamically consistent RGOPT pressure in $\beta$-equilibrium}\label{sec:thermo_consis}
This Section details the construction of the $\beta$-equilibrated and thermodynamically consistent RGOPT pressure entering the EoS employed in the compact star applications of Section \ref{sec:results}. Starting from the pressure in Eq. (\ref{Prgopt_nnlo}) (for symmetric matter) obtained upon applying the RGOPT prescription 
of Sec. \ref{Pr}, chemical equilibrium and charge neutrality are enforced by requiring the following  relations to be satisfied
\be\label{eq:BetaEq}
 \mu_u=\mu_d-\mu_e, \ \ \ \ \mu_s=\mu_d\,,
\ee
\be\label{eq:charge_neutrality}
{\rm f}_{\rm CN}\equiv\frac{2}{3}\rho_{\rm up}-\frac{1}{3}\rho_{\rm down}-\frac{1}{3}\rho_{\rm strange}-\rho_{\rm electron}=0\,,
\ee
where $\rho_i= \frac{\partial P_{\rm RGOPT}}{\partial\mu_i} |_{\Lambda}$ represents the number density for the $i$-th particle while $\mu_e$ represents the electron chemical potential. In principle, thermodynamic quantities such as pressure and number density receive electronic contributions $\sim \mu_e^4$, and $\sim \mu_e^3$, respectively. Explicit checks confirmed that $\mu_e \ll \mu$; therefore, these contributions are negligible, and we have subsequently omitted them from our analysis. Notice also that the determinant in Eq. (\ref{eq:determinant}) for the symmetric matter case 
is further generalized to incorporate the additional physical constraint of
charge neutrality or, equivalently, positive $\mu_e$. This is accounted for by a simple trick: substituting $\mu_e\equiv\mu_A^2$ and enforcing the real-valuedness of $\mu_A$ by extending Eq.~(\ref{eq:determinant}) to: 
\begin{equation}\label{eq:GeneralizedDeterminant}
    \rm det(\rm f_{RG},f_{OPT},f_{CN})_{(g_s^2,\eta,\mu_A)}=0\,,
\end{equation}
where ``$\rm f_{CN}$'' stands for the charge neutrality Eq. (\ref{eq:charge_neutrality}).

Note that, in the resulting pressure, all parameters (except $\mu_i$) 
exhibit a highly non-trivial dependence on the renormalization scale $\Lambda$: manifestly so for
$\alpha_S(\Lambda),~ m_s(\Lambda)$, dictated by RG running, but 
equally for the additional parameters $\tilde \eta, \tilde B_2$ that entail
a non-trivial $\Lambda$ dependence from the RGOPT prescription.  Thermodynamic consistency is therefore spoiled, since the relation $d P_{\rm RGOPT}/d\mu_i = \rho_i$, does not hold \cite{Gorenstein:1995vm,Biro:2001ug,Kurkela:2009gj,Lenzi:2010mz,Restrepo:2022wqn}. 
Nevertheless, one can obtain a thermodynamically consistent (TC) pressure
by considering 
\be\label{Pth}
 P_{\rm RGOPT}^{TC}(...,\mu_i) \equiv  P_{\rm RGOPT}(...,\mu_i) + \int_{\mu_0}^{\mu_i} 
 d\mu \left( \frac{\partial}{\partial_\mu} P_{\rm RGOPT}(\mu) - \frac{d}{d\mu} P_{\rm RGOPT}(\mu) \right) \,,
 \ee
 where $\mu_0$ is specified below. Note that the additional term in Eq. (\ref{Pth}) is formally of higher perturbative order: 
by construction, our NNLO $P_{\rm RGOPT}$ is RG invariant up to neglected higher order, 
i.e. $ d P_{\rm RGOPT}/d\Lambda = {\cal O}(\alpha_s^3)$, 
then from $(d/d\mu) P_{\rm RGOPT}  = 
\partial_\mu P_{\rm RGOPT} + (d\Lambda/d\mu) d P_{\rm RGOPT}/d\Lambda) $, 
the integrand above is $ (d\Lambda/d\mu)\times {\cal O}(\alpha_s^3)$,
and conventionally, one chooses a scale $\Lambda = {\cal O}(\mu)$ 
within a certain range, so that   $ (d\Lambda/d\mu)$ is just a number.
Adopting a thermodynamically consistent prescription for the compact star EoS is nonetheless essential, not only in general, but also within our construction, where Eq. (\ref{Prgopt_nnlo})
(and its generalization for non-symmetric matter)
captures
a non-trivial dependence on higher orders. The initial value $\mu_0$ in Eq. (\ref{Pth}) is chosen such that the total quark number density vanishes, $\rho_{tot}(\mu_0)=0$.\\
Thus, by solving simultaneously Eqs. (\ref{eq:GeneralizedDeterminant}) and (\ref{eq:RGredms}), we obtain in this case the triplet of solutions $(\tilde \eta,\tilde B_2, \tilde{\mu}_A$) 
(hence $\tilde{\mu}_e^2$) leading to a fully determined pressure in Eq. (\ref{Prgopt_nnlo}) for a given quark chemical potential $\mu_i$ and renormalization scale $\Lambda=X \mu_B/3$.

\subsection{Compact formula\label{cformula}}

In order to reproduce more simply the quite involved NNLO RGOPT results, in this subsection we present fitting functions for the thermodynamically consistent NNLO RGOPT pressure given in
Eq. (\ref{Pth}), which are solely functions of the scale parameter $X=
3\Lambda/\mu_B$ and the baryon chemical potential $\mu_B$. In
practice, we provide a rather accurate fit formula for
$2\le X \le 5$, with $3\le X\le 5 $ corresponding to a range essentially
relevant for QS while $2\le X\le 3$ represents the relevant range for hybrid
stars. In general, values of $X<2$ tend to match typical hadronic EoSs at very
large $\mu_B$. Consequently, the transition to quark matter occurs at chemical
potential values that lie beyond the stability limit. Therefore, by not
producing a quark core, such configurations can be safely neglected in the
present application.
\subsubsection{Current quark mass and QCD coupling input}
In our numerical applications, for the running coupling $g_s^2(\Lambda)\equiv 4\pi\as(\Lambda)$, 
we use the exact NLO result obtained for a given renormalization 
scale $\Lambda$ upon solving $g_s^2$ exactly from
\begin{equation}\label{eq:RunningNNLO}
\Lambda_{\rm \overline{MS}}= \Lambda e^{-\frac{1}{2b_0 g_s^2}}\left(\frac{b_0 g_s^2}{1+\frac{b_1}{b_0} g_s^2} \right)^{-\frac{b_1}{2b_0^2}} \,,\ 
\end{equation}
and fixing $\Lambda_{\rm \overline{MS}}=330\ $MeV \cite{ParticleDataGroup:2018ovx} so that $\alpha_s(\Lambda = 1.5 \, {\rm GeV})\simeq 0.326$ \cite{Bazavov:2012ka}.
Since the expected RG scale dependence cancellations always
occur between running coupling (or masses) at order $\alpha_s^k$, and explicit $\ln \Lambda$
dependence at order $\alpha_s^{k+1}$, 
the NLO running is sufficient in principle when considering an NNLO pressure. 
Using higher order running coupling would hardly give any visible 
differences in our results below for the relevant range of parameters considered.

We recall that, in all applications below, we approximate the 
current masses $m_u, m_d$ to zero, as usual.
Concerning the strange quark, since mass renormalization is only needed at NLO 
${\cal O}(\alpha_s^2)$ in our case,
we use the NLO running mass, given in our normalization as:
\be\label{msrun}
\mcm(\Lambda)=m_s(\Lambda_0)\(\frac{g_s^2(\Lambda)}{g_s^2(\Lambda_0)}\)^{\frac{\gamma_0}{2b_0}}\(\frac{1+\frac{b_1}{b_0}\,g_s^2(\Lambda)}{1+\frac{b_1}{b_0}\,g_s^2 (\Lambda_0)}\)^{\frac{\gamma_1}{2b_1}-\frac{\gamma_0}{2b_0}} \,,
\ee
where $m_s(\Lambda_0=\mathrm{2 GeV}) \simeq 93.5$ MeV \cite{ParticleDataGroup:2018ovx}.
Similarly to the running coupling, accounting for higher orders in the running quark mass has no significant impact on our results. \\

\subsubsection{Fit for the thermodynamically consistent pressure}
We define the following function $\mu_{B,0}^{\rm th}(X)$,
($\Lambda=X \mu_B/3$), which represents the threshold $\mu_B$ values for the
vanishing of the baryon density at different $X$ values. This quantity is given
by
\begin{equation}\label{eq:muB0}
    \begin{aligned}
        \mu_{B,0}^{\rm th}(X)=&0.1089400 + \frac{1.3335988}{X^{0.7209811}} .
    \end{aligned}
\end{equation}
For $\mu_B=\mu_{B,0}^{\rm th}(X)$, when $\rho_B(\mu^{\rm th}_{B,0}(X),X)\equiv 0$,   the pressure reads
\begin{equation}\label{eq:Pth}
    P^{\rm th}(X)= -0.0000912 - \frac{0.9840214}{X^{10}} + \frac{0.3604114}{X^8} - \frac{0.0373881}{X^6} + 9.6\times 10^{-6} X \,.
\end{equation}

To ensure the consistency of the thermodynamic relation between the pressure and
baryon density, $\partial_{\mub} P\equiv \rho_B$, both quantities have been
fitted together, yielding
\begin{equation}\label{eq:Fit}
\begin{aligned}
    \frac{P^{\rm TC}_{\rm RGOPT}}{\rm GeV^4}\equiv\ &P_{\rm fit}(\Tilde{\mu}_B=\frac{\mub}{\rm GeV},X)= P^{\rm th}(X)+(\Tilde{\mu}_B -\mu_{B,0}^{\rm th}(X))^{1.8792061} \left(\frac{a_1\ X^{a_2}}{1 + a_3\ \Tilde{\mu}_B^{a_4} X^{a_5}}\right)\\
    &\hspace{-2cm}+ (\Tilde{\mu}_B -\mu_{B,0}^{\rm th}(X))^{2.6809433} \left(\frac{a_6\ X^{a_{7}}}{1 + a_{8}\ \mub^{a_{9}} X^{a_{10}}}\right) + (\Tilde{\mu}_B -\mu_{B,0}^{\rm th}(X))^4 \left(a_{11}+\frac{a_{12}\ X^{a_{13}}}{1 + a_{14}\ \Tilde{\mu}_B^{a_{15}} X^{a_{16}}}\right) ,\\
    \end{aligned}
\end{equation}
with the coefficient values $a_i$ given in Table \ref{tableCoeff}. The validity
range of this fit is given by $\mu_{B,0}^{\rm th} \le \mub \le 2.5 {\rm GeV}$
and $2\le X \le 5$.
The baryonic density is easily obtained from Eq. (\ref{eq:Fit}) by taking the
explicit derivative with respect to $\tilde{\mu}_B$, all other parameters or
functions being independent of $\tilde{\mu}_B$. The relative
Root Mean Square Error (RMSE)\footnote{Since both $P$ and
$\rho_B$ can reach zero values, the standard RMSE cannot be appropriately
defined. Therefore, we adopt the following definition ${\rm RMSE}=\frac{1}{{\rm
Max}[y_i]-{\rm Min}[y_i]}\sqrt{\frac{1}{N}\left(f(y_i)-y_i\right)^2}$ for
$\{y_i\}_{i=1,\dots,N}$ data.} for this fit is  0.13\% for the
pressure, while for the baryonic density, it is 0.37\%.

 \begin{table}[!h]
\caption{\label{tabCoeff} Coefficients of the fit.}
\begin{center}
\begin{tabular}{||c c | c c | c c | c c ||} 
 \hline
 Coeff. & Value & Coeff. & Value & Coeff. & Value & Coeff. & Value \\ [0.5ex] 
 \hline\hline
 $a_{1}$ &  0.0256667 & $a_{5}$ & 0.6481308 & $a_{9}$ & -0.0192695 & $a_{13}$ &  1.1709795 \\
 $a_{2}$ & -2.6387134 & $a_{6}$ & -0.0560705 & $a_{10}$ & 0.1401116 & $a_{14}$ & 5.5981347 \\
 $a_{3}$ & 0.2549286 & $a_{7}$ & -0.6666632 & $a_{11}$ & 0.0009140 & $a_{15}$ & 7.7419492 \\
 $a_{4}$ & 1.8563910 & $a_{8}$ & -7.7277191 & $a_{12}$ & 11.767561 & $a_{16}$ & 9.6744514 \\[1ex] 
 \hline
\end{tabular}
\end{center}

\label{tableCoeff}
\end{table}
Finally, the energy density can be written as
\begin{equation}
    \frac{\mathcal{E}_{\rm RGOPT}}{\rm GeV^4}\equiv \mathcal{E}_{\rm fit}(\Tilde{\mu}_B,X)=\ - P_{\rm fit}(\Tilde{\mu}_B,X)+\mub\, \rho_B \,.
\end{equation}
We emphasize that in those fitted formulas all the highly nontrivial dependence on the running coupling
$\alpha_S(X \mu_B)$ and masses $\tilde m_s(X \mu_B)$, $\tilde \eta(X \mu_B)$, 
accounted for in our actual calculation through Eqs. (\ref{eq:RunningNNLO}), (\ref{msrun}),
(\ref{eq:RGredms}), (\ref{eq:determinant}), is hidden and consistently embedded in the fitted $X$-dependence.

If needed, the above pocket formulas may be compared (or normalized to) the standard
expressions for a gas of non-interacting massless quarks given in the standard Fermi-Dirac limit for the pressure
\begin{equation}
    P_{\rm FD} = \frac{N_c N_f}{12 \pi^2}\left ( \frac{\mu_B}{3} \right )^4 \,.
\end{equation}
Within this limit,  the corresponding baryonic density is just 
\begin{equation}
    \rho_{B,\rm FD} =\frac{N_c N_f}{9 \pi^2}\left ( \frac{\mu_B}{3} \right )^3\,.
\end{equation}
\section{Numerical Results}\label{sec:results}

In this Section, we present the numerical results for quark and hybrid stars. To
describe quark matter, we compare two approaches, the conventional pQCD  and the
RGOPT resummation, both at NNLO. On the other hand, to describe hadronic matter we use
relativistic mean field (RMF) models with the DD2 parametrization of Ref.
\cite{Typel:2009sy}, which produces a stiff EoS, and two soft EoSs. Namely,  the SFHo
parameterization of Ref.~\cite{Steiner:2012rk} and the DIDY parametrization of Ref. ~\cite{Frohaug:2025okz},
which considers hyperons and isospin-dependent couplings.

The RGOPT results are obtained using the fitting formula of Section \ref{sec:thermo_consis}, while for pQCD we use the fitting function of Ref. \cite{Fraga:2013qra}, which reproduces the thermodynamically consistent pressure of Ref. \cite{Kurkela:2009gj}.

\subsection{Quark stars}
\begin{figure}[t!]
    \centering
    \includegraphics[width=0.7\linewidth]{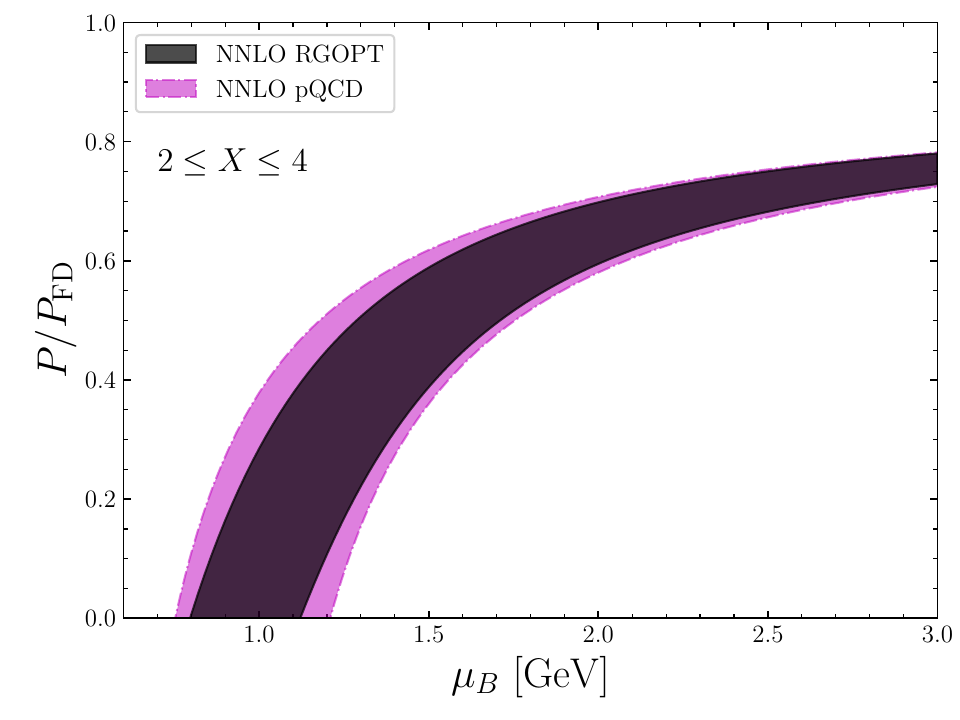}
    \caption{ Normalized pressure as a function of the baryon chemical potential
$\mu_B$ obtained with the NNLO RGOPT (gray band) and  with the
NNLO pQCD (magenta band). The lower pressure values correspond to $X=2$, while
the higher values correspond to $X=4$. }
    \label{fig:P_bands}
\end{figure}
Here, we present results for strange ($N_f=2+1$) QSs, comparing the RGOPT and pQCD predictions. 
In Fig. \ref{fig:P_bands}, we show the normalized pressure as a function of the baryon chemical potential for $2\le X \le 4$. When comparing the two approximations for the same $X$-range, it is evident that the NNLO RGOPT exhibits substantially  reduced scale dependence compared to 
NNLO pQCD. Notice that both methods yield very similar results at high $\mu_B$ values.
\begin{figure}[t!]
    \begin{subfigure}
     \centering
     \includegraphics[width=.45\linewidth]{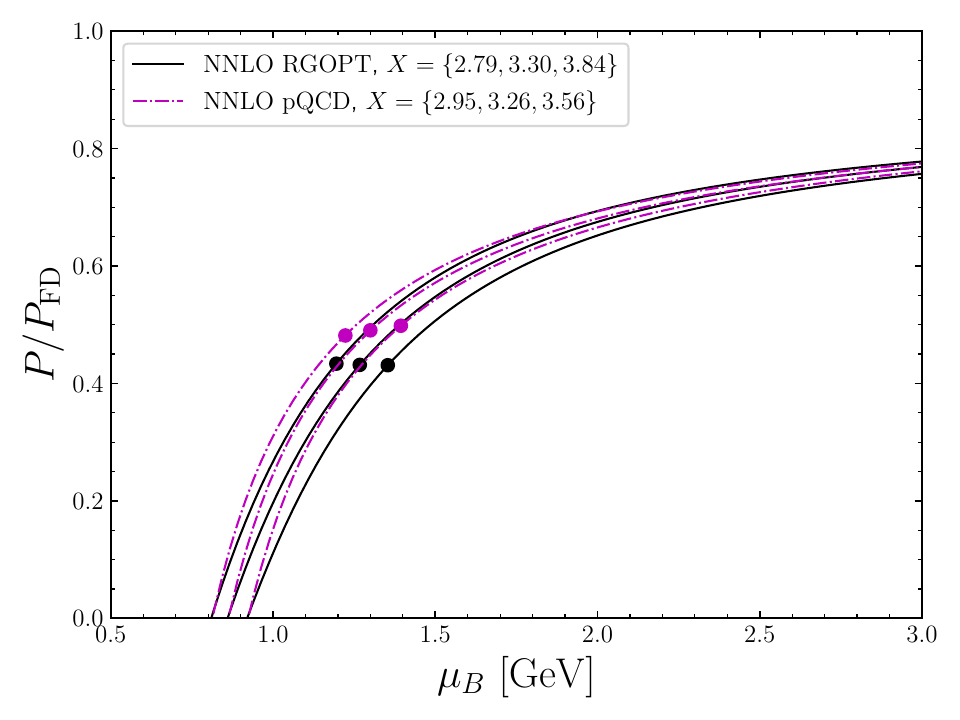}
    \end{subfigure}
    \begin{subfigure}
    \centering
    \includegraphics[width=.45\linewidth]{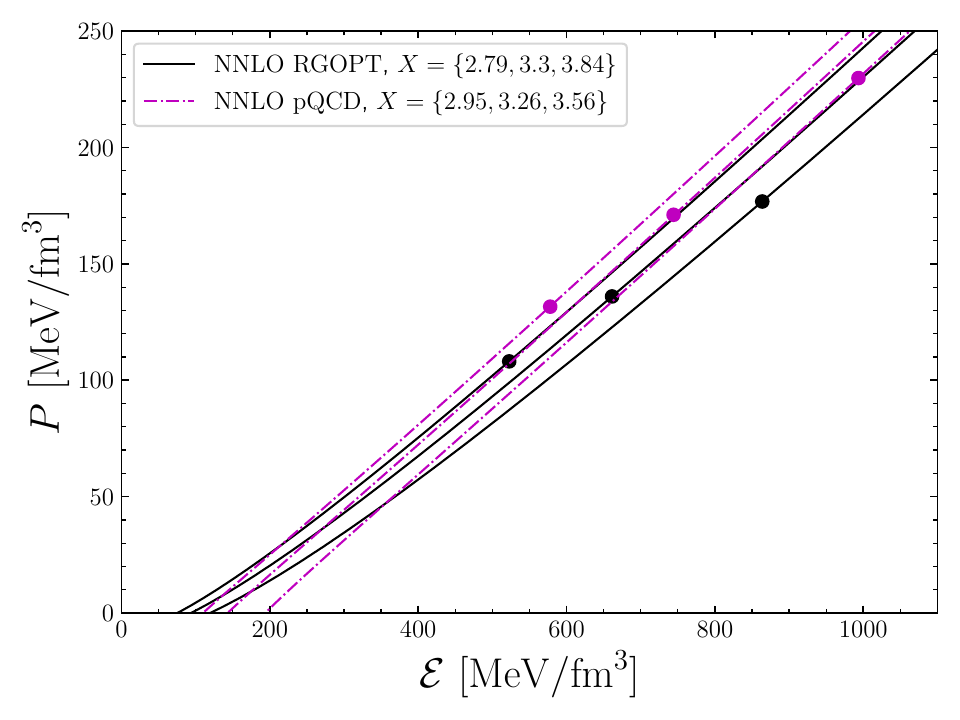}
    \end{subfigure}
\caption{ Pressure versus baryon chemical potential (left) and energy density (right) of $\beta$-equilibrated matter for the NNLO RGOPT (continuous lines),  and for NNLO pQCD (dotted lines). 
The curves were obtained with the scale values $X$ that reproduce $M_{\rm{max}}
= 2, 2.3$ and $2.6 M_\odot$, given in Table \ref{tab:tab_quarks}, respectively,
2.79, 3.30, 3.84 (NNLO RGOPT);  2.95,3.26, 3.56 (NNLO pQCD).
The upper lines corresponds to higher $X$ values.
The dots identify the values at the center of the maximum mass star configuration. }
\label{fig:P_quark_stars}
\end{figure}

Figure \ref{fig:P_quark_stars} shows the normalized pressure as a function of the baryon chemical potential and the pressure as a function of the energy density for the values of $X$  that reproduce  maximum QS  masses, $M_{\rm max}\,=2$, $2.3$ and $2.6\, M_\odot$.
The dots correspond to the values at the center for the maximum QS mass configuration.
The values of $X$ for each approximation, together with several QS properties, are given in Table \ref{tab:tab_quarks}.

\begin{table}[t!]
\caption{QS properties predicted by NNLO $\rm RGOPT \ $ with  $N_f=2+1$, and by  NNLO pQCD with $N_f=2+1$ with the $X$ scale values that reproduce  $M_{\rm max}=2,\, 2.3$ and $2.6 M_\odot$. The considered properties  are: maximum mass $M_{\rm max}$ and corresponding radius $R_{\rm max}$,  radius  of the 1.4$M_\odot$ and 0.77$M_\odot$ stars $R_{1.4}$ and $R_{0.77}$, central baryon density $\rho_B^{c,\rm max}$ and central baryon chemical potential $\mu_B^c$, 
as well as surface baryon density, baryon chemical potential and corresponding coupling constant at the surface of the maximum mass configuration, $\alpha_s^{\rm surf}=\alpha_s(X\frac{\mu_B^{\rm surf}}{3})$. 
}

\begin{center}
\begin{tabular}{c  |c  c  c  c c c c c c c} 
       \hline
       
    &  $X$ & $M_{\rm max}$ &  $R_{\rm max}$  & $R_{1.4}$ & $R_{0.77}$  & $\rho_B^{c,\rm max}$  & $\mu_B^{c,\rm max}$   & $\rho_B^{\rm surf, max}$ & $\mu_B^{\rm surf, max}$ & $\alpha_s^{\rm surf}$\\
    &  & [$M_\odot$] &  [km] & [km] & [km] & ($\rho_0 $)  & [GeV]   & [$\rho_0 $]  & [GeV] &\\
	\hline
    \multirow{2}{4.5em}{$\rm{NNLO}$ $\rm RGOPT \ $} & $2.79$ & $2.00 $ &  $12.5$ & $13.0$ & $11.3$& $4.80 $ & $1.354 $ & $0.81 $ &$0.920$ & $0.480$\\ 
  &$3.30$& $2.30 $ & $ 14.3$ &$14.4$ & $12.3$& $3.93 $ & $1.267$ & $0.68 $ & $0.860$ & $0.441$   \\
     &$3.84 $& $2.60 $ &$16.1$ & $ 15.7$ &$13.4$ & $3.30 $ & $1.195$ & $0.58 $ & $0.809 $ & $0.412$\\
     \hline
    \multirow{2}{4.5em}{$\rm{NNLO}$ $\rm pQCD \ $} & $2.95$ & $2.00 $ &  $11.2$ & $11.6$ & $9.85$& $5.48 $ & $1.394 $ & $1.31 $ & $0.923$ & $0.456$ \\ 
  &$3.26$& $2.30 $ & $ 13.3$ &$12.9$ & $11.0$& $4.40 $ & $1.300$ & $1.04 $ & $0.863$ & $0.444$ \\
     &$3.56 $& $2.60 $ &$15.1$ & $ 14.2$ &$12.0$ & $3.62 $ & $1.223$ & $0.84 $ & $0.813$ & $0.434$ \\
	\hline
\end{tabular}\label{tab:tab_quarks}
\end{center}
\end{table}
The first point to note from the table is that the RGOPT $X$ range matching
$2.0M_\odot\le M_{max}\le 2.60M_\odot$ is wider than the one of pQCD, as
expected since the former method is more resilient to scale variations.
At the same time, the values of the couplings on the surface $\alpha_s^{\rm surf}$, predicted by both approximations, are comparable and lie near the upper limit for which perturbative calculations are generally considered reliable, namely $\alpha_s(\Lambda\sim1 \rm{GeV})\sim 0.423$ \cite{Deur:2016tte}.

Regarding the scale dependence, it is interesting to note, from Table \ref {tab:tab_quarks}, that  for the  $\Delta M_{\rm max}=(2.60-2.00) M_\odot$ range considered here, the maximum masses have an almost linear dependence on $X$ with angular coefficient $\Delta M_{\rm max}/\Delta X \simeq 0.588 \,M_\odot$ per unit of $X$ for  RGOPT and $\simeq 0.983 \,M_\odot $ per unit of $X$ for pQCD, thus showing that the former is more stable to scale variations.  It is also worth pointing out that the compactness ${\cal C}_{\rm max} = M_{\rm max}/R_{\rm max}$ 
is approximately constant in both cases, namely, ${\cal C}_{\rm max} \simeq 0.24$
for the RGOPT  and ${\cal C}_{\rm max} \simeq 0.25$ 
for pQCD \footnote{We adopt geometric units, where G=c=1, so that $M_\odot=1.477
$ km.}. This result could be anticipated by noticing that the
density at the surface and at the center is lower in the case of the RGOPT as
can be seen in Table \ref{tab:tab_quarks}.

Finally, it is equally important to note  that at NNLO both approximations are consistent with the Bodmer-Witten hypothesis for stable strange matter \cite{Bodmer:1971we,Teraza1979,Witten:1984rs}, since the energy per baryon of bulk strange quark matter at zero pressure is found to be below $930$ MeV. 
\begin{figure}[t!]
    \includegraphics[width=0.7\linewidth]{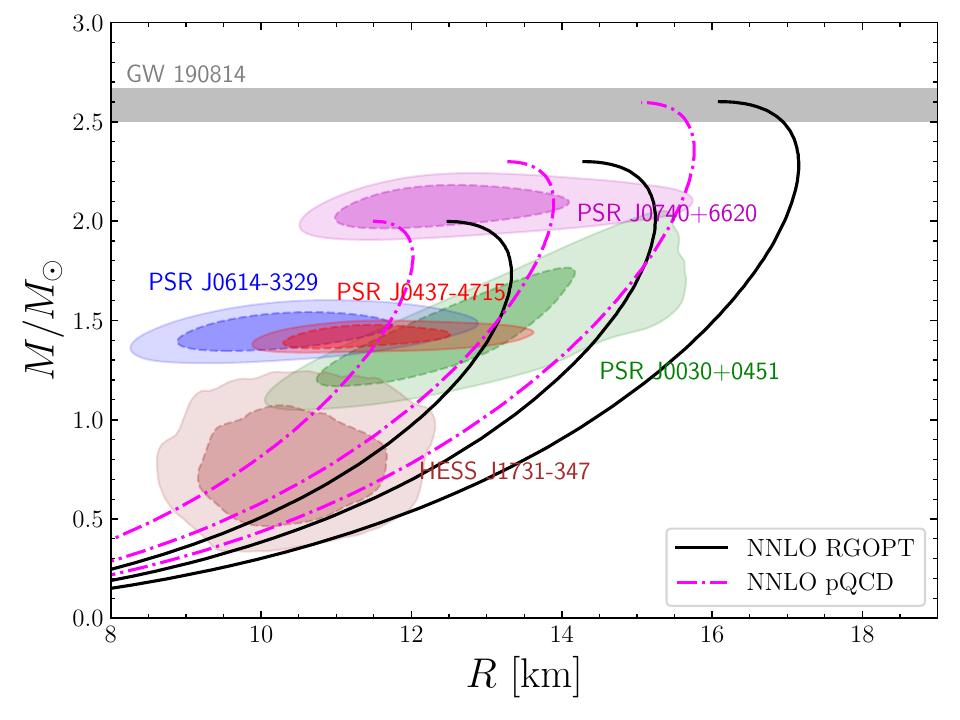}
    \caption{ Mass-radius relation of QSs obtained with NNLO pQCD (dash-dotted lines) and RGOPT (continuous lines).
    Lines reproducing maximum QS masses of $2.6M_\odot$ correspond to the higher values of $X$ of Table \ref{tab:tab_quarks}, while the ones for $2M_\odot$ correspond to the lowest $X$ values. 
    Also included are the NICER data for pulsars PSR J0740+6620
\cite{Salmi:2024aum,Riley:2021pdl, Miller:2021qha}, PSR J0614-3329
\cite{Mauviard:2025dmd}, PSR J0437-4715 \cite{Choudhury:2024xbk} and the most
recent analysis of NICER data for pulsars PSR J0030+0451 \cite{Kini:2026rjx}, as
well as the low mass compact star in the supernova remnant HESS
J1731-347 \cite{Doroshenko:2022nwp}. In particular, the ellipses represent the
68\% (dashed) and 95\% (full) confidence interval of the 2-D distribution in the
mass-radii domain, while the error bars give the 1-D marginalized posterior
distribution for the same data. A band identifying the mass of the low mass
compact object associated with GW190814 \cite{LIGOScientific:2020zkf} has also
been included. }
    \label{fig:QS}
\end{figure}

In Fig. \ref{fig:QS} we show the mass-radius relations obtained by solving the Tolman-Oppenheimer-Volkoff equations \cite{Tolman:1939jz,Oppenheimer:1939ne}, using  the EoSs of Fig. \ref{fig:P_quark_stars}.
The figure also displays observational constraints on the masses and radii of pulsars PSR J0740+6620 \cite{Riley:2021pdl,Miller:2021qha,Salmi:2024aum}, PSR J0614-3329 \cite{Mauviard:2025dmd}, PSR J0437-4715 \cite{Choudhury:2024xbk} and the most recent analysis of NICER data for pulsars PSR J0030+0451 \cite{Kini:2026rjx}.
Also shown are the compact object in the supernova remnant HESS
J1731-347 \cite{Doroshenko:2022nwp} and the band corresponding to
the low mass component of the event GW190814 \cite{LIGOScientific:2020zkf}.
In general, both approaches respect the observational constraints, with pQCD predicting the most compact QSs. 
Considering the mass-radius curves that produce $M_{\rm max}=2M_\odot$,
pQCD  predicts results which lie within the overlap region of the
$\approx 1.4M_{\odot}$ pulsars PSR J0614-3329, PSR J0437-4715 and  PSR
J0030+0451, while the RGOPT is out of the PSR J0614-3329 constraint.
For the curves yielding $M_{\rm max}=2.3M_\odot$,  all results fall outside the constraint from pulsar PSR J0614-3329. However, the pQCD curve remains within the overlap region of the  PSR J0614-3329, and PSR J0437-4715 constraints, while the  RGOPT curve is consistent only with the $95\%$ confidence region of PSR J0030+0451.
Finally, for $M_{\rm max}=2.6\, M_\odot$, the pQCD EoS still predicts
configurations consistent with the constraints of compact object
inside HESS J1731-347 and PSR J0030+0451. By contrast, the  RGOPT predictions
fall outside the PSR J0030+0451 constraints and are almost on the
boundary of the 95\% confidence region of HESS J1731-347.
\subsection{Hybrid stars}

In this Section, we discuss some properties of hybrid stars with a quark core described by the RGOPT, comparing our results with  predictions from pQCD.  To investigate the hadron-quark phase transition, a Maxwell construction is applied. For  the  hadron phase, three  different EoSs have been chosen: a soft nucleonic EoS (SFHo \cite{Steiner:2012rk}), a soft hyperonic EoS (DIDY \cite{Frohaug:2025okz}) and a stiff nucleonic EoS (DD2 \cite{Typel:2009sy}). These choices  span approximately the range of phase space that is consistent with the presently known constraints on the nuclear EoS, and also consider the possible onset of hyperons.
\begin{figure}[t!]
    \centering

\includegraphics[width=0.9\linewidth]{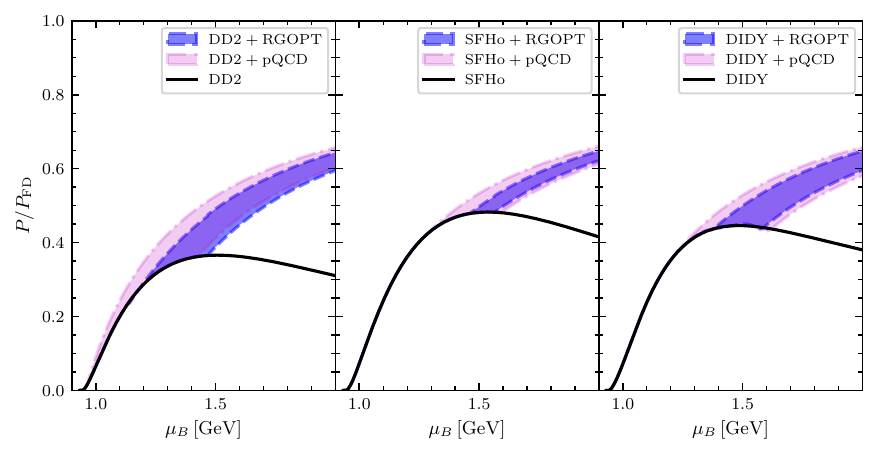}
    
\caption{Normalized pressure, $P/P_{\rm FD}$, as a function of baryon chemical potential for the hybrid EoS. Quark matter is described by the NNLO RGOPT and by the NNLO pQCD EoSs, while the hadronic matter is described by  DD2  (left), SFHo (middle) and DIDY (right) EoSs. The bands in the RGOPT/pQCD pressure indicate the variation in the scale for the values of $X$ that produces a quark core as indicated in Table \ref{tab:hybrid}, where the higher pressure corresponds to higher scale parameter $X$. }
    \label{fig:P_hybrid}
\end{figure}
For each hadronic model, we build two EoSs: i)  the star for which quark matter has started to nucleate at its center, corresponding to the smallest scale $X$ considered.  These stellar objects define the most massive hybrid stars that already contain a quark core, although quite small; ii) the hybrid star 
with the largest quark core allowed by the pair of EoSs while
satisfying the stability requirement for hybrid stars, $E(P_{\rm
quark}=0)/A>930$ . This condition is satisfied if the RGOPT and pQCD scales are
respectively given by $X<2.72$ and   $X<2.92$.
To define these two scenarios, the scale parameter $X$ for RGOPT(pQCD) increases
from 2.00(2.17) to 2.63(2.80) for DD2,  from
2.28(2.25) to 2.71(2.86) for SFHo, and from 2.01(2.00) to
2.71(2.82) for DIDY (see also Table \ref{tab:hybrid}).  The two
limiting  scale values  define the bands shown in Fig.  \ref{fig:P_hybrid},
where the  normalized pressure, $P/P_{\rm FD}$, for RGOPT and pQCD is plotted as
 a function of the baryon chemical potential.   The highest pressure corresponds
to the largest scale parameter value.

\begin{figure}[!h]
    \centering

\includegraphics[width=0.9\linewidth]{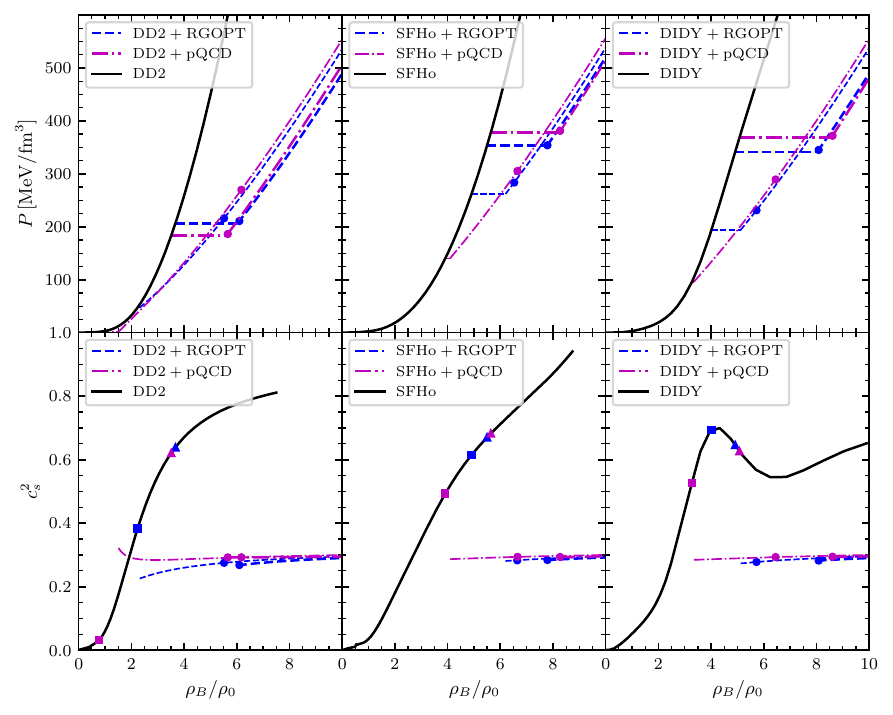
}
\caption[long]{ Pressure (top) and speed of sound (bottom) as functions of the
normalized density $\rho_B/\rho_0$. Quark matter is described by the NNLO RGOPT
(dashed lines) and by the NNLO pQCD (dash-dotted lines) EoSs for the same values
of the scale parameter $X$ as in Fig.~\ref{fig:P_hybrid}. {Thin lines represent
the largest $X$ value, while thick lines represent the smallest one.} Hadronic
matter is described by  DD2  (left),  SFHo (center) and DIDY (right) EoSs. The
dots identify the values at the center of the maximum mass star configuration,
the squares represent the values at the transition from hadron to quark matter
for the highest $X$, and the triangles for the lowest $X$ values. For the lowest
$X$ value that produces a finite but quite small quark core, the dot almost
coincides with the value of the quark pressure at the transition.
}\label{fig:vs_vsrho}
\end{figure}
 Despite being  an approximation, the Maxwell construction used to build the hadron-quark phase transition is known to produce reasonable results if the surface tension of hadronic matter in  quark matter is large \cite{Maruyama:2007ey} {although the actual value  of the surface tension remains unsettled, see also \cite{Pinto:2012aq}. } 
 We have plotted the hybrid EoS and the corresponding speed of sound squared in Fig. \ref{fig:vs_vsrho} for the three hadronic models, respectively, in the top and bottom panels. 
In this Figure, the dots identify the values at the center of the maximum mass
star configuration. For DD2, the density at the center lies within the range
$\sim$ $5.5-6.2\, \rho_0$ while for SFHo and DIDY the range spans
$5.7-8.3\, \rho_0$.
The phase transition for the largest $X$ (thin lines) occurs at  $\sim 2\,
\rho_0$ for DD2, $\sim 4\, \rho_0 $ for SFHo, and $\sim 5\, \rho_0$
for DIDY.
The strength of the phase transition, defined by the magnitude of the energy
density jump, is quite weak for pQCD, and the RGOPT presents a
visible stronger transition for SFHo and DIDY.
For the smallest scale parameter value (thick lines), the phase transition occurs at densities of the order of $4-6\, \rho_0$, the smallest value corresponding to  DD2. This feature, which can  be clearly seen in the bottom panels of  Fig. \ref{fig:vs_vsrho}, corresponds to the densities where the speed of sound jumps from the hadron branch to the quark branch. 
At these particular scales, the baryon density jump between the two phases is large, of the order or larger than $2\, \rho_0$. Notice that, for DIDY, the phase transition with the largest $X$ value occurs before the hyperon onset in the hadron branch, while for the {smallest} $X$ value, the transition occurs after the hyperon onset. 
One should remark that it is the hyperon onset that defines the bump in the  DIDY plot of the speed of sound squared (bottom right panel). Note also that, at the hadron-quark phase transition, the coupling constant  $\alpha_s^t$ is below or at the limit where perturbative expansions are considered valid (see Table \ref{tab:hybrid}). For the quark phase, the pocket formula of Ref. \cite{Fraga:2013qra} (pQCD) and the one derived in Sec. \ref{cformula} for RGOPT were used to generate the corresponding plots.


\begin{figure}[!h]
     \centering

\includegraphics[width=0.9\linewidth]{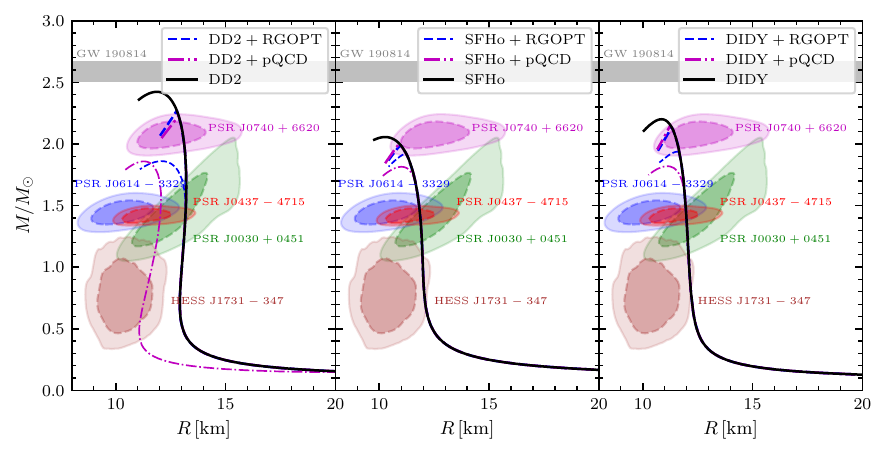}
    \caption{Hybrid NSs using NNLO RGOPT and NNLO pQCD to describe the quark degrees of freedom for the star core and the DD2 (left) SFHo (middle) and DIDY (right) EoS describing the hadronic exterior. 
Higher $X$ values produce lower maximum star masses.  The minimum value of $X$
that reproduces a finite quark core (thick lines), and the maximum $X$ that
allows the matching of the two EoSs (thin lines) are listed in Table
\ref{tab:hybrid}. Also included are the NICER data for pulsars PSR J0740+6620
\cite{Riley:2021pdl,Salmi:2024aum, Miller:2021qha}, PSR J0614-3329
\cite{Mauviard:2025dmd}, PSR J0437-4715 \cite{Choudhury:2024xbk} and the most
recent analysis of NICER data for pulsars PSR J0030+0451 \cite{Kini:2026rjx}, as
well as the low mass compact star in the supernova remnant HESS
J1731-347 \cite{Doroshenko:2022nwp}. In particular, the ellipses
represent the 68\% (dashed) and 95\% (full) confidence interval of the 2-D
distribution in the mass-radii domain, while the error bars yield the 1-D
marginalized posterior distribution for the same data. A band identifying the
mass of the low mass compact object associated with GW190814
\cite{LIGOScientific:2020zkf} has also been included. }
\label{fig:hybrid_stars}
\end{figure}
\begin{figure}[!h]
    \centering

\includegraphics[width=0.9\linewidth]{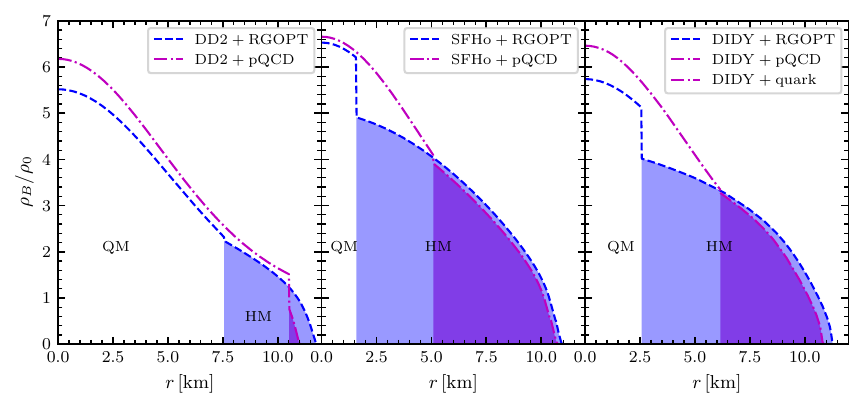}
    \caption{ Normalized baryon density  as a function of the internal radius,
$r$, for the maximum mass star configuration with the largest quark core using
NNLO RGOPT and NNLO pQCD to describe the quark core and DD2 (left), SFHo
(middle), DIDY (right) EoS to describe the hadronic exterior, obtained with
RGOPT(pQCD) scale parameter $X=2.59(2.80),\,
2.71(2.86),\, 2.71(2.82)$ respectively, for DD2,
SFHo and DIDY.
    }
    \label{fig:rho_vs_r}
\end{figure}

\begin{table}[!h]

\caption{Hybrid star properties predicted by the different models. The values of $X$ are those that reproduce a finite quark core. The considered properties for the maximum mass are:
the maximum mass 
$M_{\rm max}$, the ratio of the quark core mass $M_{\rm core}/M_{\rm max}$, the radius of the star $R_{\rm max}$ as well as the density at the center of the star, $\rho_B^{c,\rm max}$. For the predicted star of $1.4 M_\odot$ we show:  the radius $R_{1.4}$, the density at the center, $\rho_B^{c,1.4}$. Finally at the transition from hadron matter to quark matter we show: the baryon chemical potential, $\mu_B^t$, the value of the coupling $\alpha_s^t=\alpha_s(X\frac{\mu_B^t}{3})$, the density of hadronic matter at the transition, $\rho_B^{t,\rm had}$, the difference of pressure between quark matter and hadron matter, $\Delta \rho_B^{t}=\rho_B^{\rm quark}-\rho_B^{\rm had}$ and the mass predicted by the configuration at the transition, $M_t$.}

\begin{center}
\begin{tabular}{c  |c  c  c  c c c c c c c c c c} 
       \hline
    &  $X$ & $M_{\rm max}$  & $M_{\rm core}/M_{\rm max}$ &  $R_{\rm max}$ & $R_{\rm core}$  & $\rho_B^{c,\rm max}$   & $R_{1.4}$  & $\rho_B^{c,1.4}$ & $\mu_B^t$ & $\alpha_s^t$ & $\rho_B^{t,\rm had}$   & $\Delta \rho_B^{t}$ & $M_{t}$ \\
    &  & [$M_\odot$] & &  [km] & [km] & [$\rho_0 $] & [km]  & [$\rho_0 $] & [GeV] & & [$\rho_0$]   & [$\rho_0 $]  & [$M_\odot$]\\
	\hline
    \multirow{2}{4.5em}{$\rm{DD2\ +}$ $\rm 
    RGOPT\ $} & $2.00$  & $2.27$ & $2\times 10^{-3}$ &  $12.8$ & $0.793$ & $6.10$ & $13.2$ & $2.19 $ & $1.467$ & $0.430$ & $3.67 $ & $2.36 $ &$2.27$\\ 
  &$2.63$& $1.86 $ & $0.496$& $ 12.1$ & $7.56$ &$5.52$  & $13.2$& $2.19 $& $1.134$ & $0.425$ & $2.24$ & $0.0837 $ & $1.44$   \\
    \hline
    \multirow{2}{4.5em}{$\rm{DD2\ +}$ $\rm pQCD \ $} & $2.17$ & $2.23 $ &  $1\times 10^{-3}$ & $12.9$ & $0.693$& $5.66 $ & $13.2 $ & $2.19 $ & $1.427$ & $0.413$ & $3.51$ & $2.10$ & $2.23$ \\ 
  &$2.80$& $1.86 $ & $ 0.980$ &$11.3$ & $10.5$& $6.17 $ & $12.0$ & $2.69 $ & $0.962$ & $0.460$ & $0.76$ & $0.75$ & $0.15$ \\
  
     \hline
	\multirow{2}{4.5em}{$\rm{SFHo \ +}$ $\rm RGOPT \ $} & $2.28$ & $1.99 $ &
    $< 10^{-4}$&  $11.0$ & $0.231$ & $7.63$ & $11.9$& $3.20 $ & $1.565 $ & $0.375$ & $5.50 $ &$2.28$ & $1.99$\\ 
  &$2.71$& $1.92 $ & $0.009$ & $ 11.2$ & $  1.59$ &$6.53$ & $11.9$& $3.20 $ & $1.455$ & $0.352$ & $4.91 $ & $1.29$ & $1.92$   \\
     \hline

    \multirow{2}{4.5em}{$\rm{SFHo\ +}$ $\rm pQCD \ $} & $2.25$ & $2.00 $ &  $2\times 10^{-4}$ & $10.9$ & $0.462$& $8.27 $ & $11.9 $ & $3.20 $ & $1.593$ & $0.373$ & $5.64$ & $2.58$ & $2.00$ \\ 
  &$2.86$& $1.81 $ & $ 0.243$ &$11.0$ & $5.09$& $6.65 $ & $11.9$ & $3.20 $ & $1.283$ & $0.368$ & $3.91$ & $0.17$ & $1.68$ \\
    \hline
          \multirow{2}{4.5em}{$\rm{DIDY \ +}$ $\rm RGOPT \ $} & $2.00$ & $2.13  $ &
    $3\times 10^{-4}$&  $11.3$ & $0.503$ & $8.08$ & $12.0$& $3.04 $ & $1.587 $ & $0.406$ & $4.91 $ &$3.12$ & $2.13$\\ 
  &$2.71$& $1.94 $ & $0.033$ & $ 11.6$ & $2.57$ &$5.74$ & $12.0$& $3.04 $ & $1.380$ & $0.364$ & $4.02 $ & $1.10$ & $1.92$ \\
     \hline
     \multirow{2}{4.5em}{$\rm{DIDY\ +}$ $\rm pQCD \ $} & $2.00$ & $2.15 $ &  $1\times 10^{-4}$ & $11.3$ & $0.314$& $8.61 $ & $12.0 $ & $3.04 $ & $1.621$ & $0.400$ & $5.08$ & $3.49$ & $2.15$ \\ 
  &$2.82$& $1.82 $ & $ 0.368$ &$11.1$ & $6.24$& $6.47 $ & $12.0$ & $3.05 $ & $1.212$ & $0.385$ & $3.27$ & $0.071$ & $1.55$ \\
   
     \hline
\end{tabular}\label{tab:hybrid}

\end{center}

\end{table}

The mass-radius curves for hybrid stars, together with the observational
constraints from NICER and the compact object inside HESS J1731 −
347 are displayed in Fig. \ref{fig:hybrid_stars},  while some of the  main
associated  properties are summarized in Table \ref{tab:hybrid}.
The maximum mass configurations, corresponding to the smallest  $X$ value and the  smallest
quark core, are essentially defined by the properties of the hadronic EoS (namely SFHo, DIDY or DD2). 
Although the curves present a cusp like behavior, they contain  a finite but small core of quark matter. 
In contrast, the configurations with the highest $X$, correspond to the largest quark core predicted by each specific pair of EoS.
In the case of pQCD, these configurations produce maximum masses
that are less than $0.1\,M_\odot$ below the 95\% confidence region of the
two-dimensional mass-radius distribution of PSR J0740+6620, similarly to the
RGOPT + DD2 case. On the other hand, for RGOPT + SFHo and RGOPT + DIDY, the
stability condition imposed by the Bodmer--Witten hypothesis constrains the
hybrid stars to have maximum masses consistent with this observational
constraint.
Interestingly, the EoS with the largest quark core, corresponding to DD2+pQCD is in agreement with the observational constraint given by HESS J1731 - 347, as can be seen in the left panel of Fig. \ref{fig:hybrid_stars}.

In Fig. \ref{fig:rho_vs_r}, the density profiles of the stars defined by the EoS with the highest $X$ values are represented as  functions of the distance to the surface. The star corresponding to the highest scale contains a core of quarks with a thickness 
of 7.56 km (DD2),  2.57 km (DIDY) and
1.59 km (SFHo), see Table \ref{tab:hybrid}. The difference
between DD2 and the other two models is easily understood since a stiff EoS
favors a transition to quark matter at lower densities, allowing a large quark
core to be more easily supported.

{We have also calculated the density profiles of stars described by the quark}   pQCD EoS  
and the same models for  hadron matter. 
In the case of pQCD + DD2, the $X$ parameter is about
0.15 to 0.20  larger for the pQCD quark cores, and the star with the largest
quark core only has a small crust of hadron matter with a thickness of the order
of 0.5 km. The onset of quark matter at a very low density has a direct impact
on the radius of low mass stars, and  the radius of a  1.4$M_\odot$ star is
about 1 km shorter for stars with a pQCD core.

\section{Conclusions}\label{sec:conclusions}
We have extended a recent renormalization-group-induced resummation of pQCD,
RGOPT at  NNLO~\cite{Fernandez:2024ilg}, to incorporate $\beta$-equilibrium and
charge neutrality in a thermodynamically consistent  EoS, tailored to describe
massive quarks in the $N_f =2+1$ case. Having such an EoS, from a first
principles QCD evaluation, we have investigated the properties of (pure) QSs as
well as the quark core of hybrid stars.
Maximum masses corresponding to $2, 2.3$ and 2.6 solar masses have been considered as phenomenological inputs to possibly constrain the corresponding arbitrary renormalization scale. 
In the case of pure QSs, our results show that the NNLO RGOPT predictions are less sensitive to scale variations than those generated by pQCD evaluations at the same perturbative order. 
The resummed RGOPT EoS final expression being involved,  we have provided a compact formula  that can be readily used in applications aiming to describe the quark sector associated with compact stellar objects simply by using $\mu_B$ and $X$ as inputs. 
As our results indicate, this formula incorporates all important RG properties  so that the uncertainties (due to scale dependence) observed within its  pQCD  counterpart can be mitigated.

Our numerical analysis started by considering the case of pure QSs. 
The renormalization scale parameter was chosen by enforcing the corresponding
QSs to be compatible with  astrophysical observations and, as a consequence,
values in the range $X=2.71-3.30$ were obtained.
It should be remarked that a larger value ($X=3.84$) needs to be
considered if the secondary component of GW190814 is a NS.
At the same time, the range of possible $X$ values obtained with a pQCD quark EoS, upon imposing the same observations, turned out to be 2.95 to 3.26 (and  3.56, if GW190814 is considered), as a direct consequence of a larger scale dependence. 

 To describe hybrid stars, we have used the RGOPT or pQCD EoS for the quark phase and merged it with three different  EoSs that describe the hadron phase with different levels of stiffness (SFHo, DIDY, and DD2) and  hadron content  (since DIDY includes hyperons, in contrast to SFHo and DD2). 
 To take into account the hadron-quark phase transition,  the standard Maxwell construction has been adopted. 
 Then, for each model, we generated two hybrid EoS extremes by varying $X$: i) the  lowest value describes a hybrid star with a small quark core and gives the most massive star, which is essentially defined by the hadron model; ii) the other extreme, which defines the largest $X$, describes a hybrid star with the largest quark core. 
 These two extremes set the range of interest for the scale parameter: { DD2
with RGOPT (pQCD), gives $X=2.63-2.00$ ($X=2.80-2.17$) for maximum
masses in the range $1.86-2.27\, M_\odot$ (1.86 and $2.23\,
M_\odot$).
 In the case of the hadronic SFHo and DIDY EoSs, together with the quark  RGOPT
or pQCD  EoS, the RGOPT produce more constrained results,
producing maximum star masses that are compatible with PSR J0740+6620, while
pQCD still produces maximum star masses around $0.1\, M\odot$ smaller than the
lower limit if the 95\% confidence region.
 
 Considering pQCD+DD2, hybrid stars with a small hadronic crust about 1 km thick
were obtained. These stars are compatible with the compact object inside HESS
J1731-347; similar conclusions were drawn in
\cite{Sagun:2023rzp,DiClemente:2022wqp}.
 Otherwise, stars with the largest quark core compatible with PSR J0740+6620
have a hadronic shell that is about 3 to 5 km thick and are not compatible with
the object inside HESS J1731-347. The  highest value of the speed
of sound  occurs in the hadronic branch and goes above the conformal limit,
$\sqrt{1/3}$. After the hadron-quark phase transition, the speed of sound drops
to values below $\sqrt{1/3}$ in the quark phase.}

 In the future, the possible existence of colorsuperconducting phases will be
investigated. It will also be interesting to analyze the consequences of
constraining the NS EoS with improved
 renormalization scale uncertainties as obtained in the present study, when
thermodynamical  constraints, as described in \cite{Komoltsev:2021jzg}, are
imposed.

\begin{acknowledgements}
This material is based upon work supported by the National Science Foundation under grants No. PHY-2208724,
PHY-2116686 and PHY-2514763, and within the framework of the MUSES collaboration, under Grant No. OAC-2103680. This material is also based upon work supported by the U.S. Department of Energy, Office of Science, Office of Nuclear Physics, under Award Numbers DE-SC0022023, and DE- SC0023861, as well as by
the National Aeronautics and Space Agency (NASA) under Award Number 80NSSC24K0767.
 M.B.P. is partially supported by Conselho
Nacional de Desenvolvimento Cient\'{\i}fico e Tecnol\'{o}gico (CNPq, Brazil),
Process  No.  307261/2021-2  and 403016/2024-0. The work is also part of the project
Instituto Nacional de Ci\^encia e Tecnologia - F\'isica Nuclear e
Aplica\c{c}\~oes (INCT - FNA), Process CNPq 408419/2024-5. L.F. has been supported in part by the Research Council of Finland grants Nos. 353772 and 354533. C.P was partially supported by national funds from FCT (Fundação para a Ciência e a Tecnologia, I.P, Portugal) under project UID/04564/2025, identified by DOI 10.54499/UIDB/04564/2025.

\end{acknowledgements}

\newpage

\appendix
\section{List of contributions to the $N_f=2^*+1^*$ RGOPT pressure}\label{Appendix:A}

For completeness, rather than repeating many long expressions, we simply 
indicate precisely where to find the relevant contributions of the complete
NNLO pressure of Eq. (\ref{Ptotal}) in \cite{Fernandez:2024ilg}, where the equation numbers
refer to the later reference: 
\begin{itemize}
\item
$P^{m}_{\rm LO, NLO}(m_i,\mu_i)$, $P^{v}_{\rm LO, NLO}(m_i)$,  
 respectively in Eqs. (2), (3), (23), (24);
\item
$P_{\rm 2GI}^{N_f=2^*+1^*}(m_i,\mu_i)$, $P_{\rm VM}^{N_f=2^*+1^*}(m_i,\mu_i)$ in Eqs. (8), (9), (18), (46),(47); 
\item
$P_{\rm Ring}^{N_f=2^*+1^*}(m)\equiv -\Omega_{\rm Ring}^{N_f=2^*+1^*}(m)$ in Eq. (21);
\item
$P^v_{\rm NNLO,i}(m_i)$ in Eqs. (25), (26), (44); 
\item Finally
$P_{sub,i }$, $P_{sub}^{nd}$ in Eqs. (45), (A.9), (A.10).
\end{itemize}

\section{Renormalization group material}\label{Appendix:B}
Since it is a central quantity in our approach, we recall the expression of the massive (homogeneous) renormalization group operator
\begin{align}
 \Lambda \frac{d}{d \Lambda}\equiv \Lambda \frac{\partial}{\partial \Lambda} + \beta(g_s^2 )\frac{\partial}{\partial g_s^2 } - 
\gamma_m(g_s^2 ) m \frac{\partial}{\partial m} \;
\label{RG}
\end{align}
where in our NNLO analysis we use the beta-function $\beta(g_s^2)$ and 
quark mass anomalous dimension gamma-function $\gamma_m(g_s^2)$ at three-loop order, defined in our conventions as
\begin{equation}
 \beta\left(g_s^2 \right)=-2g_s^4\left(b_0 +b_1g_s^2+b_2 g_s^4+ 
 \mathcal{O}\left(g^6\right) \right)\;,
 \end{equation}
 \begin{equation}
 \gamma_m\left(g_s^2\right)=g_s^2\left( \gamma_0+\gamma_1g_s^2 +\gamma_2 g_s^4 +
 \mathcal{O}\left(g^6\right)\right)\;,
\end{equation}
with the coefficients, for the relevant QCD case with $C_A=N_c$, $C_F=4/3$, $N_c=3$, 
and $N_f$ quark flavors:
\begin{equation}\label{eq:BetaCoef}
\begin{aligned}
b_0= & \frac{1}{(4\pi)^2}\left(\fr{11}{3}C_A-\frac{2}{3}N_f\right) \ \ ,\ \ 
\end{aligned}
\end{equation}
\begin{equation}\label{eq:GammaCoef}
\begin{aligned}
\gamma_0 = & \frac{2}{(4\pi)^2} (N_c C_F ), \ \ \ 
\end{aligned}
\end{equation}
and higher order QCD RG coefficients up to relevant three-loop order given in our normalization in \cite{Fernandez:2024ilg} 
(see e.g. refs\cite{Baikov:2016tgj,Luthe:2016ima,Herzog:2017ohr} for 
QCD RG coefficients up to five-loop order). 
In addition, as mentioned in Sec. II, upon including massive vacuum terms the renormalization group invariance of the massive pressure requires additional zero-point subtraction contributions, $P_{sub}\sim \sum_k s_k (g_s^2)^{(k-1)}$ in Eq.(\ref{Ptotal}), directly related to the
fact that RG invariance actually involves \cite{Chetyrkin:1994ex,Baikov:2018nzi,Kneur:2015dda} the vacuum energy anomalous dimension $\hat \Gamma^0(g_s^2)$, i.e.:
\be
 \Lambda \frac{d}{d \Lambda} P(m,g_s^2) -m^4 \hat\Gamma^0(g_s^2) \equiv 0 
\ee
with 
\be 
\hat \Gamma^0(g_s^2) \equiv \sum_k \Gamma^0_k (g_s^2)^{k}
\ee
In the normalization of the pressure in Eq. (\ref{Ptotal}), the resulting subtraction coefficients $s_k$ are
\begin{equation}\label{eq:s_i Quark}
\begin{aligned}
s_0 \equiv &\frac{\Gamma^0_0}{2\big(b_0-2 \gamma_0\big)}=\frac{3}{7} \ \ , \ \ 
\end{aligned}
\end{equation}
with higher order coefficients given in \cite{Fernandez:2024ilg}.\\
Finally, one should bear in mind that performing a renormalization scheme change (RSC) according
to Eq. (\ref{eq:RSC}) also implies consistent $B_2$-dependent modifications \cite{Fernandez:2024ilg} in the higher order RG coefficients $\gamma_k, s_k$ above.

\bibliography{bibliography}

\begin{thebibliography}{106}%
\makeatletter
\providecommand \@ifxundefined [1]{%
 \@ifx{#1\undefined}
}%
\providecommand \@ifnum [1]{%
 \ifnum #1\expandafter \@firstoftwo
 \else \expandafter \@secondoftwo
 \fi
}%
\providecommand \@ifx [1]{%
 \ifx #1\expandafter \@firstoftwo
 \else \expandafter \@secondoftwo
 \fi
}%
\providecommand \natexlab [1]{#1}%
\providecommand \enquote  [1]{``#1''}%
\providecommand \bibnamefont  [1]{#1}%
\providecommand \bibfnamefont [1]{#1}%
\providecommand \citenamefont [1]{#1}%
\providecommand \href@noop [0]{\@secondoftwo}%
\providecommand \href [0]{\begingroup \@sanitize@url \@href}%
\providecommand \@href[1]{\@@startlink{#1}\@@href}%
\providecommand \@@href[1]{\endgroup#1\@@endlink}%
\providecommand \@sanitize@url [0]{\catcode `\\12\catcode `\$12\catcode
  `\&12\catcode `\#12\catcode `\^12\catcode `\_12\catcode `\%12\relax}%
\providecommand \@@startlink[1]{}%
\providecommand \@@endlink[0]{}%
\providecommand \url  [0]{\begingroup\@sanitize@url \@url }%
\providecommand \@url [1]{\endgroup\@href {#1}{\urlprefix }}%
\providecommand \urlprefix  [0]{URL }%
\providecommand \Eprint [0]{\href }%
\providecommand \doibase [0]{https://doi.org/}%
\providecommand \selectlanguage [0]{\@gobble}%
\providecommand \bibinfo  [0]{\@secondoftwo}%
\providecommand \bibfield  [0]{\@secondoftwo}%
\providecommand \translation [1]{[#1]}%
\providecommand \BibitemOpen [0]{}%
\providecommand \bibitemStop [0]{}%
\providecommand \bibitemNoStop [0]{.\EOS\space}%
\providecommand \EOS [0]{\spacefactor3000\relax}%
\providecommand \BibitemShut  [1]{\csname bibitem#1\endcsname}%
\let\auto@bib@innerbib\@empty
\bibitem [{\citenamefont {Demorest}\ \emph {et~al.}(2010)\citenamefont
  {Demorest}, \citenamefont {Pennucci}, \citenamefont {Ransom}, \citenamefont
  {Roberts},\ and\ \citenamefont {Hessels}}]{Demorest:2010bx}%
  \BibitemOpen
  \bibfield  {author} {\bibinfo {author} {\bibfnamefont {P.}~\bibnamefont
  {Demorest}}, \bibinfo {author} {\bibfnamefont {T.}~\bibnamefont {Pennucci}},
  \bibinfo {author} {\bibfnamefont {S.}~\bibnamefont {Ransom}}, \bibinfo
  {author} {\bibfnamefont {M.}~\bibnamefont {Roberts}},\ and\ \bibinfo {author}
  {\bibfnamefont {J.}~\bibnamefont {Hessels}},\ }\bibfield  {title} {\bibinfo
  {title} {{A two-solar-mass neutron star measured using Shapiro delay}},\
  }\href {https://doi.org/10.1038/nature09466} {\bibfield  {journal} {\bibinfo
  {journal} {Nature}\ }\textbf {\bibinfo {volume} {467}},\ \bibinfo {pages}
  {1081} (\bibinfo {year} {2010})}\BibitemShut {NoStop}%
\bibitem [{\citenamefont {Fonseca}\ \emph {et~al.}(2021)\citenamefont {Fonseca}
  \emph {et~al.}}]{Fonseca:2021wxt}%
  \BibitemOpen
  \bibfield  {author} {\bibinfo {author} {\bibfnamefont {E.}~\bibnamefont
  {Fonseca}} \emph {et~al.},\ }\bibfield  {title} {\bibinfo {title} {{Refined
  Mass and Geometric Measurements of the High-mass PSR J0740+6620}},\ }\href
  {https://doi.org/10.3847/2041-8213/ac03b8} {\bibfield  {journal} {\bibinfo
  {journal} {Astrophys. J. Lett.}\ }\textbf {\bibinfo {volume} {915}},\
  \bibinfo {pages} {L12} (\bibinfo {year} {2021})},\ \Eprint
  {https://arxiv.org/abs/2104.00880} {arXiv:2104.00880 [astro-ph.HE]}
  \BibitemShut {NoStop}%
\bibitem [{\citenamefont {Antoniadis}\ \emph {et~al.}(2013)\citenamefont
  {Antoniadis} \emph {et~al.}}]{Antoniadis:2013pzd}%
  \BibitemOpen
  \bibfield  {author} {\bibinfo {author} {\bibfnamefont {J.}~\bibnamefont
  {Antoniadis}} \emph {et~al.},\ }\bibfield  {title} {\bibinfo {title} {{A
  Massive Pulsar in a Compact Relativistic Binary}},\ }\href
  {https://doi.org/10.1126/science.1233232} {\bibfield  {journal} {\bibinfo
  {journal} {Science}\ }\textbf {\bibinfo {volume} {340}},\ \bibinfo {pages}
  {6131} (\bibinfo {year} {2013})}\BibitemShut {NoStop}%
\bibitem [{\citenamefont {Annala}\ \emph {et~al.}(2023)\citenamefont {Annala},
  \citenamefont {Gorda}, \citenamefont {Hirvonen}, \citenamefont {Komoltsev},
  \citenamefont {Kurkela}, \citenamefont {N{\"a}ttil{\"a}},\ and\ \citenamefont
  {Vuorinen}}]{Annala:2023cwx}%
  \BibitemOpen
  \bibfield  {author} {\bibinfo {author} {\bibfnamefont {E.}~\bibnamefont
  {Annala}}, \bibinfo {author} {\bibfnamefont {T.}~\bibnamefont {Gorda}},
  \bibinfo {author} {\bibfnamefont {J.}~\bibnamefont {Hirvonen}}, \bibinfo
  {author} {\bibfnamefont {O.}~\bibnamefont {Komoltsev}}, \bibinfo {author}
  {\bibfnamefont {A.}~\bibnamefont {Kurkela}}, \bibinfo {author} {\bibfnamefont
  {J.}~\bibnamefont {N{\"a}ttil{\"a}}},\ and\ \bibinfo {author} {\bibfnamefont
  {A.}~\bibnamefont {Vuorinen}},\ }\bibfield  {title} {\bibinfo {title}
  {{Strongly interacting matter exhibits deconfined behavior in massive neutron
  stars}},\ }\href {https://doi.org/10.1038/s41467-023-44051-y} {\bibfield
  {journal} {\bibinfo  {journal} {Nature Commun.}\ }\textbf {\bibinfo {volume}
  {14}},\ \bibinfo {pages} {8451} (\bibinfo {year} {2023})},\ \Eprint
  {https://arxiv.org/abs/2303.11356} {arXiv:2303.11356 [astro-ph.HE]}
  \BibitemShut {NoStop}%
\bibitem [{\citenamefont {Albino}\ \emph {et~al.}(2026)\citenamefont {Albino},
  \citenamefont {Malik}, \citenamefont {Ferreira},\ and\ \citenamefont
  {Provid{\^e}ncia}}]{Albino:2025puc}%
  \BibitemOpen
  \bibfield  {author} {\bibinfo {author} {\bibfnamefont {M.}~\bibnamefont
  {Albino}}, \bibinfo {author} {\bibfnamefont {T.}~\bibnamefont {Malik}},
  \bibinfo {author} {\bibfnamefont {M.}~\bibnamefont {Ferreira}},\ and\
  \bibinfo {author} {\bibfnamefont {C.}~\bibnamefont {Provid{\^e}ncia}},\
  }\bibfield  {title} {\bibinfo {title} {{Bayesian inference of hybrid stars
  with large quark cores}},\ }\href {https://doi.org/10.1103/jrz4-zjq1}
  {\bibfield  {journal} {\bibinfo  {journal} {Phys. Rev. D}\ }\textbf {\bibinfo
  {volume} {113}},\ \bibinfo {pages} {083019} (\bibinfo {year} {2026})},\
  \Eprint {https://arxiv.org/abs/2511.02653} {arXiv:2511.02653 [nucl-th]}
  \BibitemShut {NoStop}%
\bibitem [{\citenamefont {de~Forcrand}(2009)}]{deForcrand:2009zkb}%
  \BibitemOpen
  \bibfield  {author} {\bibinfo {author} {\bibfnamefont {P.}~\bibnamefont
  {de~Forcrand}},\ }\bibfield  {title} {\bibinfo {title} {{Simulating QCD at
  finite density}},\ }\href {https://doi.org/10.22323/1.091.0010} {\bibfield
  {journal} {\bibinfo  {journal} {PoS}\ }\textbf {\bibinfo {volume}
  {LAT2009}},\ \bibinfo {pages} {010} (\bibinfo {year} {2009})},\ \Eprint
  {https://arxiv.org/abs/1005.0539} {arXiv:1005.0539 [hep-lat]} \BibitemShut
  {NoStop}%
\bibitem [{\citenamefont {Drischler}\ \emph {et~al.}(2019)\citenamefont
  {Drischler}, \citenamefont {Hebeler},\ and\ \citenamefont
  {Schwenk}}]{Drischler:2017wtt}%
  \BibitemOpen
  \bibfield  {author} {\bibinfo {author} {\bibfnamefont {C.}~\bibnamefont
  {Drischler}}, \bibinfo {author} {\bibfnamefont {K.}~\bibnamefont {Hebeler}},\
  and\ \bibinfo {author} {\bibfnamefont {A.}~\bibnamefont {Schwenk}},\
  }\bibfield  {title} {\bibinfo {title} {{Chiral interactions up to
  next-to-next-to-next-to-leading order and nuclear saturation}},\ }\href
  {https://doi.org/10.1103/PhysRevLett.122.042501} {\bibfield  {journal}
  {\bibinfo  {journal} {Phys. Rev. Lett.}\ }\textbf {\bibinfo {volume} {122}},\
  \bibinfo {pages} {042501} (\bibinfo {year} {2019})},\ \Eprint
  {https://arxiv.org/abs/1710.08220} {arXiv:1710.08220 [nucl-th]} \BibitemShut
  {NoStop}%
\bibitem [{\citenamefont {Hebeler}(2021)}]{Hebeler:2020ocj}%
  \BibitemOpen
  \bibfield  {author} {\bibinfo {author} {\bibfnamefont {K.}~\bibnamefont
  {Hebeler}},\ }\bibfield  {title} {\bibinfo {title} {{Three-nucleon forces:
  Implementation and applications to atomic nuclei and dense matter}},\ }\href
  {https://doi.org/10.1016/j.physrep.2020.08.009} {\bibfield  {journal}
  {\bibinfo  {journal} {Phys. Rept.}\ }\textbf {\bibinfo {volume} {890}},\
  \bibinfo {pages} {1} (\bibinfo {year} {2021})},\ \Eprint
  {https://arxiv.org/abs/2002.09548} {arXiv:2002.09548 [nucl-th]} \BibitemShut
  {NoStop}%
\bibitem [{\citenamefont {Oertel}\ \emph {et~al.}(2017)\citenamefont {Oertel},
  \citenamefont {Hempel}, \citenamefont {Kl\"ahn},\ and\ \citenamefont
  {Typel}}]{Oertel:2016bki}%
  \BibitemOpen
  \bibfield  {author} {\bibinfo {author} {\bibfnamefont {M.}~\bibnamefont
  {Oertel}}, \bibinfo {author} {\bibfnamefont {M.}~\bibnamefont {Hempel}},
  \bibinfo {author} {\bibfnamefont {T.}~\bibnamefont {Kl\"ahn}},\ and\ \bibinfo
  {author} {\bibfnamefont {S.}~\bibnamefont {Typel}},\ }\bibfield  {title}
  {\bibinfo {title} {{Equations of state for supernovae and compact stars}},\
  }\href {https://doi.org/10.1103/RevModPhys.89.015007} {\bibfield  {journal}
  {\bibinfo  {journal} {Rev. Mod. Phys.}\ }\textbf {\bibinfo {volume} {89}},\
  \bibinfo {pages} {015007} (\bibinfo {year} {2017})},\ \Eprint
  {https://arxiv.org/abs/1610.03361} {arXiv:1610.03361 [astro-ph.HE]}
  \BibitemShut {NoStop}%
\bibitem [{\citenamefont {Dutra}\ \emph {et~al.}(2014)\citenamefont {Dutra},
  \citenamefont {Louren{\c{c}}o}, \citenamefont {Avancini}, \citenamefont
  {Carlson}, \citenamefont {Delfino}, \citenamefont {Menezes}, \citenamefont
  {Provid{\^e}ncia}, \citenamefont {Typel},\ and\ \citenamefont
  {Stone}}]{Dutra:2014qga}%
  \BibitemOpen
  \bibfield  {author} {\bibinfo {author} {\bibfnamefont {M.}~\bibnamefont
  {Dutra}}, \bibinfo {author} {\bibfnamefont {O.}~\bibnamefont
  {Louren{\c{c}}o}}, \bibinfo {author} {\bibfnamefont {S.~S.}\ \bibnamefont
  {Avancini}}, \bibinfo {author} {\bibfnamefont {B.~V.}\ \bibnamefont
  {Carlson}}, \bibinfo {author} {\bibfnamefont {A.}~\bibnamefont {Delfino}},
  \bibinfo {author} {\bibfnamefont {D.~P.}\ \bibnamefont {Menezes}}, \bibinfo
  {author} {\bibfnamefont {C.}~\bibnamefont {Provid{\^e}ncia}}, \bibinfo
  {author} {\bibfnamefont {S.}~\bibnamefont {Typel}},\ and\ \bibinfo {author}
  {\bibfnamefont {J.~R.}\ \bibnamefont {Stone}},\ }\bibfield  {title} {\bibinfo
  {title} {{Relativistic Mean-Field Hadronic Models under Nuclear Matter
  Constraints}},\ }\href {https://doi.org/10.1103/PhysRevC.90.055203}
  {\bibfield  {journal} {\bibinfo  {journal} {Phys. Rev. C}\ }\textbf {\bibinfo
  {volume} {90}},\ \bibinfo {pages} {055203} (\bibinfo {year} {2014})},\
  \Eprint {https://arxiv.org/abs/1405.3633} {arXiv:1405.3633 [nucl-th]}
  \BibitemShut {NoStop}%
\bibitem [{\citenamefont {Cartaxo}\ \emph {et~al.}(2026)\citenamefont
  {Cartaxo}, \citenamefont {Huang}, \citenamefont {Malik}, \citenamefont
  {Sourav}, \citenamefont {Yuan}, \citenamefont {Zhou}, \citenamefont {Liu},\
  and\ \citenamefont {Provid{\^e}ncia}}]{Cartaxo:2025jpi}%
  \BibitemOpen
  \bibfield  {author} {\bibinfo {author} {\bibfnamefont {J.}~\bibnamefont
  {Cartaxo}}, \bibinfo {author} {\bibfnamefont {C.}~\bibnamefont {Huang}},
  \bibinfo {author} {\bibfnamefont {T.}~\bibnamefont {Malik}}, \bibinfo
  {author} {\bibfnamefont {S.}~\bibnamefont {Sourav}}, \bibinfo {author}
  {\bibfnamefont {W.-L.}\ \bibnamefont {Yuan}}, \bibinfo {author}
  {\bibfnamefont {T.}~\bibnamefont {Zhou}}, \bibinfo {author} {\bibfnamefont
  {X.}~\bibnamefont {Liu}},\ and\ \bibinfo {author} {\bibfnamefont
  {C.}~\bibnamefont {Provid{\^e}ncia}},\ }\bibfield  {title} {\bibinfo {title}
  {{Covariant Energy Density Functionals for Modeling the Equation of State of
  Neutron Star Matter: Cross-comparison Analysis Using CompactObject}},\ }\href
  {https://doi.org/10.3847/1538-4365/ae2310} {\bibfield  {journal} {\bibinfo
  {journal} {Astrophys. J. Suppl.}\ }\textbf {\bibinfo {volume} {282}},\
  \bibinfo {pages} {33} (\bibinfo {year} {2026})},\ \Eprint
  {https://arxiv.org/abs/2506.03112} {arXiv:2506.03112 [nucl-th]} \BibitemShut
  {NoStop}%
\bibitem [{\citenamefont {Freedman}\ and\ \citenamefont
  {McLerran}(1977)}]{Freedman:1976ub}%
  \BibitemOpen
  \bibfield  {author} {\bibinfo {author} {\bibfnamefont {B.~A.}\ \bibnamefont
  {Freedman}}\ and\ \bibinfo {author} {\bibfnamefont {L.~D.}\ \bibnamefont
  {McLerran}},\ }\bibfield  {title} {\bibinfo {title} {{Fermions and Gauge
  Vector Mesons at Finite Temperature and Density. 3. The Ground State Energy
  of a Relativistic Quark Gas}},\ }\href
  {https://doi.org/10.1103/PhysRevD.16.1169} {\bibfield  {journal} {\bibinfo
  {journal} {Phys. Rev. D}\ }\textbf {\bibinfo {volume} {16}},\ \bibinfo
  {pages} {1169} (\bibinfo {year} {1977})}\BibitemShut {NoStop}%
\bibitem [{\citenamefont {Kurkela}\ \emph {et~al.}(2010)\citenamefont
  {Kurkela}, \citenamefont {Romatschke},\ and\ \citenamefont
  {Vuorinen}}]{Kurkela:2009gj}%
  \BibitemOpen
  \bibfield  {author} {\bibinfo {author} {\bibfnamefont {A.}~\bibnamefont
  {Kurkela}}, \bibinfo {author} {\bibfnamefont {P.}~\bibnamefont
  {Romatschke}},\ and\ \bibinfo {author} {\bibfnamefont {A.}~\bibnamefont
  {Vuorinen}},\ }\bibfield  {title} {\bibinfo {title} {{Cold Quark Matter}},\
  }\href {https://doi.org/10.1103/PhysRevD.81.105021} {\bibfield  {journal}
  {\bibinfo  {journal} {Phys. Rev. D}\ }\textbf {\bibinfo {volume} {81}},\
  \bibinfo {pages} {105021} (\bibinfo {year} {2010})},\ \Eprint
  {https://arxiv.org/abs/0912.1856} {arXiv:0912.1856 [hep-ph]} \BibitemShut
  {NoStop}%
\bibitem [{\citenamefont {Fraga}\ \emph {et~al.}(2014)\citenamefont {Fraga},
  \citenamefont {Kurkela},\ and\ \citenamefont {Vuorinen}}]{Fraga:2013qra}%
  \BibitemOpen
  \bibfield  {author} {\bibinfo {author} {\bibfnamefont {E.~S.}\ \bibnamefont
  {Fraga}}, \bibinfo {author} {\bibfnamefont {A.}~\bibnamefont {Kurkela}},\
  and\ \bibinfo {author} {\bibfnamefont {A.}~\bibnamefont {Vuorinen}},\
  }\bibfield  {title} {\bibinfo {title} {{Interacting quark matter equation of
  state for compact stars}},\ }\href
  {https://doi.org/10.1088/2041-8205/781/2/L25} {\bibfield  {journal} {\bibinfo
   {journal} {Astrophys. J. Lett.}\ }\textbf {\bibinfo {volume} {781}},\
  \bibinfo {pages} {L25} (\bibinfo {year} {2014})},\ \Eprint
  {https://arxiv.org/abs/1311.5154} {arXiv:1311.5154 [nucl-th]} \BibitemShut
  {NoStop}%
\bibitem [{\citenamefont {Gorda}\ \emph {et~al.}(2018)\citenamefont {Gorda},
  \citenamefont {Kurkela}, \citenamefont {Romatschke}, \citenamefont
  {S\"appi},\ and\ \citenamefont {Vuorinen}}]{Gorda:2018gpy}%
  \BibitemOpen
  \bibfield  {author} {\bibinfo {author} {\bibfnamefont {T.}~\bibnamefont
  {Gorda}}, \bibinfo {author} {\bibfnamefont {A.}~\bibnamefont {Kurkela}},
  \bibinfo {author} {\bibfnamefont {P.}~\bibnamefont {Romatschke}}, \bibinfo
  {author} {\bibfnamefont {S.}~\bibnamefont {S\"appi}},\ and\ \bibinfo {author}
  {\bibfnamefont {A.}~\bibnamefont {Vuorinen}},\ }\bibfield  {title} {\bibinfo
  {title} {{Next-to-Next-to-Next-to-Leading Order Pressure of Cold Quark
  Matter: Leading Logarithm}},\ }\href
  {https://doi.org/10.1103/PhysRevLett.121.202701} {\bibfield  {journal}
  {\bibinfo  {journal} {Phys. Rev. Lett.}\ }\textbf {\bibinfo {volume} {121}},\
  \bibinfo {pages} {202701} (\bibinfo {year} {2018})},\ \Eprint
  {https://arxiv.org/abs/1807.04120} {arXiv:1807.04120 [hep-ph]} \BibitemShut
  {NoStop}%
\bibitem [{\citenamefont {Gorda}\ \emph
  {et~al.}(2021{\natexlab{a}})\citenamefont {Gorda}, \citenamefont {Kurkela},
  \citenamefont {Paatelainen}, \citenamefont {S\"appi},\ and\ \citenamefont
  {Vuorinen}}]{Gorda:2021kme}%
  \BibitemOpen
  \bibfield  {author} {\bibinfo {author} {\bibfnamefont {T.}~\bibnamefont
  {Gorda}}, \bibinfo {author} {\bibfnamefont {A.}~\bibnamefont {Kurkela}},
  \bibinfo {author} {\bibfnamefont {R.}~\bibnamefont {Paatelainen}}, \bibinfo
  {author} {\bibfnamefont {S.}~\bibnamefont {S\"appi}},\ and\ \bibinfo {author}
  {\bibfnamefont {A.}~\bibnamefont {Vuorinen}},\ }\bibfield  {title} {\bibinfo
  {title} {{Cold quark matter at N3LO: Soft contributions}},\ }\href
  {https://doi.org/10.1103/PhysRevD.104.074015} {\bibfield  {journal} {\bibinfo
   {journal} {Phys. Rev. D}\ }\textbf {\bibinfo {volume} {104}},\ \bibinfo
  {pages} {074015} (\bibinfo {year} {2021}{\natexlab{a}})},\ \Eprint
  {https://arxiv.org/abs/2103.07427} {arXiv:2103.07427 [hep-ph]} \BibitemShut
  {NoStop}%
\bibitem [{\citenamefont {Klevansky}(1992)}]{Klevansky:1992qe}%
  \BibitemOpen
  \bibfield  {author} {\bibinfo {author} {\bibfnamefont {S.~P.}\ \bibnamefont
  {Klevansky}},\ }\bibfield  {title} {\bibinfo {title} {{The Nambu-Jona-Lasinio
  model of quantum chromodynamics}},\ }\href
  {https://doi.org/10.1103/RevModPhys.64.649} {\bibfield  {journal} {\bibinfo
  {journal} {Rev. Mod. Phys.}\ }\textbf {\bibinfo {volume} {64}},\ \bibinfo
  {pages} {649} (\bibinfo {year} {1992})}\BibitemShut {NoStop}%
\bibitem [{\citenamefont {Hatsuda}\ and\ \citenamefont
  {Kunihiro}(1994)}]{Hatsuda:1994pi}%
  \BibitemOpen
  \bibfield  {author} {\bibinfo {author} {\bibfnamefont {T.}~\bibnamefont
  {Hatsuda}}\ and\ \bibinfo {author} {\bibfnamefont {T.}~\bibnamefont
  {Kunihiro}},\ }\bibfield  {title} {\bibinfo {title} {{QCD phenomenology based
  on a chiral effective Lagrangian}},\ }\href
  {https://doi.org/10.1016/0370-1573(94)90022-1} {\bibfield  {journal}
  {\bibinfo  {journal} {Phys. Rept.}\ }\textbf {\bibinfo {volume} {247}},\
  \bibinfo {pages} {221} (\bibinfo {year} {1994})},\ \Eprint
  {https://arxiv.org/abs/hep-ph/9401310} {arXiv:hep-ph/9401310} \BibitemShut
  {NoStop}%
\bibitem [{\citenamefont {Rehberg}\ \emph {et~al.}(1996)\citenamefont
  {Rehberg}, \citenamefont {Klevansky},\ and\ \citenamefont
  {Hufner}}]{Rehberg:1995kh}%
  \BibitemOpen
  \bibfield  {author} {\bibinfo {author} {\bibfnamefont {P.}~\bibnamefont
  {Rehberg}}, \bibinfo {author} {\bibfnamefont {S.~P.}\ \bibnamefont
  {Klevansky}},\ and\ \bibinfo {author} {\bibfnamefont {J.}~\bibnamefont
  {Hufner}},\ }\bibfield  {title} {\bibinfo {title} {{Hadronization in the
  SU(3) Nambu-Jona-Lasinio model}},\ }\href
  {https://doi.org/10.1103/PhysRevC.53.410} {\bibfield  {journal} {\bibinfo
  {journal} {Phys. Rev. C}\ }\textbf {\bibinfo {volume} {53}},\ \bibinfo
  {pages} {410} (\bibinfo {year} {1996})},\ \Eprint
  {https://arxiv.org/abs/hep-ph/9506436} {arXiv:hep-ph/9506436} \BibitemShut
  {NoStop}%
\bibitem [{\citenamefont {Buballa}(2005)}]{Buballa:2003qv}%
  \BibitemOpen
  \bibfield  {author} {\bibinfo {author} {\bibfnamefont {M.}~\bibnamefont
  {Buballa}},\ }\bibfield  {title} {\bibinfo {title} {{NJL model analysis of
  quark matter at large density}},\ }\href
  {https://doi.org/10.1016/j.physrep.2004.11.004} {\bibfield  {journal}
  {\bibinfo  {journal} {Phys. Rept.}\ }\textbf {\bibinfo {volume} {407}},\
  \bibinfo {pages} {205} (\bibinfo {year} {2005})},\ \Eprint
  {https://arxiv.org/abs/hep-ph/0402234} {arXiv:hep-ph/0402234} \BibitemShut
  {NoStop}%
\bibitem [{\citenamefont {Lenaghan}\ \emph {et~al.}(2000)\citenamefont
  {Lenaghan}, \citenamefont {Rischke},\ and\ \citenamefont
  {Schaffner-Bielich}}]{Lenaghan:2000ey}%
  \BibitemOpen
  \bibfield  {author} {\bibinfo {author} {\bibfnamefont {J.~T.}\ \bibnamefont
  {Lenaghan}}, \bibinfo {author} {\bibfnamefont {D.~H.}\ \bibnamefont
  {Rischke}},\ and\ \bibinfo {author} {\bibfnamefont {J.}~\bibnamefont
  {Schaffner-Bielich}},\ }\bibfield  {title} {\bibinfo {title} {{Chiral
  symmetry restoration at nonzero temperature in the SU(3)(r) x SU(3)(l) linear
  sigma model}},\ }\href {https://doi.org/10.1103/PhysRevD.62.085008}
  {\bibfield  {journal} {\bibinfo  {journal} {Phys. Rev. D}\ }\textbf {\bibinfo
  {volume} {62}},\ \bibinfo {pages} {085008} (\bibinfo {year} {2000})},\
  \Eprint {https://arxiv.org/abs/nucl-th/0004006} {arXiv:nucl-th/0004006}
  \BibitemShut {NoStop}%
\bibitem [{\citenamefont {Gorda}\ \emph
  {et~al.}(2023{\natexlab{a}})\citenamefont {Gorda}, \citenamefont
  {Paatelainen}, \citenamefont {S\"appi},\ and\ \citenamefont
  {Sepp\"anen}}]{Gorda:2023mkk}%
  \BibitemOpen
  \bibfield  {author} {\bibinfo {author} {\bibfnamefont {T.}~\bibnamefont
  {Gorda}}, \bibinfo {author} {\bibfnamefont {R.}~\bibnamefont {Paatelainen}},
  \bibinfo {author} {\bibfnamefont {S.}~\bibnamefont {S\"appi}},\ and\ \bibinfo
  {author} {\bibfnamefont {K.}~\bibnamefont {Sepp\"anen}},\ }\bibfield  {title}
  {\bibinfo {title} {{Equation of State of Cold Quark Matter to $O(\alpha_s^3
  \ln \alpha_s)$}},\ }\href {https://doi.org/10.1103/PhysRevLett.131.181902}
  {\bibfield  {journal} {\bibinfo  {journal} {Phys. Rev. Lett.}\ }\textbf
  {\bibinfo {volume} {131}},\ \bibinfo {pages} {181902} (\bibinfo {year}
  {2023}{\natexlab{a}})},\ \Eprint {https://arxiv.org/abs/2307.08734}
  {arXiv:2307.08734 [hep-ph]} \BibitemShut {NoStop}%
\bibitem [{\citenamefont {Baym}\ \emph {et~al.}(2018)\citenamefont {Baym},
  \citenamefont {Hatsuda}, \citenamefont {Kojo}, \citenamefont {Powell},
  \citenamefont {Song},\ and\ \citenamefont {Takatsuka}}]{Baym:2017whm}%
  \BibitemOpen
  \bibfield  {author} {\bibinfo {author} {\bibfnamefont {G.}~\bibnamefont
  {Baym}}, \bibinfo {author} {\bibfnamefont {T.}~\bibnamefont {Hatsuda}},
  \bibinfo {author} {\bibfnamefont {T.}~\bibnamefont {Kojo}}, \bibinfo {author}
  {\bibfnamefont {P.~D.}\ \bibnamefont {Powell}}, \bibinfo {author}
  {\bibfnamefont {Y.}~\bibnamefont {Song}},\ and\ \bibinfo {author}
  {\bibfnamefont {T.}~\bibnamefont {Takatsuka}},\ }\bibfield  {title} {\bibinfo
  {title} {{From hadrons to quarks in neutron stars: a review}},\ }\href
  {https://doi.org/10.1088/1361-6633/aaae14} {\bibfield  {journal} {\bibinfo
  {journal} {Rept. Prog. Phys.}\ }\textbf {\bibinfo {volume} {81}},\ \bibinfo
  {pages} {056902} (\bibinfo {year} {2018})},\ \Eprint
  {https://arxiv.org/abs/1707.04966} {arXiv:1707.04966 [astro-ph.HE]}
  \BibitemShut {NoStop}%
\bibitem [{\citenamefont {Kurkela}\ \emph {et~al.}(2014)\citenamefont
  {Kurkela}, \citenamefont {Fraga}, \citenamefont {Schaffner-Bielich},\ and\
  \citenamefont {Vuorinen}}]{Kurkela:2014vha}%
  \BibitemOpen
  \bibfield  {author} {\bibinfo {author} {\bibfnamefont {A.}~\bibnamefont
  {Kurkela}}, \bibinfo {author} {\bibfnamefont {E.~S.}\ \bibnamefont {Fraga}},
  \bibinfo {author} {\bibfnamefont {J.}~\bibnamefont {Schaffner-Bielich}},\
  and\ \bibinfo {author} {\bibfnamefont {A.}~\bibnamefont {Vuorinen}},\
  }\bibfield  {title} {\bibinfo {title} {{Constraining neutron star matter with
  Quantum Chromodynamics}},\ }\href
  {https://doi.org/10.1088/0004-637X/789/2/127} {\bibfield  {journal} {\bibinfo
   {journal} {Astrophys. J.}\ }\textbf {\bibinfo {volume} {789}},\ \bibinfo
  {pages} {127} (\bibinfo {year} {2014})},\ \Eprint
  {https://arxiv.org/abs/1402.6618} {arXiv:1402.6618 [astro-ph.HE]}
  \BibitemShut {NoStop}%
\bibitem [{\citenamefont {Annala}\ \emph {et~al.}(2020)\citenamefont {Annala},
  \citenamefont {Gorda}, \citenamefont {Kurkela}, \citenamefont
  {N{\"a}ttil{\"a}},\ and\ \citenamefont {Vuorinen}}]{Annala:2019puf}%
  \BibitemOpen
  \bibfield  {author} {\bibinfo {author} {\bibfnamefont {E.}~\bibnamefont
  {Annala}}, \bibinfo {author} {\bibfnamefont {T.}~\bibnamefont {Gorda}},
  \bibinfo {author} {\bibfnamefont {A.}~\bibnamefont {Kurkela}}, \bibinfo
  {author} {\bibfnamefont {J.}~\bibnamefont {N{\"a}ttil{\"a}}},\ and\ \bibinfo
  {author} {\bibfnamefont {A.}~\bibnamefont {Vuorinen}},\ }\bibfield  {title}
  {\bibinfo {title} {{Evidence for quark-matter cores in massive neutron
  stars}},\ }\href {https://doi.org/10.1038/s41567-020-0914-9} {\bibfield
  {journal} {\bibinfo  {journal} {Nature Phys.}\ }\textbf {\bibinfo {volume}
  {16}},\ \bibinfo {pages} {907} (\bibinfo {year} {2020})},\ \Eprint
  {https://arxiv.org/abs/1903.09121} {arXiv:1903.09121 [astro-ph.HE]}
  \BibitemShut {NoStop}%
\bibitem [{\citenamefont {Annala}\ \emph {et~al.}(2022)\citenamefont {Annala},
  \citenamefont {Gorda}, \citenamefont {Katerini}, \citenamefont {Kurkela},
  \citenamefont {N{\"a}ttil{\"a}}, \citenamefont {Paschalidis},\ and\
  \citenamefont {Vuorinen}}]{Annala:2021gom}%
  \BibitemOpen
  \bibfield  {author} {\bibinfo {author} {\bibfnamefont {E.}~\bibnamefont
  {Annala}}, \bibinfo {author} {\bibfnamefont {T.}~\bibnamefont {Gorda}},
  \bibinfo {author} {\bibfnamefont {E.}~\bibnamefont {Katerini}}, \bibinfo
  {author} {\bibfnamefont {A.}~\bibnamefont {Kurkela}}, \bibinfo {author}
  {\bibfnamefont {J.}~\bibnamefont {N{\"a}ttil{\"a}}}, \bibinfo {author}
  {\bibfnamefont {V.}~\bibnamefont {Paschalidis}},\ and\ \bibinfo {author}
  {\bibfnamefont {A.}~\bibnamefont {Vuorinen}},\ }\bibfield  {title} {\bibinfo
  {title} {{Multimessenger Constraints for Ultradense Matter}},\ }\href
  {https://doi.org/10.1103/PhysRevX.12.011058} {\bibfield  {journal} {\bibinfo
  {journal} {Phys. Rev. X}\ }\textbf {\bibinfo {volume} {12}},\ \bibinfo
  {pages} {011058} (\bibinfo {year} {2022})},\ \Eprint
  {https://arxiv.org/abs/2105.05132} {arXiv:2105.05132 [astro-ph.HE]}
  \BibitemShut {NoStop}%
\bibitem [{\citenamefont {Altiparmak}\ \emph {et~al.}(2022)\citenamefont
  {Altiparmak}, \citenamefont {Ecker},\ and\ \citenamefont
  {Rezzolla}}]{Altiparmak:2022bke}%
  \BibitemOpen
  \bibfield  {author} {\bibinfo {author} {\bibfnamefont {S.}~\bibnamefont
  {Altiparmak}}, \bibinfo {author} {\bibfnamefont {C.}~\bibnamefont {Ecker}},\
  and\ \bibinfo {author} {\bibfnamefont {L.}~\bibnamefont {Rezzolla}},\
  }\bibfield  {title} {\bibinfo {title} {{On the Sound Speed in Neutron
  Stars}},\ }\href {https://doi.org/10.3847/2041-8213/ac9b2a} {\bibfield
  {journal} {\bibinfo  {journal} {Astrophys. J. Lett.}\ }\textbf {\bibinfo
  {volume} {939}},\ \bibinfo {pages} {L34} (\bibinfo {year} {2022})},\ \Eprint
  {https://arxiv.org/abs/2203.14974} {arXiv:2203.14974 [astro-ph.HE]}
  \BibitemShut {NoStop}%
\bibitem [{\citenamefont {Gorda}\ \emph
  {et~al.}(2023{\natexlab{b}})\citenamefont {Gorda}, \citenamefont {Komoltsev},
  \citenamefont {Kurkela},\ and\ \citenamefont {Mazeliauskas}}]{Gorda:2023usm}%
  \BibitemOpen
  \bibfield  {author} {\bibinfo {author} {\bibfnamefont {T.}~\bibnamefont
  {Gorda}}, \bibinfo {author} {\bibfnamefont {O.}~\bibnamefont {Komoltsev}},
  \bibinfo {author} {\bibfnamefont {A.}~\bibnamefont {Kurkela}},\ and\ \bibinfo
  {author} {\bibfnamefont {A.}~\bibnamefont {Mazeliauskas}},\ }\bibfield
  {title} {\bibinfo {title} {{Bayesian uncertainty quantification of
  perturbative QCD input to the neutron-star equation of state}},\ }\href
  {https://doi.org/10.1007/JHEP06(2023)002} {\bibfield  {journal} {\bibinfo
  {journal} {JHEP}\ }\textbf {\bibinfo {volume} {06}},\ \bibinfo {pages}
  {002}},\ \Eprint {https://arxiv.org/abs/2303.02175} {arXiv:2303.02175
  [hep-ph]} \BibitemShut {NoStop}%
\bibitem [{\citenamefont {Komoltsev}\ \emph {et~al.}(2024)\citenamefont
  {Komoltsev}, \citenamefont {Somasundaram}, \citenamefont {Gorda},
  \citenamefont {Kurkela}, \citenamefont {Margueron},\ and\ \citenamefont
  {Tews}}]{Komoltsev:2023zor}%
  \BibitemOpen
  \bibfield  {author} {\bibinfo {author} {\bibfnamefont {O.}~\bibnamefont
  {Komoltsev}}, \bibinfo {author} {\bibfnamefont {R.}~\bibnamefont
  {Somasundaram}}, \bibinfo {author} {\bibfnamefont {T.}~\bibnamefont {Gorda}},
  \bibinfo {author} {\bibfnamefont {A.}~\bibnamefont {Kurkela}}, \bibinfo
  {author} {\bibfnamefont {J.}~\bibnamefont {Margueron}},\ and\ \bibinfo
  {author} {\bibfnamefont {I.}~\bibnamefont {Tews}},\ }\bibfield  {title}
  {\bibinfo {title} {{Equation of state at neutron-star densities and beyond
  from perturbative QCD}},\ }\href
  {https://doi.org/10.1103/PhysRevD.109.094030} {\bibfield  {journal} {\bibinfo
   {journal} {Phys. Rev. D}\ }\textbf {\bibinfo {volume} {109}},\ \bibinfo
  {pages} {094030} (\bibinfo {year} {2024})},\ \Eprint
  {https://arxiv.org/abs/2312.14127} {arXiv:2312.14127 [nucl-th]} \BibitemShut
  {NoStop}%
\bibitem [{\citenamefont {Komoltsev}\ and\ \citenamefont
  {Kurkela}(2022)}]{Komoltsev:2021jzg}%
  \BibitemOpen
  \bibfield  {author} {\bibinfo {author} {\bibfnamefont {O.}~\bibnamefont
  {Komoltsev}}\ and\ \bibinfo {author} {\bibfnamefont {A.}~\bibnamefont
  {Kurkela}},\ }\bibfield  {title} {\bibinfo {title} {{How Perturbative QCD
  Constrains the Equation of State at Neutron-Star Densities}},\ }\href
  {https://doi.org/10.1103/PhysRevLett.128.202701} {\bibfield  {journal}
  {\bibinfo  {journal} {Phys. Rev. Lett.}\ }\textbf {\bibinfo {volume} {128}},\
  \bibinfo {pages} {202701} (\bibinfo {year} {2022})},\ \Eprint
  {https://arxiv.org/abs/2111.05350} {arXiv:2111.05350 [nucl-th]} \BibitemShut
  {NoStop}%
\bibitem [{\citenamefont {Lim}\ and\ \citenamefont {Holt}(2019)}]{Lim:2019som}%
  \BibitemOpen
  \bibfield  {author} {\bibinfo {author} {\bibfnamefont {Y.}~\bibnamefont
  {Lim}}\ and\ \bibinfo {author} {\bibfnamefont {J.~W.}\ \bibnamefont {Holt}},\
  }\bibfield  {title} {\bibinfo {title} {{Bayesian modeling of the nuclear
  equation of state for neutron star tidal deformabilities and GW170817}},\
  }\href {https://doi.org/10.1140/epja/i2019-12917-9} {\bibfield  {journal}
  {\bibinfo  {journal} {Eur. Phys. J. A}\ }\textbf {\bibinfo {volume} {55}},\
  \bibinfo {pages} {209} (\bibinfo {year} {2019})},\ \Eprint
  {https://arxiv.org/abs/1902.05502} {arXiv:1902.05502 [nucl-th]} \BibitemShut
  {NoStop}%
\bibitem [{\citenamefont {Traversi}\ \emph {et~al.}(2020)\citenamefont
  {Traversi}, \citenamefont {Char},\ and\ \citenamefont
  {Pagliara}}]{Traversi:2020aaa}%
  \BibitemOpen
  \bibfield  {author} {\bibinfo {author} {\bibfnamefont {S.}~\bibnamefont
  {Traversi}}, \bibinfo {author} {\bibfnamefont {P.}~\bibnamefont {Char}},\
  and\ \bibinfo {author} {\bibfnamefont {G.}~\bibnamefont {Pagliara}},\
  }\bibfield  {title} {\bibinfo {title} {{Bayesian Inference of Dense Matter
  Equation of State within Relativistic Mean Field Models using Astrophysical
  Measurements}},\ }\href {https://doi.org/10.3847/1538-4357/ab99c1} {\bibfield
   {journal} {\bibinfo  {journal} {Astrophys. J.}\ }\textbf {\bibinfo {volume}
  {897}},\ \bibinfo {pages} {165} (\bibinfo {year} {2020})},\ \Eprint
  {https://arxiv.org/abs/2002.08951} {arXiv:2002.08951 [astro-ph.HE]}
  \BibitemShut {NoStop}%
\bibitem [{\citenamefont {Zhu}\ \emph {et~al.}(2023)\citenamefont {Zhu},
  \citenamefont {Li},\ and\ \citenamefont {Liu}}]{Zhu:2022ibs}%
  \BibitemOpen
  \bibfield  {author} {\bibinfo {author} {\bibfnamefont {Z.}~\bibnamefont
  {Zhu}}, \bibinfo {author} {\bibfnamefont {A.}~\bibnamefont {Li}},\ and\
  \bibinfo {author} {\bibfnamefont {T.}~\bibnamefont {Liu}},\ }\bibfield
  {title} {\bibinfo {title} {{A Bayesian Inference of a Relativistic Mean-field
  Model of Neutron Star Matter from Observations of NICER and
  GW170817/AT2017gfo}},\ }\href {https://doi.org/10.3847/1538-4357/acac1f}
  {\bibfield  {journal} {\bibinfo  {journal} {Astrophys. J.}\ }\textbf
  {\bibinfo {volume} {943}},\ \bibinfo {pages} {163} (\bibinfo {year}
  {2023})},\ \Eprint {https://arxiv.org/abs/2211.02007} {arXiv:2211.02007
  [astro-ph.HE]} \BibitemShut {NoStop}%
\bibitem [{\citenamefont {Malik}\ and\ \citenamefont
  {Provid\^encia}(2022)}]{Malik:2022jqc}%
  \BibitemOpen
  \bibfield  {author} {\bibinfo {author} {\bibfnamefont {T.}~\bibnamefont
  {Malik}}\ and\ \bibinfo {author} {\bibfnamefont {C.}~\bibnamefont
  {Provid\^encia}},\ }\bibfield  {title} {\bibinfo {title} {{Bayesian inference
  of signatures of hyperons inside neutron stars}},\ }\href
  {https://doi.org/10.1103/PhysRevD.106.063024} {\bibfield  {journal} {\bibinfo
   {journal} {Phys. Rev. D}\ }\textbf {\bibinfo {volume} {106}},\ \bibinfo
  {pages} {063024} (\bibinfo {year} {2022})},\ \Eprint
  {https://arxiv.org/abs/2205.15843} {arXiv:2205.15843 [nucl-th]} \BibitemShut
  {NoStop}%
\bibitem [{\citenamefont {Malik}\ \emph {et~al.}(2023)\citenamefont {Malik},
  \citenamefont {Ferreira}, \citenamefont {Albino},\ and\ \citenamefont
  {Provid{\^e}ncia}}]{Malik:2023mnx}%
  \BibitemOpen
  \bibfield  {author} {\bibinfo {author} {\bibfnamefont {T.}~\bibnamefont
  {Malik}}, \bibinfo {author} {\bibfnamefont {M.}~\bibnamefont {Ferreira}},
  \bibinfo {author} {\bibfnamefont {M.~B.}\ \bibnamefont {Albino}},\ and\
  \bibinfo {author} {\bibfnamefont {C.}~\bibnamefont {Provid{\^e}ncia}},\
  }\bibfield  {title} {\bibinfo {title} {{Spanning the full range of neutron
  star properties within a microscopic description}},\ }\href
  {https://doi.org/10.1103/PhysRevD.107.103018} {\bibfield  {journal} {\bibinfo
   {journal} {Phys. Rev. D}\ }\textbf {\bibinfo {volume} {107}},\ \bibinfo
  {pages} {103018} (\bibinfo {year} {2023})},\ \Eprint
  {https://arxiv.org/abs/2301.08169} {arXiv:2301.08169 [nucl-th]} \BibitemShut
  {NoStop}%
\bibitem [{\citenamefont {Takatsy}\ \emph {et~al.}(2023)\citenamefont
  {Takatsy}, \citenamefont {Kovacs}, \citenamefont {Wolf},\ and\ \citenamefont
  {Schaffner-Bielich}}]{Takatsy:2023xzf}%
  \BibitemOpen
  \bibfield  {author} {\bibinfo {author} {\bibfnamefont {J.}~\bibnamefont
  {Takatsy}}, \bibinfo {author} {\bibfnamefont {P.}~\bibnamefont {Kovacs}},
  \bibinfo {author} {\bibfnamefont {G.}~\bibnamefont {Wolf}},\ and\ \bibinfo
  {author} {\bibfnamefont {J.}~\bibnamefont {Schaffner-Bielich}},\ }\bibfield
  {title} {\bibinfo {title} {{What neutron stars tell about the hadron-quark
  phase transition: A Bayesian study}},\ }\href
  {https://doi.org/10.1103/PhysRevD.108.043002} {\bibfield  {journal} {\bibinfo
   {journal} {Phys. Rev. D}\ }\textbf {\bibinfo {volume} {108}},\ \bibinfo
  {pages} {043002} (\bibinfo {year} {2023})},\ \Eprint
  {https://arxiv.org/abs/2303.00013} {arXiv:2303.00013 [astro-ph.HE]}
  \BibitemShut {NoStop}%
\bibitem [{\citenamefont {Zhou}(2025)}]{Zhou:2023zrm}%
  \BibitemOpen
  \bibfield  {author} {\bibinfo {author} {\bibfnamefont {D.}~\bibnamefont
  {Zhou}},\ }\bibfield  {title} {\bibinfo {title} {{Reexamining constraints on
  neutron star properties from perturbative QCD}},\ }\href
  {https://doi.org/10.1103/PhysRevC.111.015810} {\bibfield  {journal} {\bibinfo
   {journal} {Phys. Rev. C}\ }\textbf {\bibinfo {volume} {111}},\ \bibinfo
  {pages} {015810} (\bibinfo {year} {2025})},\ \Eprint
  {https://arxiv.org/abs/2307.11125} {arXiv:2307.11125 [astro-ph.HE]}
  \BibitemShut {NoStop}%
\bibitem [{\citenamefont {Gupta}\ \emph {et~al.}(2011)\citenamefont {Gupta},
  \citenamefont {Luo}, \citenamefont {Mohanty}, \citenamefont {Ritter},\ and\
  \citenamefont {Xu}}]{Gupta:2011wh}%
  \BibitemOpen
  \bibfield  {author} {\bibinfo {author} {\bibfnamefont {S.}~\bibnamefont
  {Gupta}}, \bibinfo {author} {\bibfnamefont {X.}~\bibnamefont {Luo}}, \bibinfo
  {author} {\bibfnamefont {B.}~\bibnamefont {Mohanty}}, \bibinfo {author}
  {\bibfnamefont {H.~G.}\ \bibnamefont {Ritter}},\ and\ \bibinfo {author}
  {\bibfnamefont {N.}~\bibnamefont {Xu}},\ }\bibfield  {title} {\bibinfo
  {title} {{Scale for the Phase Diagram of Quantum Chromodynamics}},\ }\href
  {https://doi.org/10.1126/science.1204621} {\bibfield  {journal} {\bibinfo
  {journal} {Science}\ }\textbf {\bibinfo {volume} {332}},\ \bibinfo {pages}
  {1525} (\bibinfo {year} {2011})},\ \Eprint {https://arxiv.org/abs/1105.3934}
  {arXiv:1105.3934 [hep-ph]} \BibitemShut {NoStop}%
\bibitem [{\citenamefont {Blaizot}\ \emph {et~al.}(2003)\citenamefont
  {Blaizot}, \citenamefont {Iancu},\ and\ \citenamefont
  {Rebhan}}]{Blaizot:2003tw}%
  \BibitemOpen
  \bibfield  {author} {\bibinfo {author} {\bibfnamefont {J.-P.}\ \bibnamefont
  {Blaizot}}, \bibinfo {author} {\bibfnamefont {E.}~\bibnamefont {Iancu}},\
  and\ \bibinfo {author} {\bibfnamefont {A.}~\bibnamefont {Rebhan}},\
  }\bibfield  {title} {\bibinfo {title} {{Thermodynamics of the high
  temperature quark gluon plasma}}\ }(\bibinfo {year} {2003})\ pp.\ \bibinfo
  {pages} {60--122},\ \Eprint {https://arxiv.org/abs/hep-ph/0303185}
  {arXiv:hep-ph/0303185} \BibitemShut {NoStop}%
\bibitem [{\citenamefont {Kraemmer}\ and\ \citenamefont
  {Rebhan}(2004)}]{Kraemmer:2003gd}%
  \BibitemOpen
  \bibfield  {author} {\bibinfo {author} {\bibfnamefont {U.}~\bibnamefont
  {Kraemmer}}\ and\ \bibinfo {author} {\bibfnamefont {A.}~\bibnamefont
  {Rebhan}},\ }\bibfield  {title} {\bibinfo {title} {{Advances in perturbative
  thermal field theory}},\ }\href {https://doi.org/10.1088/0034-4885/67/3/R05}
  {\bibfield  {journal} {\bibinfo  {journal} {Rept. Prog. Phys.}\ }\textbf
  {\bibinfo {volume} {67}},\ \bibinfo {pages} {351} (\bibinfo {year} {2004})},\
  \Eprint {https://arxiv.org/abs/hep-ph/0310337} {arXiv:hep-ph/0310337}
  \BibitemShut {NoStop}%
\bibitem [{\citenamefont {Ghiglieri}\ \emph {et~al.}(2020)\citenamefont
  {Ghiglieri}, \citenamefont {Kurkela}, \citenamefont {Strickland},\ and\
  \citenamefont {Vuorinen}}]{Ghiglieri:2020dpq}%
  \BibitemOpen
  \bibfield  {author} {\bibinfo {author} {\bibfnamefont {J.}~\bibnamefont
  {Ghiglieri}}, \bibinfo {author} {\bibfnamefont {A.}~\bibnamefont {Kurkela}},
  \bibinfo {author} {\bibfnamefont {M.}~\bibnamefont {Strickland}},\ and\
  \bibinfo {author} {\bibfnamefont {A.}~\bibnamefont {Vuorinen}},\ }\bibfield
  {title} {\bibinfo {title} {{Perturbative Thermal QCD: Formalism and
  Applications}},\ }\href {https://doi.org/10.1016/j.physrep.2020.07.004}
  {\bibfield  {journal} {\bibinfo  {journal} {Phys. Rept.}\ }\textbf {\bibinfo
  {volume} {880}},\ \bibinfo {pages} {1} (\bibinfo {year} {2020})},\ \Eprint
  {https://arxiv.org/abs/2002.10188} {arXiv:2002.10188 [hep-ph]} \BibitemShut
  {NoStop}%
\bibitem [{\citenamefont {Braaten}\ and\ \citenamefont
  {Pisarski}(1992)}]{Braaten:1991gm}%
  \BibitemOpen
  \bibfield  {author} {\bibinfo {author} {\bibfnamefont {E.}~\bibnamefont
  {Braaten}}\ and\ \bibinfo {author} {\bibfnamefont {R.~D.}\ \bibnamefont
  {Pisarski}},\ }\bibfield  {title} {\bibinfo {title} {{Simple effective
  Lagrangian for hard thermal loops}},\ }\href
  {https://doi.org/10.1103/PhysRevD.45.R1827} {\bibfield  {journal} {\bibinfo
  {journal} {Phys. Rev. D}\ }\textbf {\bibinfo {volume} {45}},\ \bibinfo
  {pages} {R1827} (\bibinfo {year} {1992})}\BibitemShut {NoStop}%
\bibitem [{\citenamefont {Gorda}\ \emph
  {et~al.}(2021{\natexlab{b}})\citenamefont {Gorda}, \citenamefont {Kurkela},
  \citenamefont {Paatelainen}, \citenamefont {S\"appi},\ and\ \citenamefont
  {Vuorinen}}]{Gorda:2021znl}%
  \BibitemOpen
  \bibfield  {author} {\bibinfo {author} {\bibfnamefont {T.}~\bibnamefont
  {Gorda}}, \bibinfo {author} {\bibfnamefont {A.}~\bibnamefont {Kurkela}},
  \bibinfo {author} {\bibfnamefont {R.}~\bibnamefont {Paatelainen}}, \bibinfo
  {author} {\bibfnamefont {S.}~\bibnamefont {S\"appi}},\ and\ \bibinfo {author}
  {\bibfnamefont {A.}~\bibnamefont {Vuorinen}},\ }\bibfield  {title} {\bibinfo
  {title} {{Soft Interactions in Cold Quark Matter}},\ }\href
  {https://doi.org/10.1103/PhysRevLett.127.162003} {\bibfield  {journal}
  {\bibinfo  {journal} {Phys. Rev. Lett.}\ }\textbf {\bibinfo {volume} {127}},\
  \bibinfo {pages} {162003} (\bibinfo {year} {2021}{\natexlab{b}})},\ \Eprint
  {https://arxiv.org/abs/2103.05658} {arXiv:2103.05658 [hep-ph]} \BibitemShut
  {NoStop}%
\bibitem [{\citenamefont {Fernandez}\ and\ \citenamefont
  {Kneur}(2022)}]{Fernandez:2021jfr}%
  \BibitemOpen
  \bibfield  {author} {\bibinfo {author} {\bibfnamefont {L.}~\bibnamefont
  {Fernandez}}\ and\ \bibinfo {author} {\bibfnamefont {J.-L.}\ \bibnamefont
  {Kneur}},\ }\bibfield  {title} {\bibinfo {title} {{All Order Resummed Leading
  and Next-to-Leading Soft Modes of Dense QCD Pressure}},\ }\href
  {https://doi.org/10.1103/PhysRevLett.129.212001} {\bibfield  {journal}
  {\bibinfo  {journal} {Phys. Rev. Lett.}\ }\textbf {\bibinfo {volume} {129}},\
  \bibinfo {pages} {212001} (\bibinfo {year} {2022})},\ \Eprint
  {https://arxiv.org/abs/2109.02410} {arXiv:2109.02410 [hep-ph]} \BibitemShut
  {NoStop}%
\bibitem [{\citenamefont {Fraga}\ and\ \citenamefont
  {Romatschke}(2005)}]{Fraga:2004gz}%
  \BibitemOpen
  \bibfield  {author} {\bibinfo {author} {\bibfnamefont {E.~S.}\ \bibnamefont
  {Fraga}}\ and\ \bibinfo {author} {\bibfnamefont {P.}~\bibnamefont
  {Romatschke}},\ }\bibfield  {title} {\bibinfo {title} {{The Role of quark
  mass in cold and dense perturbative QCD}},\ }\href
  {https://doi.org/10.1103/PhysRevD.71.105014} {\bibfield  {journal} {\bibinfo
  {journal} {Phys. Rev. D}\ }\textbf {\bibinfo {volume} {71}},\ \bibinfo
  {pages} {105014} (\bibinfo {year} {2005})},\ \Eprint
  {https://arxiv.org/abs/hep-ph/0412298} {arXiv:hep-ph/0412298} \BibitemShut
  {NoStop}%
\bibitem [{\citenamefont {Laine}\ and\ \citenamefont
  {Schroder}(2006)}]{Laine:2006cp}%
  \BibitemOpen
  \bibfield  {author} {\bibinfo {author} {\bibfnamefont {M.}~\bibnamefont
  {Laine}}\ and\ \bibinfo {author} {\bibfnamefont {Y.}~\bibnamefont
  {Schroder}},\ }\bibfield  {title} {\bibinfo {title} {{Quark mass thresholds
  in QCD thermodynamics}},\ }\href {https://doi.org/10.1103/PhysRevD.73.085009}
  {\bibfield  {journal} {\bibinfo  {journal} {Phys. Rev. D}\ }\textbf {\bibinfo
  {volume} {73}},\ \bibinfo {pages} {085009} (\bibinfo {year} {2006})},\
  \Eprint {https://arxiv.org/abs/hep-ph/0603048} {arXiv:hep-ph/0603048}
  \BibitemShut {NoStop}%
\bibitem [{\citenamefont {Graf}\ \emph {et~al.}(2016)\citenamefont {Graf},
  \citenamefont {Schaffner-Bielich},\ and\ \citenamefont
  {Fraga}}]{Graf:2015tda}%
  \BibitemOpen
  \bibfield  {author} {\bibinfo {author} {\bibfnamefont {T.}~\bibnamefont
  {Graf}}, \bibinfo {author} {\bibfnamefont {J.}~\bibnamefont
  {Schaffner-Bielich}},\ and\ \bibinfo {author} {\bibfnamefont {E.~S.}\
  \bibnamefont {Fraga}},\ }\bibfield  {title} {\bibinfo {title} {{The impact of
  quark masses on pQCD thermodynamics}},\ }\href
  {https://doi.org/10.1140/epja/i2016-16208-9} {\bibfield  {journal} {\bibinfo
  {journal} {Eur. Phys. J. A}\ }\textbf {\bibinfo {volume} {52}},\ \bibinfo
  {pages} {208} (\bibinfo {year} {2016})},\ \Eprint
  {https://arxiv.org/abs/1507.08941} {arXiv:1507.08941 [hep-ph]} \BibitemShut
  {NoStop}%
\bibitem [{\citenamefont {Ipp}\ \emph {et~al.}(2006)\citenamefont {Ipp},
  \citenamefont {Kajantie}, \citenamefont {Rebhan},\ and\ \citenamefont
  {Vuorinen}}]{Ipp:2006ij}%
  \BibitemOpen
  \bibfield  {author} {\bibinfo {author} {\bibfnamefont {A.}~\bibnamefont
  {Ipp}}, \bibinfo {author} {\bibfnamefont {K.}~\bibnamefont {Kajantie}},
  \bibinfo {author} {\bibfnamefont {A.}~\bibnamefont {Rebhan}},\ and\ \bibinfo
  {author} {\bibfnamefont {A.}~\bibnamefont {Vuorinen}},\ }\bibfield  {title}
  {\bibinfo {title} {{The Pressure of deconfined QCD for all temperatures and
  quark chemical potentials}},\ }\href
  {https://doi.org/10.1103/PhysRevD.74.045016} {\bibfield  {journal} {\bibinfo
  {journal} {Phys. Rev. D}\ }\textbf {\bibinfo {volume} {74}},\ \bibinfo
  {pages} {045016} (\bibinfo {year} {2006})},\ \Eprint
  {https://arxiv.org/abs/hep-ph/0604060} {arXiv:hep-ph/0604060} \BibitemShut
  {NoStop}%
\bibitem [{\citenamefont {Kurkela}\ and\ \citenamefont
  {Vuorinen}(2016)}]{Kurkela:2016was}%
  \BibitemOpen
  \bibfield  {author} {\bibinfo {author} {\bibfnamefont {A.}~\bibnamefont
  {Kurkela}}\ and\ \bibinfo {author} {\bibfnamefont {A.}~\bibnamefont
  {Vuorinen}},\ }\bibfield  {title} {\bibinfo {title} {{Cool quark matter}},\
  }\href {https://doi.org/10.1103/PhysRevLett.117.042501} {\bibfield  {journal}
  {\bibinfo  {journal} {Phys. Rev. Lett.}\ }\textbf {\bibinfo {volume} {117}},\
  \bibinfo {pages} {042501} (\bibinfo {year} {2016})},\ \Eprint
  {https://arxiv.org/abs/1603.00750} {arXiv:1603.00750 [hep-ph]} \BibitemShut
  {NoStop}%
\bibitem [{\citenamefont {K{\"a}rkk{\"a}inen}\ \emph
  {et~al.}(2025)\citenamefont {K{\"a}rkk{\"a}inen}, \citenamefont {Navarrete},
  \citenamefont {Nurmela}, \citenamefont {Paatelainen}, \citenamefont
  {Sepp{\"a}nen},\ and\ \citenamefont {Vuorinen}}]{Karkkainen:2025nkz}%
  \BibitemOpen
  \bibfield  {author} {\bibinfo {author} {\bibfnamefont {A.}~\bibnamefont
  {K{\"a}rkk{\"a}inen}}, \bibinfo {author} {\bibfnamefont {P.}~\bibnamefont
  {Navarrete}}, \bibinfo {author} {\bibfnamefont {M.}~\bibnamefont {Nurmela}},
  \bibinfo {author} {\bibfnamefont {R.}~\bibnamefont {Paatelainen}}, \bibinfo
  {author} {\bibfnamefont {K.}~\bibnamefont {Sepp{\"a}nen}},\ and\ \bibinfo
  {author} {\bibfnamefont {A.}~\bibnamefont {Vuorinen}},\ }\bibfield  {title}
  {\bibinfo {title} {{Quark Matter at Four Loops: Hardships and How to Overcome
  Them}},\ }\href {https://doi.org/10.1103/627n-5g6l} {\bibfield  {journal}
  {\bibinfo  {journal} {Phys. Rev. Lett.}\ }\textbf {\bibinfo {volume} {135}},\
  \bibinfo {pages} {021901} (\bibinfo {year} {2025})},\ \Eprint
  {https://arxiv.org/abs/2501.17921} {arXiv:2501.17921 [hep-ph]} \BibitemShut
  {NoStop}%
\bibitem [{\citenamefont {Kneur}\ and\ \citenamefont
  {Neveu}(2013)}]{Kneur:2013coa}%
  \BibitemOpen
  \bibfield  {author} {\bibinfo {author} {\bibfnamefont {J.-L.}\ \bibnamefont
  {Kneur}}\ and\ \bibinfo {author} {\bibfnamefont {A.}~\bibnamefont {Neveu}},\
  }\bibfield  {title} {\bibinfo {title} {{$\alpha_S$ from $F_\pi$ and
  Renormalization Group Optimized Perturbation Theory}},\ }\href
  {https://doi.org/10.1103/PhysRevD.88.074025} {\bibfield  {journal} {\bibinfo
  {journal} {Phys. Rev. D}\ }\textbf {\bibinfo {volume} {88}},\ \bibinfo
  {pages} {074025} (\bibinfo {year} {2013})},\ \Eprint
  {https://arxiv.org/abs/1305.6910} {arXiv:1305.6910 [hep-ph]} \BibitemShut
  {NoStop}%
\bibitem [{\citenamefont {Kneur}\ and\ \citenamefont
  {Neveu}(2015)}]{Kneur:2015dda}%
  \BibitemOpen
  \bibfield  {author} {\bibinfo {author} {\bibfnamefont {J.-L.}\ \bibnamefont
  {Kneur}}\ and\ \bibinfo {author} {\bibfnamefont {A.}~\bibnamefont {Neveu}},\
  }\bibfield  {title} {\bibinfo {title} {{Chiral condensate from
  renormalization group optimized perturbation}},\ }\href
  {https://doi.org/10.1103/PhysRevD.92.074027} {\bibfield  {journal} {\bibinfo
  {journal} {Phys. Rev. D}\ }\textbf {\bibinfo {volume} {92}},\ \bibinfo
  {pages} {074027} (\bibinfo {year} {2015})},\ \Eprint
  {https://arxiv.org/abs/1506.07506} {arXiv:1506.07506 [hep-ph]} \BibitemShut
  {NoStop}%
\bibitem [{\citenamefont {Kneur}\ and\ \citenamefont
  {Pinto}(2016)}]{Kneur:2015uha}%
  \BibitemOpen
  \bibfield  {author} {\bibinfo {author} {\bibfnamefont {J.~L.}\ \bibnamefont
  {Kneur}}\ and\ \bibinfo {author} {\bibfnamefont {M.~B.}\ \bibnamefont
  {Pinto}},\ }\bibfield  {title} {\bibinfo {title} {{Scale Invariant Resummed
  Perturbation at Finite Temperatures}},\ }\href
  {https://doi.org/10.1103/PhysRevLett.116.031601} {\bibfield  {journal}
  {\bibinfo  {journal} {Phys. Rev. Lett.}\ }\textbf {\bibinfo {volume} {116}},\
  \bibinfo {pages} {031601} (\bibinfo {year} {2016})},\ \Eprint
  {https://arxiv.org/abs/1507.03508} {arXiv:1507.03508 [hep-ph]} \BibitemShut
  {NoStop}%
\bibitem [{\citenamefont {Parwani}(1992)}]{Parwani:1991gq}%
  \BibitemOpen
  \bibfield  {author} {\bibinfo {author} {\bibfnamefont {R.~R.}\ \bibnamefont
  {Parwani}},\ }\bibfield  {title} {\bibinfo {title} {{Resummation in a hot
  scalar field theory}},\ }\href {https://doi.org/10.1103/PhysRevD.45.4695}
  {\bibfield  {journal} {\bibinfo  {journal} {Phys. Rev. D}\ }\textbf {\bibinfo
  {volume} {45}},\ \bibinfo {pages} {4695} (\bibinfo {year} {1992})},\ \bibinfo
  {note} {[Erratum: Phys.Rev.D 48, 5965 (1993)]},\ \Eprint
  {https://arxiv.org/abs/hep-ph/9204216} {arXiv:hep-ph/9204216} \BibitemShut
  {NoStop}%
\bibitem [{\citenamefont {Karsch}\ \emph {et~al.}(1997)\citenamefont {Karsch},
  \citenamefont {Patkos},\ and\ \citenamefont {Petreczky}}]{Karsch1997}%
  \BibitemOpen
  \bibfield  {author} {\bibinfo {author} {\bibfnamefont {F.}~\bibnamefont
  {Karsch}}, \bibinfo {author} {\bibfnamefont {A.}~\bibnamefont {Patkos}},\
  and\ \bibinfo {author} {\bibfnamefont {P.}~\bibnamefont {Petreczky}},\
  }\bibfield  {title} {\bibinfo {title} {{Screened perturbation theory}},\
  }\href {https://doi.org/10.1016/S0370-2693(97)00392-4} {\bibfield  {journal}
  {\bibinfo  {journal} {Phys. Lett. B}\ }\textbf {\bibinfo {volume} {401}},\
  \bibinfo {pages} {69} (\bibinfo {year} {1997})},\ \Eprint
  {https://arxiv.org/abs/hep-ph/9702376} {arXiv:hep-ph/9702376} \BibitemShut
  {NoStop}%
\bibitem [{\citenamefont {Andersen}\ \emph {et~al.}(2001)\citenamefont
  {Andersen}, \citenamefont {Braaten},\ and\ \citenamefont
  {Strickland}}]{Andersen:2000yj}%
  \BibitemOpen
  \bibfield  {author} {\bibinfo {author} {\bibfnamefont {J.~O.}\ \bibnamefont
  {Andersen}}, \bibinfo {author} {\bibfnamefont {E.}~\bibnamefont {Braaten}},\
  and\ \bibinfo {author} {\bibfnamefont {M.}~\bibnamefont {Strickland}},\
  }\bibfield  {title} {\bibinfo {title} {{Screened perturbation theory to three
  loops}},\ }\href {https://doi.org/10.1103/PhysRevD.63.105008} {\bibfield
  {journal} {\bibinfo  {journal} {Phys. Rev. D}\ }\textbf {\bibinfo {volume}
  {63}},\ \bibinfo {pages} {105008} (\bibinfo {year} {2001})},\ \Eprint
  {https://arxiv.org/abs/hep-ph/0007159} {arXiv:hep-ph/0007159} \BibitemShut
  {NoStop}%
\bibitem [{\citenamefont {Andersen}\ \emph {et~al.}(1999)\citenamefont
  {Andersen}, \citenamefont {Braaten},\ and\ \citenamefont
  {Strickland}}]{Andersen:1999fw}%
  \BibitemOpen
  \bibfield  {author} {\bibinfo {author} {\bibfnamefont {J.~O.}\ \bibnamefont
  {Andersen}}, \bibinfo {author} {\bibfnamefont {E.}~\bibnamefont {Braaten}},\
  and\ \bibinfo {author} {\bibfnamefont {M.}~\bibnamefont {Strickland}},\
  }\bibfield  {title} {\bibinfo {title} {{Hard thermal loop resummation of the
  free energy of a hot gluon plasma}},\ }\href
  {https://doi.org/10.1103/PhysRevLett.83.2139} {\bibfield  {journal} {\bibinfo
   {journal} {Phys. Rev. Lett.}\ }\textbf {\bibinfo {volume} {83}},\ \bibinfo
  {pages} {2139} (\bibinfo {year} {1999})},\ \Eprint
  {https://arxiv.org/abs/hep-ph/9902327} {arXiv:hep-ph/9902327} \BibitemShut
  {NoStop}%
\bibitem [{\citenamefont {Andersen}\ \emph {et~al.}(2000)\citenamefont
  {Andersen}, \citenamefont {Braaten},\ and\ \citenamefont
  {Strickland}}]{Andersen:1999va}%
  \BibitemOpen
  \bibfield  {author} {\bibinfo {author} {\bibfnamefont {J.~O.}\ \bibnamefont
  {Andersen}}, \bibinfo {author} {\bibfnamefont {E.}~\bibnamefont {Braaten}},\
  and\ \bibinfo {author} {\bibfnamefont {M.}~\bibnamefont {Strickland}},\
  }\bibfield  {title} {\bibinfo {title} {{Hard thermal loop resummation of the
  free energy of a hot quark - gluon plasma}},\ }\href
  {https://doi.org/10.1103/PhysRevD.61.074016} {\bibfield  {journal} {\bibinfo
  {journal} {Phys. Rev. D}\ }\textbf {\bibinfo {volume} {61}},\ \bibinfo
  {pages} {074016} (\bibinfo {year} {2000})},\ \Eprint
  {https://arxiv.org/abs/hep-ph/9908323} {arXiv:hep-ph/9908323} \BibitemShut
  {NoStop}%
\bibitem [{\citenamefont {Kneur}\ and\ \citenamefont
  {Pinto}(2015)}]{Kneur:2015moa}%
  \BibitemOpen
  \bibfield  {author} {\bibinfo {author} {\bibfnamefont {J.~L.}\ \bibnamefont
  {Kneur}}\ and\ \bibinfo {author} {\bibfnamefont {M.~B.}\ \bibnamefont
  {Pinto}},\ }\bibfield  {title} {\bibinfo {title} {{Renormalization Group
  Optimized Perturbation Theory at Finite Temperatures}},\ }\href
  {https://doi.org/10.1103/PhysRevD.92.116008} {\bibfield  {journal} {\bibinfo
  {journal} {Phys. Rev. D}\ }\textbf {\bibinfo {volume} {92}},\ \bibinfo
  {pages} {116008} (\bibinfo {year} {2015})},\ \Eprint
  {https://arxiv.org/abs/1508.02610} {arXiv:1508.02610 [hep-ph]} \BibitemShut
  {NoStop}%
\bibitem [{\citenamefont {Kneur}\ \emph
  {et~al.}(2021{\natexlab{a}})\citenamefont {Kneur}, \citenamefont {Pinto},\
  and\ \citenamefont {Restrepo}}]{Kneur:2021dfo}%
  \BibitemOpen
  \bibfield  {author} {\bibinfo {author} {\bibfnamefont {J.-L.}\ \bibnamefont
  {Kneur}}, \bibinfo {author} {\bibfnamefont {M.~B.}\ \bibnamefont {Pinto}},\
  and\ \bibinfo {author} {\bibfnamefont {T.~E.}\ \bibnamefont {Restrepo}},\
  }\bibfield  {title} {\bibinfo {title} {{Renormalization group improved
  pressure for hot and dense quark matter}},\ }\href
  {https://doi.org/10.1103/PhysRevD.104.034003} {\bibfield  {journal} {\bibinfo
   {journal} {Phys. Rev. D}\ }\textbf {\bibinfo {volume} {104}},\ \bibinfo
  {pages} {034003} (\bibinfo {year} {2021}{\natexlab{a}})},\ \Eprint
  {https://arxiv.org/abs/2101.08240} {arXiv:2101.08240 [hep-ph]} \BibitemShut
  {NoStop}%
\bibitem [{\citenamefont {Kneur}\ \emph
  {et~al.}(2021{\natexlab{b}})\citenamefont {Kneur}, \citenamefont {Pinto},\
  and\ \citenamefont {Restrepo}}]{Kneur:2021feo}%
  \BibitemOpen
  \bibfield  {author} {\bibinfo {author} {\bibfnamefont {J.-L.}\ \bibnamefont
  {Kneur}}, \bibinfo {author} {\bibfnamefont {M.~B.}\ \bibnamefont {Pinto}},\
  and\ \bibinfo {author} {\bibfnamefont {T.~E.}\ \bibnamefont {Restrepo}},\
  }\bibfield  {title} {\bibinfo {title} {{QCD pressure: Renormalization group
  optimized perturbation theory confronts lattice}},\ }\href
  {https://doi.org/10.1103/PhysRevD.104.L031502} {\bibfield  {journal}
  {\bibinfo  {journal} {Phys. Rev. D}\ }\textbf {\bibinfo {volume} {104}},\
  \bibinfo {pages} {L031502} (\bibinfo {year} {2021}{\natexlab{b}})},\ \Eprint
  {https://arxiv.org/abs/2101.02124} {arXiv:2101.02124 [hep-ph]} \BibitemShut
  {NoStop}%
\bibitem [{\citenamefont {Kneur}\ \emph {et~al.}(2019)\citenamefont {Kneur},
  \citenamefont {Pinto},\ and\ \citenamefont {Restrepo}}]{Kneur:2019tao}%
  \BibitemOpen
  \bibfield  {author} {\bibinfo {author} {\bibfnamefont {J.-L.}\ \bibnamefont
  {Kneur}}, \bibinfo {author} {\bibfnamefont {M.~B.}\ \bibnamefont {Pinto}},\
  and\ \bibinfo {author} {\bibfnamefont {T.~E.}\ \bibnamefont {Restrepo}},\
  }\bibfield  {title} {\bibinfo {title} {{Renormalization group improved
  pressure for cold and dense QCD}},\ }\href
  {https://doi.org/10.1103/PhysRevD.100.114006} {\bibfield  {journal} {\bibinfo
   {journal} {Phys. Rev. D}\ }\textbf {\bibinfo {volume} {100}},\ \bibinfo
  {pages} {114006} (\bibinfo {year} {2019})},\ \Eprint
  {https://arxiv.org/abs/1908.08363} {arXiv:1908.08363 [hep-ph]} \BibitemShut
  {NoStop}%
\bibitem [{\citenamefont {Restrepo}\ \emph {et~al.}(2023)\citenamefont
  {Restrepo}, \citenamefont {Provid\^encia},\ and\ \citenamefont
  {Pinto}}]{Restrepo:2022wqn}%
  \BibitemOpen
  \bibfield  {author} {\bibinfo {author} {\bibfnamefont {T.~E.}\ \bibnamefont
  {Restrepo}}, \bibinfo {author} {\bibfnamefont {C.}~\bibnamefont
  {Provid\^encia}},\ and\ \bibinfo {author} {\bibfnamefont {M.~B.}\
  \bibnamefont {Pinto}},\ }\bibfield  {title} {\bibinfo {title} {{Nonstrange
  quark stars within resummed QCD}},\ }\href
  {https://doi.org/10.1103/PhysRevD.107.114015} {\bibfield  {journal} {\bibinfo
   {journal} {Phys. Rev. D}\ }\textbf {\bibinfo {volume} {107}},\ \bibinfo
  {pages} {114015} (\bibinfo {year} {2023})},\ \Eprint
  {https://arxiv.org/abs/2212.11184} {arXiv:2212.11184 [hep-ph]} \BibitemShut
  {NoStop}%
\bibitem [{\citenamefont {Restrepo}\ \emph {et~al.}(2025)\citenamefont
  {Restrepo}, \citenamefont {Kneur}, \citenamefont {Provid{\^e}ncia},\ and\
  \citenamefont {Pinto}}]{Restrepo:2025qgp}%
  \BibitemOpen
  \bibfield  {author} {\bibinfo {author} {\bibfnamefont {T.~E.}\ \bibnamefont
  {Restrepo}}, \bibinfo {author} {\bibfnamefont {J.-L.}\ \bibnamefont {Kneur}},
  \bibinfo {author} {\bibfnamefont {C.}~\bibnamefont {Provid{\^e}ncia}},\ and\
  \bibinfo {author} {\bibfnamefont {M.~B.}\ \bibnamefont {Pinto}},\ }\bibfield
  {title} {\bibinfo {title} {{Comparing strange and nonstrange quark stars
  within resummed QCD at NLO}},\ }\href {https://doi.org/10.1103/7x41-j7mv}
  {\bibfield  {journal} {\bibinfo  {journal} {Phys. Rev. D}\ }\textbf {\bibinfo
  {volume} {112}},\ \bibinfo {pages} {054013} (\bibinfo {year} {2025})},\
  \Eprint {https://arxiv.org/abs/2501.14935} {arXiv:2501.14935 [hep-ph]}
  \BibitemShut {NoStop}%
\bibitem [{\citenamefont {Fernandez}\ and\ \citenamefont
  {Kneur}(2025)}]{Fernandez:2024ilg}%
  \BibitemOpen
  \bibfield  {author} {\bibinfo {author} {\bibfnamefont {L.}~\bibnamefont
  {Fernandez}}\ and\ \bibinfo {author} {\bibfnamefont {J.-L.}\ \bibnamefont
  {Kneur}},\ }\bibfield  {title} {\bibinfo {title} {{Cold quark matter:
  Renormalization group improvement at next-to-next-to leading order}},\ }\href
  {https://doi.org/10.1103/PhysRevD.111.034020} {\bibfield  {journal} {\bibinfo
   {journal} {Phys. Rev. D}\ }\textbf {\bibinfo {volume} {111}},\ \bibinfo
  {pages} {034020} (\bibinfo {year} {2025})},\ \Eprint
  {https://arxiv.org/abs/2408.16674} {arXiv:2408.16674 [hep-ph]} \BibitemShut
  {NoStop}%
\bibitem [{\citenamefont {Riley}\ \emph {et~al.}(2019)\citenamefont {Riley}
  \emph {et~al.}}]{Riley:2019yda}%
  \BibitemOpen
  \bibfield  {author} {\bibinfo {author} {\bibfnamefont {T.~E.}\ \bibnamefont
  {Riley}} \emph {et~al.},\ }\bibfield  {title} {\bibinfo {title} {{A $NICER$
  View of PSR J0030+0451: Millisecond Pulsar Parameter Estimation}},\ }\href
  {https://doi.org/10.3847/2041-8213/ab481c} {\bibfield  {journal} {\bibinfo
  {journal} {Astrophys. J. Lett.}\ }\textbf {\bibinfo {volume} {887}},\
  \bibinfo {pages} {L21} (\bibinfo {year} {2019})},\ \Eprint
  {https://arxiv.org/abs/1912.05702} {arXiv:1912.05702 [astro-ph.HE]}
  \BibitemShut {NoStop}%
\bibitem [{\citenamefont {Miller}\ \emph {et~al.}(2019)\citenamefont {Miller}
  \emph {et~al.}}]{Miller:2019cac}%
  \BibitemOpen
  \bibfield  {author} {\bibinfo {author} {\bibfnamefont {M.~C.}\ \bibnamefont
  {Miller}} \emph {et~al.},\ }\bibfield  {title} {\bibinfo {title} {{PSR
  J0030+0451 Mass and Radius from $NICER$ Data and Implications for the
  Properties of Neutron Star Matter}},\ }\href
  {https://doi.org/10.3847/2041-8213/ab50c5} {\bibfield  {journal} {\bibinfo
  {journal} {Astrophys. J. Lett.}\ }\textbf {\bibinfo {volume} {887}},\
  \bibinfo {pages} {L24} (\bibinfo {year} {2019})},\ \Eprint
  {https://arxiv.org/abs/1912.05705} {arXiv:1912.05705 [astro-ph.HE]}
  \BibitemShut {NoStop}%
\bibitem [{\citenamefont {Riley}\ \emph {et~al.}(2021)\citenamefont {Riley}
  \emph {et~al.}}]{Riley:2021pdl}%
  \BibitemOpen
  \bibfield  {author} {\bibinfo {author} {\bibfnamefont {T.~E.}\ \bibnamefont
  {Riley}} \emph {et~al.},\ }\bibfield  {title} {\bibinfo {title} {{A NICER
  View of the Massive Pulsar PSR J0740+6620 Informed by Radio Timing and
  XMM-Newton Spectroscopy}},\ }\href {https://doi.org/10.3847/2041-8213/ac0a81}
  {\bibfield  {journal} {\bibinfo  {journal} {Astrophys. J. Lett.}\ }\textbf
  {\bibinfo {volume} {918}},\ \bibinfo {pages} {L27} (\bibinfo {year}
  {2021})},\ \Eprint {https://arxiv.org/abs/2105.06980} {arXiv:2105.06980
  [astro-ph.HE]} \BibitemShut {NoStop}%
\bibitem [{\citenamefont {Raaijmakers}\ \emph {et~al.}(2021)\citenamefont
  {Raaijmakers}, \citenamefont {Greif}, \citenamefont {Hebeler}, \citenamefont
  {Hinderer}, \citenamefont {Nissanke}, \citenamefont {Schwenk}, \citenamefont
  {Riley}, \citenamefont {Watts}, \citenamefont {Lattimer},\ and\ \citenamefont
  {Ho}}]{Raaijmakers:2021uju}%
  \BibitemOpen
  \bibfield  {author} {\bibinfo {author} {\bibfnamefont {G.}~\bibnamefont
  {Raaijmakers}}, \bibinfo {author} {\bibfnamefont {S.~K.}\ \bibnamefont
  {Greif}}, \bibinfo {author} {\bibfnamefont {K.}~\bibnamefont {Hebeler}},
  \bibinfo {author} {\bibfnamefont {T.}~\bibnamefont {Hinderer}}, \bibinfo
  {author} {\bibfnamefont {S.}~\bibnamefont {Nissanke}}, \bibinfo {author}
  {\bibfnamefont {A.}~\bibnamefont {Schwenk}}, \bibinfo {author} {\bibfnamefont
  {T.~E.}\ \bibnamefont {Riley}}, \bibinfo {author} {\bibfnamefont {A.~L.}\
  \bibnamefont {Watts}}, \bibinfo {author} {\bibfnamefont {J.~M.}\ \bibnamefont
  {Lattimer}},\ and\ \bibinfo {author} {\bibfnamefont {W.~C.~G.}\ \bibnamefont
  {Ho}},\ }\bibfield  {title} {\bibinfo {title} {{Constraints on the Dense
  Matter Equation of State and Neutron Star Properties from
  NICER{\textquoteright}s Mass{\textendash}Radius Estimate of PSR J0740+6620
  and Multimessenger Observations}},\ }\href
  {https://doi.org/10.3847/2041-8213/ac089a} {\bibfield  {journal} {\bibinfo
  {journal} {Astrophys. J. Lett.}\ }\textbf {\bibinfo {volume} {918}},\
  \bibinfo {pages} {L29} (\bibinfo {year} {2021})},\ \Eprint
  {https://arxiv.org/abs/2105.06981} {arXiv:2105.06981 [astro-ph.HE]}
  \BibitemShut {NoStop}%
\bibitem [{\citenamefont {Miller}\ \emph {et~al.}(2021)\citenamefont {Miller}
  \emph {et~al.}}]{Miller:2021qha}%
  \BibitemOpen
  \bibfield  {author} {\bibinfo {author} {\bibfnamefont {M.~C.}\ \bibnamefont
  {Miller}} \emph {et~al.},\ }\bibfield  {title} {\bibinfo {title} {{The Radius
  of PSR J0740+6620 from NICER and XMM-Newton Data}},\ }\href
  {https://doi.org/10.3847/2041-8213/ac089b} {\bibfield  {journal} {\bibinfo
  {journal} {Astrophys. J. Lett.}\ }\textbf {\bibinfo {volume} {918}},\
  \bibinfo {pages} {L28} (\bibinfo {year} {2021})},\ \Eprint
  {https://arxiv.org/abs/2105.06979} {arXiv:2105.06979 [astro-ph.HE]}
  \BibitemShut {NoStop}%
\bibitem [{\citenamefont {Salmi}\ \emph {et~al.}(2024)\citenamefont {Salmi}
  \emph {et~al.}}]{Salmi:2024aum}%
  \BibitemOpen
  \bibfield  {author} {\bibinfo {author} {\bibfnamefont {T.}~\bibnamefont
  {Salmi}} \emph {et~al.},\ }\bibfield  {title} {\bibinfo {title} {{The Radius
  of the High-mass Pulsar PSR J0740+6620 with 3.6 yr of NICER Data}},\ }\href
  {https://doi.org/10.3847/1538-4357/ad5f1f} {\bibfield  {journal} {\bibinfo
  {journal} {Astrophys. J.}\ }\textbf {\bibinfo {volume} {974}},\ \bibinfo
  {pages} {294} (\bibinfo {year} {2024})},\ \Eprint
  {https://arxiv.org/abs/2406.14466} {arXiv:2406.14466 [astro-ph.HE]}
  \BibitemShut {NoStop}%
\bibitem [{\citenamefont {Choudhury}\ \emph {et~al.}(2024)\citenamefont
  {Choudhury} \emph {et~al.}}]{Choudhury:2024xbk}%
  \BibitemOpen
  \bibfield  {author} {\bibinfo {author} {\bibfnamefont {D.}~\bibnamefont
  {Choudhury}} \emph {et~al.},\ }\bibfield  {title} {\bibinfo {title} {{A NICER
  View of the Nearest and Brightest Millisecond Pulsar: PSR
  J0437{\textendash}4715}},\ }\href {https://doi.org/10.3847/2041-8213/ad5a6f}
  {\bibfield  {journal} {\bibinfo  {journal} {Astrophys. J. Lett.}\ }\textbf
  {\bibinfo {volume} {971}},\ \bibinfo {pages} {L20} (\bibinfo {year}
  {2024})},\ \Eprint {https://arxiv.org/abs/2407.06789} {arXiv:2407.06789
  [astro-ph.HE]} \BibitemShut {NoStop}%
\bibitem [{\citenamefont {Mauviard}\ \emph {et~al.}(2025)\citenamefont
  {Mauviard} \emph {et~al.}}]{Mauviard:2025dmd}%
  \BibitemOpen
  \bibfield  {author} {\bibinfo {author} {\bibfnamefont {L.}~\bibnamefont
  {Mauviard}} \emph {et~al.},\ }\bibfield  {title} {\bibinfo {title} {{A NICER
  View of the 1.4 M$_{\odot}$ Edge-on Pulsar PSR J0614-3329}},\ }\href
  {https://doi.org/10.3847/1538-4357/ae145d} {\bibfield  {journal} {\bibinfo
  {journal} {Astrophys. J.}\ }\textbf {\bibinfo {volume} {995}},\ \bibinfo
  {pages} {60} (\bibinfo {year} {2025})},\ \Eprint
  {https://arxiv.org/abs/2506.14883} {arXiv:2506.14883 [astro-ph.HE]}
  \BibitemShut {NoStop}%
\bibitem [{\citenamefont {Doroshenko}\ \emph {et~al.}(2022)\citenamefont
  {Doroshenko}, \citenamefont {Suleimanov}, \citenamefont {P{\"u}hlhofer},\
  and\ \citenamefont {Santangelo}}]{Doroshenko:2022nwp}%
  \BibitemOpen
  \bibfield  {author} {\bibinfo {author} {\bibfnamefont {V.}~\bibnamefont
  {Doroshenko}}, \bibinfo {author} {\bibfnamefont {V.}~\bibnamefont
  {Suleimanov}}, \bibinfo {author} {\bibfnamefont {G.}~\bibnamefont
  {P{\"u}hlhofer}},\ and\ \bibinfo {author} {\bibfnamefont {A.}~\bibnamefont
  {Santangelo}},\ }\bibfield  {title} {\bibinfo {title} {{A strangely light
  neutron star within a supernova remnant}},\ }\href
  {https://doi.org/10.1038/s41550-022-01800-1} {\bibfield  {journal} {\bibinfo
  {journal} {Nature Astron.}\ }\textbf {\bibinfo {volume} {6}},\ \bibinfo
  {pages} {1444} (\bibinfo {year} {2022})}\BibitemShut {NoStop}%
\bibitem [{\citenamefont {Romani}\ \emph {et~al.}(2022)\citenamefont {Romani},
  \citenamefont {Kandel}, \citenamefont {Filippenko}, \citenamefont {Brink},\
  and\ \citenamefont {Zheng}}]{Romani:2022jhd}%
  \BibitemOpen
  \bibfield  {author} {\bibinfo {author} {\bibfnamefont {R.~W.}\ \bibnamefont
  {Romani}}, \bibinfo {author} {\bibfnamefont {D.}~\bibnamefont {Kandel}},
  \bibinfo {author} {\bibfnamefont {A.~V.}\ \bibnamefont {Filippenko}},
  \bibinfo {author} {\bibfnamefont {T.~G.}\ \bibnamefont {Brink}},\ and\
  \bibinfo {author} {\bibfnamefont {W.}~\bibnamefont {Zheng}},\ }\bibfield
  {title} {\bibinfo {title} {{PSR J0952{\ensuremath{-}}0607: The Fastest and
  Heaviest Known Galactic Neutron Star}},\ }\href
  {https://doi.org/10.3847/2041-8213/ac8007} {\bibfield  {journal} {\bibinfo
  {journal} {Astrophys. J. Lett.}\ }\textbf {\bibinfo {volume} {934}},\
  \bibinfo {pages} {L17} (\bibinfo {year} {2022})},\ \Eprint
  {https://arxiv.org/abs/2207.05124} {arXiv:2207.05124 [astro-ph.HE]}
  \BibitemShut {NoStop}%
\bibitem [{\citenamefont {Romani}\ \emph {et~al.}(2026)\citenamefont {Romani},
  \citenamefont {Beleznay}, \citenamefont {Filippenko}, \citenamefont {Brink},\
  and\ \citenamefont {Zheng}}]{Romani:2025ytn}%
  \BibitemOpen
  \bibfield  {author} {\bibinfo {author} {\bibfnamefont {R.~W.}\ \bibnamefont
  {Romani}}, \bibinfo {author} {\bibfnamefont {M.}~\bibnamefont {Beleznay}},
  \bibinfo {author} {\bibfnamefont {A.~V.}\ \bibnamefont {Filippenko}},
  \bibinfo {author} {\bibfnamefont {T.~G.}\ \bibnamefont {Brink}},\ and\
  \bibinfo {author} {\bibfnamefont {W.}~\bibnamefont {Zheng}},\ }\bibfield
  {title} {\bibinfo {title} {{PSR J0952-0607: Tightening a Record-high Neutron
  Star Mass}},\ }\href {https://doi.org/10.3847/1538-4357/ae28c5} {\bibfield
  {journal} {\bibinfo  {journal} {Astrophys. J.}\ }\textbf {\bibinfo {volume}
  {996}},\ \bibinfo {pages} {101} (\bibinfo {year} {2026})},\ \Eprint
  {https://arxiv.org/abs/2512.05099} {arXiv:2512.05099 [astro-ph.HE]}
  \BibitemShut {NoStop}%
\bibitem [{\citenamefont {Abbott}\ \emph {et~al.}(2017)\citenamefont {Abbott}
  \emph {et~al.}}]{LIGOScientific:2017vwq}%
  \BibitemOpen
  \bibfield  {author} {\bibinfo {author} {\bibfnamefont {B.~P.}\ \bibnamefont
  {Abbott}} \emph {et~al.} (\bibinfo {collaboration} {LIGO Scientific,
  Virgo}),\ }\bibfield  {title} {\bibinfo {title} {{GW170817: Observation of
  Gravitational Waves from a Binary Neutron Star Inspiral}},\ }\href
  {https://doi.org/10.1103/PhysRevLett.119.161101} {\bibfield  {journal}
  {\bibinfo  {journal} {Phys. Rev. Lett.}\ }\textbf {\bibinfo {volume} {119}},\
  \bibinfo {pages} {161101} (\bibinfo {year} {2017})},\ \Eprint
  {https://arxiv.org/abs/1710.05832} {arXiv:1710.05832 [gr-qc]} \BibitemShut
  {NoStop}%
\bibitem [{\citenamefont {Abbott}\ \emph {et~al.}(2018)\citenamefont {Abbott}
  \emph {et~al.}}]{LIGOScientific:2018cki}%
  \BibitemOpen
  \bibfield  {author} {\bibinfo {author} {\bibfnamefont {B.~P.}\ \bibnamefont
  {Abbott}} \emph {et~al.} (\bibinfo {collaboration} {LIGO Scientific,
  Virgo}),\ }\bibfield  {title} {\bibinfo {title} {{GW170817: Measurements of
  neutron star radii and equation of state}},\ }\href
  {https://doi.org/10.1103/PhysRevLett.121.161101} {\bibfield  {journal}
  {\bibinfo  {journal} {Phys. Rev. Lett.}\ }\textbf {\bibinfo {volume} {121}},\
  \bibinfo {pages} {161101} (\bibinfo {year} {2018})},\ \Eprint
  {https://arxiv.org/abs/1805.11581} {arXiv:1805.11581 [gr-qc]} \BibitemShut
  {NoStop}%
\bibitem [{\citenamefont {Steiner}\ \emph {et~al.}(2013)\citenamefont
  {Steiner}, \citenamefont {Hempel},\ and\ \citenamefont
  {Fischer}}]{Steiner:2012rk}%
  \BibitemOpen
  \bibfield  {author} {\bibinfo {author} {\bibfnamefont {A.~W.}\ \bibnamefont
  {Steiner}}, \bibinfo {author} {\bibfnamefont {M.}~\bibnamefont {Hempel}},\
  and\ \bibinfo {author} {\bibfnamefont {T.}~\bibnamefont {Fischer}},\
  }\bibfield  {title} {\bibinfo {title} {{Core-collapse supernova equations of
  state based on neutron star observations}},\ }\href
  {https://doi.org/10.1088/0004-637X/774/1/17} {\bibfield  {journal} {\bibinfo
  {journal} {Astrophys. J.}\ }\textbf {\bibinfo {volume} {774}},\ \bibinfo
  {pages} {17} (\bibinfo {year} {2013})},\ \Eprint
  {https://arxiv.org/abs/1207.2184} {arXiv:1207.2184 [astro-ph.SR]}
  \BibitemShut {NoStop}%
\bibitem [{\citenamefont {Frohaug}\ \emph {et~al.}(2026)\citenamefont
  {Frohaug}, \citenamefont {Maslov}, \citenamefont {Dexheimer}, \citenamefont
  {Grefa}, \citenamefont {Jahan}, \citenamefont {Ratti},\ and\ \citenamefont
  {Restrepo}}]{Frohaug:2025okz}%
  \BibitemOpen
  \bibfield  {author} {\bibinfo {author} {\bibfnamefont {G.}~\bibnamefont
  {Frohaug}}, \bibinfo {author} {\bibfnamefont {K.}~\bibnamefont {Maslov}},
  \bibinfo {author} {\bibfnamefont {V.}~\bibnamefont {Dexheimer}}, \bibinfo
  {author} {\bibfnamefont {J.}~\bibnamefont {Grefa}}, \bibinfo {author}
  {\bibfnamefont {J.}~\bibnamefont {Jahan}}, \bibinfo {author} {\bibfnamefont
  {C.}~\bibnamefont {Ratti}},\ and\ \bibinfo {author} {\bibfnamefont {T.~E.}\
  \bibnamefont {Restrepo}},\ }\bibfield  {title} {\bibinfo {title}
  {{Relativistic mean-field model with density- and isospin-density-dependent
  couplings}},\ }\href {https://doi.org/10.1103/txsy-tmcf} {\bibfield
  {journal} {\bibinfo  {journal} {Phys. Rev. D}\ }\textbf {\bibinfo {volume}
  {113}},\ \bibinfo {pages} {123049} (\bibinfo {year} {2026})},\ \Eprint
  {https://arxiv.org/abs/2511.15646} {arXiv:2511.15646 [nucl-th]} \BibitemShut
  {NoStop}%
\bibitem [{\citenamefont {Typel}\ \emph {et~al.}(2010)\citenamefont {Typel},
  \citenamefont {Ropke}, \citenamefont {Klahn}, \citenamefont {Blaschke},\ and\
  \citenamefont {Wolter}}]{Typel:2009sy}%
  \BibitemOpen
  \bibfield  {author} {\bibinfo {author} {\bibfnamefont {S.}~\bibnamefont
  {Typel}}, \bibinfo {author} {\bibfnamefont {G.}~\bibnamefont {Ropke}},
  \bibinfo {author} {\bibfnamefont {T.}~\bibnamefont {Klahn}}, \bibinfo
  {author} {\bibfnamefont {D.}~\bibnamefont {Blaschke}},\ and\ \bibinfo
  {author} {\bibfnamefont {H.~H.}\ \bibnamefont {Wolter}},\ }\bibfield  {title}
  {\bibinfo {title} {{Composition and thermodynamics of nuclear matter with
  light clusters}},\ }\href {https://doi.org/10.1103/PhysRevC.81.015803}
  {\bibfield  {journal} {\bibinfo  {journal} {Phys. Rev. C}\ }\textbf {\bibinfo
  {volume} {81}},\ \bibinfo {pages} {015803} (\bibinfo {year} {2010})},\
  \Eprint {https://arxiv.org/abs/0908.2344} {arXiv:0908.2344 [nucl-th]}
  \BibitemShut {NoStop}%
\bibitem [{\citenamefont {Chetyrkin}\ and\ \citenamefont
  {Kuhn}(1994)}]{Chetyrkin:1994ex}%
  \BibitemOpen
  \bibfield  {author} {\bibinfo {author} {\bibfnamefont {K.~G.}\ \bibnamefont
  {Chetyrkin}}\ and\ \bibinfo {author} {\bibfnamefont {J.~H.}\ \bibnamefont
  {Kuhn}},\ }\bibfield  {title} {\bibinfo {title} {{Quartic mass corrections to
  $R_\text{had}$}},\ }\href {https://doi.org/10.1016/0550-3213(94)90605-X}
  {\bibfield  {journal} {\bibinfo  {journal} {Nucl. Phys. B}\ }\textbf
  {\bibinfo {volume} {432}},\ \bibinfo {pages} {337} (\bibinfo {year}
  {1994})},\ \Eprint {https://arxiv.org/abs/hep-ph/9406299}
  {arXiv:hep-ph/9406299} \BibitemShut {NoStop}%
\bibitem [{\citenamefont {Baikov}\ and\ \citenamefont
  {Chetyrkin}(2018)}]{Baikov:2018nzi}%
  \BibitemOpen
  \bibfield  {author} {\bibinfo {author} {\bibfnamefont {P.~A.}\ \bibnamefont
  {Baikov}}\ and\ \bibinfo {author} {\bibfnamefont {K.~G.}\ \bibnamefont
  {Chetyrkin}},\ }\bibfield  {title} {\bibinfo {title} {{QCD vacuum energy in 5
  loops}},\ }\href {https://doi.org/10.22323/1.290.0025} {\bibfield  {journal}
  {\bibinfo  {journal} {PoS}\ }\textbf {\bibinfo {volume} {RADCOR2017}},\
  \bibinfo {pages} {025} (\bibinfo {year} {2018})}\BibitemShut {NoStop}%
\bibitem [{\citenamefont {Andersen}\ \emph {et~al.}(2002)\citenamefont
  {Andersen}, \citenamefont {Braaten}, \citenamefont {Petitgirard},\ and\
  \citenamefont {Strickland}}]{Andersen:2002ey}%
  \BibitemOpen
  \bibfield  {author} {\bibinfo {author} {\bibfnamefont {J.~O.}\ \bibnamefont
  {Andersen}}, \bibinfo {author} {\bibfnamefont {E.}~\bibnamefont {Braaten}},
  \bibinfo {author} {\bibfnamefont {E.}~\bibnamefont {Petitgirard}},\ and\
  \bibinfo {author} {\bibfnamefont {M.}~\bibnamefont {Strickland}},\ }\bibfield
   {title} {\bibinfo {title} {{HTL perturbation theory to two loops}},\ }\href
  {https://doi.org/10.1103/PhysRevD.66.085016} {\bibfield  {journal} {\bibinfo
  {journal} {Phys. Rev. D}\ }\textbf {\bibinfo {volume} {66}},\ \bibinfo
  {pages} {085016} (\bibinfo {year} {2002})},\ \Eprint
  {https://arxiv.org/abs/hep-ph/0205085} {arXiv:hep-ph/0205085} \BibitemShut
  {NoStop}%
\bibitem [{\citenamefont {Haque}\ \emph {et~al.}(2013)\citenamefont {Haque},
  \citenamefont {Mustafa},\ and\ \citenamefont {Strickland}}]{Haque:2012my}%
  \BibitemOpen
  \bibfield  {author} {\bibinfo {author} {\bibfnamefont {N.}~\bibnamefont
  {Haque}}, \bibinfo {author} {\bibfnamefont {M.~G.}\ \bibnamefont {Mustafa}},\
  and\ \bibinfo {author} {\bibfnamefont {M.}~\bibnamefont {Strickland}},\
  }\bibfield  {title} {\bibinfo {title} {{Two-loop hard thermal loop pressure
  at finite temperature and chemical potential}},\ }\href
  {https://doi.org/10.1103/PhysRevD.87.105007} {\bibfield  {journal} {\bibinfo
  {journal} {Phys. Rev. D}\ }\textbf {\bibinfo {volume} {87}},\ \bibinfo
  {pages} {105007} (\bibinfo {year} {2013})},\ \Eprint
  {https://arxiv.org/abs/1212.1797} {arXiv:1212.1797 [hep-ph]} \BibitemShut
  {NoStop}%
\bibitem [{\citenamefont {Kneur}\ and\ \citenamefont
  {Neveu}(2020)}]{Kneur:2020bph}%
  \BibitemOpen
  \bibfield  {author} {\bibinfo {author} {\bibfnamefont {J.-L.}\ \bibnamefont
  {Kneur}}\ and\ \bibinfo {author} {\bibfnamefont {A.}~\bibnamefont {Neveu}},\
  }\bibfield  {title} {\bibinfo {title} {{Chiral condensate and spectral
  density at full five-loop and partial six-loop orders of renormalization
  group optimized perturbation theory}},\ }\href
  {https://doi.org/10.1103/PhysRevD.101.074009} {\bibfield  {journal} {\bibinfo
   {journal} {Phys. Rev. D}\ }\textbf {\bibinfo {volume} {101}},\ \bibinfo
  {pages} {074009} (\bibinfo {year} {2020})},\ \Eprint
  {https://arxiv.org/abs/2001.11670} {arXiv:2001.11670 [hep-ph]} \BibitemShut
  {NoStop}%
\bibitem [{\citenamefont {Gorenstein}\ and\ \citenamefont
  {Yang}(1995)}]{Gorenstein:1995vm}%
  \BibitemOpen
  \bibfield  {author} {\bibinfo {author} {\bibfnamefont {M.~I.}\ \bibnamefont
  {Gorenstein}}\ and\ \bibinfo {author} {\bibfnamefont {S.-N.}\ \bibnamefont
  {Yang}},\ }\bibfield  {title} {\bibinfo {title} {{Gluon plasma with a medium
  dependent dispersion relation}},\ }\href
  {https://doi.org/10.1103/PhysRevD.52.5206} {\bibfield  {journal} {\bibinfo
  {journal} {Phys. Rev. D}\ }\textbf {\bibinfo {volume} {52}},\ \bibinfo
  {pages} {5206} (\bibinfo {year} {1995})}\BibitemShut {NoStop}%
\bibitem [{\citenamefont {Biro}\ \emph {et~al.}(2003)\citenamefont {Biro},
  \citenamefont {Shanenko},\ and\ \citenamefont {Toneev}}]{Biro:2001ug}%
  \BibitemOpen
  \bibfield  {author} {\bibinfo {author} {\bibfnamefont {T.~S.}\ \bibnamefont
  {Biro}}, \bibinfo {author} {\bibfnamefont {A.~A.}\ \bibnamefont {Shanenko}},\
  and\ \bibinfo {author} {\bibfnamefont {V.~D.}\ \bibnamefont {Toneev}},\
  }\bibfield  {title} {\bibinfo {title} {{Towards thermodynamical consistency
  of quasiparticle picture}},\ }\href {https://doi.org/10.1134/1.1577921}
  {\bibfield  {journal} {\bibinfo  {journal} {Phys. Atom. Nucl.}\ }\textbf
  {\bibinfo {volume} {66}},\ \bibinfo {pages} {982} (\bibinfo {year} {2003})},\
  \Eprint {https://arxiv.org/abs/nucl-th/0102027} {arXiv:nucl-th/0102027}
  \BibitemShut {NoStop}%
\bibitem [{\citenamefont {Lenzi}\ \emph {et~al.}(2010)\citenamefont {Lenzi},
  \citenamefont {Schneider}, \citenamefont {Provid\^encia},\ and\ \citenamefont
  {Marinho}}]{Lenzi:2010mz}%
  \BibitemOpen
  \bibfield  {author} {\bibinfo {author} {\bibfnamefont {C.~H.}\ \bibnamefont
  {Lenzi}}, \bibinfo {author} {\bibfnamefont {A.~S.}\ \bibnamefont
  {Schneider}}, \bibinfo {author} {\bibfnamefont {C.}~\bibnamefont
  {Provid\^encia}},\ and\ \bibinfo {author} {\bibfnamefont {R.~M.}\
  \bibnamefont {Marinho}},\ }\bibfield  {title} {\bibinfo {title} {{Compact
  stars with a quark core within NJL model}},\ }\href
  {https://doi.org/10.1103/PhysRevC.82.015809} {\bibfield  {journal} {\bibinfo
  {journal} {Phys. Rev. C}\ }\textbf {\bibinfo {volume} {82}},\ \bibinfo
  {pages} {015809} (\bibinfo {year} {2010})},\ \Eprint
  {https://arxiv.org/abs/1001.3169} {arXiv:1001.3169 [nucl-th]} \BibitemShut
  {NoStop}%
\bibitem [{\citenamefont {Tanabashi}\ \emph {et~al.}(2018)\citenamefont
  {Tanabashi} \emph {et~al.}}]{ParticleDataGroup:2018ovx}%
  \BibitemOpen
  \bibfield  {author} {\bibinfo {author} {\bibfnamefont {M.}~\bibnamefont
  {Tanabashi}} \emph {et~al.} (\bibinfo {collaboration} {Particle Data
  Group}),\ }\bibfield  {title} {\bibinfo {title} {{Review of Particle
  Physics}},\ }\href {https://doi.org/10.1103/PhysRevD.98.030001} {\bibfield
  {journal} {\bibinfo  {journal} {Phys. Rev. D}\ }\textbf {\bibinfo {volume}
  {98}},\ \bibinfo {pages} {030001} (\bibinfo {year} {2018})}\BibitemShut
  {NoStop}%
\bibitem [{\citenamefont {Bazavov}\ \emph {et~al.}(2012)\citenamefont
  {Bazavov}, \citenamefont {Brambilla}, \citenamefont {Garcia~i Tormo},
  \citenamefont {Petreczky}, \citenamefont {Soto},\ and\ \citenamefont
  {Vairo}}]{Bazavov:2012ka}%
  \BibitemOpen
  \bibfield  {author} {\bibinfo {author} {\bibfnamefont {A.}~\bibnamefont
  {Bazavov}}, \bibinfo {author} {\bibfnamefont {N.}~\bibnamefont {Brambilla}},
  \bibinfo {author} {\bibfnamefont {X.}~\bibnamefont {Garcia~i Tormo}},
  \bibinfo {author} {\bibfnamefont {P.}~\bibnamefont {Petreczky}}, \bibinfo
  {author} {\bibfnamefont {J.}~\bibnamefont {Soto}},\ and\ \bibinfo {author}
  {\bibfnamefont {A.}~\bibnamefont {Vairo}},\ }\bibfield  {title} {\bibinfo
  {title} {{Determination of $\alpha_s$ from the QCD static energy}},\ }\href
  {https://doi.org/10.1103/PhysRevD.86.114031} {\bibfield  {journal} {\bibinfo
  {journal} {Phys. Rev. D}\ }\textbf {\bibinfo {volume} {86}},\ \bibinfo
  {pages} {114031} (\bibinfo {year} {2012})},\ \Eprint
  {https://arxiv.org/abs/1205.6155} {arXiv:1205.6155 [hep-ph]} \BibitemShut
  {NoStop}%
\bibitem [{\citenamefont {Deur}\ \emph {et~al.}(2016)\citenamefont {Deur},
  \citenamefont {Brodsky},\ and\ \citenamefont {de~Teramond}}]{Deur:2016tte}%
  \BibitemOpen
  \bibfield  {author} {\bibinfo {author} {\bibfnamefont {A.}~\bibnamefont
  {Deur}}, \bibinfo {author} {\bibfnamefont {S.~J.}\ \bibnamefont {Brodsky}},\
  and\ \bibinfo {author} {\bibfnamefont {G.~F.}\ \bibnamefont {de~Teramond}},\
  }\bibfield  {title} {\bibinfo {title} {{The QCD Running Coupling}},\ }\href
  {https://doi.org/10.1016/j.ppnp.2016.04.003} {\bibfield  {journal} {\bibinfo
  {journal} {Prog. Part. Nucl. Phys.}\ }\textbf {\bibinfo {volume} {90}},\
  \bibinfo {pages} {1} (\bibinfo {year} {2016})},\ \Eprint
  {https://arxiv.org/abs/1604.08082} {arXiv:1604.08082 [hep-ph]} \BibitemShut
  {NoStop}%
\bibitem [{\citenamefont {Bodmer}(1971)}]{Bodmer:1971we}%
  \BibitemOpen
  \bibfield  {author} {\bibinfo {author} {\bibfnamefont {A.~R.}\ \bibnamefont
  {Bodmer}},\ }\bibfield  {title} {\bibinfo {title} {{Collapsed nuclei}},\
  }\href {https://doi.org/10.1103/PhysRevD.4.1601} {\bibfield  {journal}
  {\bibinfo  {journal} {Phys. Rev. D}\ }\textbf {\bibinfo {volume} {4}},\
  \bibinfo {pages} {1601} (\bibinfo {year} {1971})}\BibitemShut {NoStop}%
\bibitem [{\citenamefont {Terazawa}(1979)}]{Teraza1979}%
  \BibitemOpen
  \bibfield  {author} {\bibinfo {author} {\bibfnamefont {H.}~\bibnamefont
  {Terazawa}},\ }\href@noop {} {\emph {\bibinfo {title} {Quark shell model and
  superheavy hyper-nucleus}}},\ \bibinfo {type} {INS Report}\ \bibinfo {number}
  {336}\ (\bibinfo  {institution} {University of Tokyo},\ \bibinfo {year}
  {1979})\BibitemShut {NoStop}%
\bibitem [{\citenamefont {Witten}(1984)}]{Witten:1984rs}%
  \BibitemOpen
  \bibfield  {author} {\bibinfo {author} {\bibfnamefont {E.}~\bibnamefont
  {Witten}},\ }\bibfield  {title} {\bibinfo {title} {{Cosmic Separation of
  Phases}},\ }\href {https://doi.org/10.1103/PhysRevD.30.272} {\bibfield
  {journal} {\bibinfo  {journal} {Phys. Rev. D}\ }\textbf {\bibinfo {volume}
  {30}},\ \bibinfo {pages} {272} (\bibinfo {year} {1984})}\BibitemShut
  {NoStop}%
\bibitem [{\citenamefont {Kini}\ \emph {et~al.}(2026)\citenamefont {Kini} \emph
  {et~al.}}]{Kini:2026rjx}%
  \BibitemOpen
  \bibfield  {author} {\bibinfo {author} {\bibfnamefont {Y.}~\bibnamefont
  {Kini}} \emph {et~al.},\ }\bibfield  {title} {\bibinfo {title} {{A NICER View
  of PSR J0030+0451: Updated Constraints from Six Years of NICER
  Observations}},\ }\href@noop {} {\  (\bibinfo {year} {2026})},\ \Eprint
  {https://arxiv.org/abs/2602.23743} {arXiv:2602.23743 [astro-ph.HE]}
  \BibitemShut {NoStop}%
\bibitem [{\citenamefont {Abbott}\ \emph {et~al.}(2020)\citenamefont {Abbott}
  \emph {et~al.}}]{LIGOScientific:2020zkf}%
  \BibitemOpen
  \bibfield  {author} {\bibinfo {author} {\bibfnamefont {R.}~\bibnamefont
  {Abbott}} \emph {et~al.} (\bibinfo {collaboration} {LIGO Scientific,
  Virgo}),\ }\bibfield  {title} {\bibinfo {title} {{GW190814: Gravitational
  Waves from the Coalescence of a 23 Solar Mass Black Hole with a 2.6 Solar
  Mass Compact Object}},\ }\href {https://doi.org/10.3847/2041-8213/ab960f}
  {\bibfield  {journal} {\bibinfo  {journal} {Astrophys. J. Lett.}\ }\textbf
  {\bibinfo {volume} {896}},\ \bibinfo {pages} {L44} (\bibinfo {year}
  {2020})},\ \Eprint {https://arxiv.org/abs/2006.12611} {arXiv:2006.12611
  [astro-ph.HE]} \BibitemShut {NoStop}%
\bibitem [{\citenamefont {Tolman}(1939)}]{Tolman:1939jz}%
  \BibitemOpen
  \bibfield  {author} {\bibinfo {author} {\bibfnamefont {R.~C.}\ \bibnamefont
  {Tolman}},\ }\bibfield  {title} {\bibinfo {title} {{Static solutions of
  Einstein's field equations for spheres of fluid}},\ }\href
  {https://doi.org/10.1103/PhysRev.55.364} {\bibfield  {journal} {\bibinfo
  {journal} {Phys. Rev.}\ }\textbf {\bibinfo {volume} {55}},\ \bibinfo {pages}
  {364} (\bibinfo {year} {1939})}\BibitemShut {NoStop}%
\bibitem [{\citenamefont {Oppenheimer}\ and\ \citenamefont
  {Volkoff}(1939)}]{Oppenheimer:1939ne}%
  \BibitemOpen
  \bibfield  {author} {\bibinfo {author} {\bibfnamefont {J.~R.}\ \bibnamefont
  {Oppenheimer}}\ and\ \bibinfo {author} {\bibfnamefont {G.~M.}\ \bibnamefont
  {Volkoff}},\ }\bibfield  {title} {\bibinfo {title} {{On massive neutron
  cores}},\ }\href {https://doi.org/10.1103/PhysRev.55.374} {\bibfield
  {journal} {\bibinfo  {journal} {Phys. Rev.}\ }\textbf {\bibinfo {volume}
  {55}},\ \bibinfo {pages} {374} (\bibinfo {year} {1939})}\BibitemShut
  {NoStop}%
\bibitem [{\citenamefont {Maruyama}\ \emph {et~al.}(2007)\citenamefont
  {Maruyama}, \citenamefont {Chiba}, \citenamefont {Schulze},\ and\
  \citenamefont {Tatsumi}}]{Maruyama:2007ey}%
  \BibitemOpen
  \bibfield  {author} {\bibinfo {author} {\bibfnamefont {T.}~\bibnamefont
  {Maruyama}}, \bibinfo {author} {\bibfnamefont {S.}~\bibnamefont {Chiba}},
  \bibinfo {author} {\bibfnamefont {H.-J.}\ \bibnamefont {Schulze}},\ and\
  \bibinfo {author} {\bibfnamefont {T.}~\bibnamefont {Tatsumi}},\ }\bibfield
  {title} {\bibinfo {title} {{Hadron-quark mixed phase in hyperon stars}},\
  }\href {https://doi.org/10.1103/PhysRevD.76.123015} {\bibfield  {journal}
  {\bibinfo  {journal} {Phys. Rev. D}\ }\textbf {\bibinfo {volume} {76}},\
  \bibinfo {pages} {123015} (\bibinfo {year} {2007})},\ \Eprint
  {https://arxiv.org/abs/0708.3277} {arXiv:0708.3277 [nucl-th]} \BibitemShut
  {NoStop}%
\bibitem [{\citenamefont {Pinto}\ \emph {et~al.}(2012)\citenamefont {Pinto},
  \citenamefont {Koch},\ and\ \citenamefont {Randrup}}]{Pinto:2012aq}%
  \BibitemOpen
  \bibfield  {author} {\bibinfo {author} {\bibfnamefont {M.~B.}\ \bibnamefont
  {Pinto}}, \bibinfo {author} {\bibfnamefont {V.}~\bibnamefont {Koch}},\ and\
  \bibinfo {author} {\bibfnamefont {J.}~\bibnamefont {Randrup}},\ }\bibfield
  {title} {\bibinfo {title} {{The Surface Tension of Quark Matter in a
  Geometrical Approach}},\ }\href {https://doi.org/10.1103/PhysRevC.86.025203}
  {\bibfield  {journal} {\bibinfo  {journal} {Phys. Rev. C}\ }\textbf {\bibinfo
  {volume} {86}},\ \bibinfo {pages} {025203} (\bibinfo {year} {2012})},\
  \Eprint {https://arxiv.org/abs/1207.5186} {arXiv:1207.5186 [hep-ph]}
  \BibitemShut {NoStop}%
\bibitem [{\citenamefont {Sagun}\ \emph {et~al.}(2023)\citenamefont {Sagun},
  \citenamefont {Giangrandi}, \citenamefont {Dietrich}, \citenamefont
  {Ivanytskyi}, \citenamefont {Negreiros},\ and\ \citenamefont
  {Provid{\^e}ncia}}]{Sagun:2023rzp}%
  \BibitemOpen
  \bibfield  {author} {\bibinfo {author} {\bibfnamefont {V.}~\bibnamefont
  {Sagun}}, \bibinfo {author} {\bibfnamefont {E.}~\bibnamefont {Giangrandi}},
  \bibinfo {author} {\bibfnamefont {T.}~\bibnamefont {Dietrich}}, \bibinfo
  {author} {\bibfnamefont {O.}~\bibnamefont {Ivanytskyi}}, \bibinfo {author}
  {\bibfnamefont {R.}~\bibnamefont {Negreiros}},\ and\ \bibinfo {author}
  {\bibfnamefont {C.}~\bibnamefont {Provid{\^e}ncia}},\ }\bibfield  {title}
  {\bibinfo {title} {{What Is the Nature of the HESS J1731-347 Compact
  Object?}},\ }\href {https://doi.org/10.3847/1538-4357/acfc9e} {\bibfield
  {journal} {\bibinfo  {journal} {Astrophys. J.}\ }\textbf {\bibinfo {volume}
  {958}},\ \bibinfo {pages} {49} (\bibinfo {year} {2023})},\ \Eprint
  {https://arxiv.org/abs/2306.12326} {arXiv:2306.12326 [astro-ph.HE]}
  \BibitemShut {NoStop}%
\bibitem [{\citenamefont {Di~Clemente}\ \emph {et~al.}(2024)\citenamefont
  {Di~Clemente}, \citenamefont {Drago},\ and\ \citenamefont
  {Pagliara}}]{DiClemente:2022wqp}%
  \BibitemOpen
  \bibfield  {author} {\bibinfo {author} {\bibfnamefont {F.}~\bibnamefont
  {Di~Clemente}}, \bibinfo {author} {\bibfnamefont {A.}~\bibnamefont {Drago}},\
  and\ \bibinfo {author} {\bibfnamefont {G.}~\bibnamefont {Pagliara}},\
  }\bibfield  {title} {\bibinfo {title} {{Is the Compact Object Associated with
  HESS J1731-347 a Strange Quark Star? A Possible Astrophysical Scenario for
  Its Formation}},\ }\href {https://doi.org/10.3847/1538-4357/ad445b}
  {\bibfield  {journal} {\bibinfo  {journal} {Astrophys. J.}\ }\textbf
  {\bibinfo {volume} {967}},\ \bibinfo {pages} {159} (\bibinfo {year}
  {2024})},\ \Eprint {https://arxiv.org/abs/2211.07485} {arXiv:2211.07485
  [astro-ph.HE]} \BibitemShut {NoStop}%
\bibitem [{\citenamefont {Baikov}\ \emph {et~al.}(2017)\citenamefont {Baikov},
  \citenamefont {Chetyrkin},\ and\ \citenamefont {K{\"u}hn}}]{Baikov:2016tgj}%
  \BibitemOpen
  \bibfield  {author} {\bibinfo {author} {\bibfnamefont {P.~A.}\ \bibnamefont
  {Baikov}}, \bibinfo {author} {\bibfnamefont {K.~G.}\ \bibnamefont
  {Chetyrkin}},\ and\ \bibinfo {author} {\bibfnamefont {J.~H.}\ \bibnamefont
  {K{\"u}hn}},\ }\bibfield  {title} {\bibinfo {title} {{Five-Loop Running of
  the QCD Coupling Constant}},\ }\href
  {https://doi.org/10.1103/PhysRevLett.118.082002} {\bibfield  {journal}
  {\bibinfo  {journal} {Phys. Rev. Lett.}\ }\textbf {\bibinfo {volume} {118}},\
  \bibinfo {pages} {082002} (\bibinfo {year} {2017})},\ \Eprint
  {https://arxiv.org/abs/1606.08659} {arXiv:1606.08659 [hep-ph]} \BibitemShut
  {NoStop}%
\bibitem [{\citenamefont {Luthe}\ \emph {et~al.}(2016)\citenamefont {Luthe},
  \citenamefont {Maier}, \citenamefont {Marquard},\ and\ \citenamefont
  {Schr{\"o}der}}]{Luthe:2016ima}%
  \BibitemOpen
  \bibfield  {author} {\bibinfo {author} {\bibfnamefont {T.}~\bibnamefont
  {Luthe}}, \bibinfo {author} {\bibfnamefont {A.}~\bibnamefont {Maier}},
  \bibinfo {author} {\bibfnamefont {P.}~\bibnamefont {Marquard}},\ and\
  \bibinfo {author} {\bibfnamefont {Y.}~\bibnamefont {Schr{\"o}der}},\
  }\bibfield  {title} {\bibinfo {title} {{Towards the five-loop Beta function
  for a general gauge group}},\ }\href
  {https://doi.org/10.1007/JHEP07(2016)127} {\bibfield  {journal} {\bibinfo
  {journal} {JHEP}\ }\textbf {\bibinfo {volume} {07}},\ \bibinfo {pages}
  {127}},\ \Eprint {https://arxiv.org/abs/1606.08662} {arXiv:1606.08662
  [hep-ph]} \BibitemShut {NoStop}%
\bibitem [{\citenamefont {Herzog}\ \emph {et~al.}(2017)\citenamefont {Herzog},
  \citenamefont {Ruijl}, \citenamefont {Ueda}, \citenamefont {Vermaseren},\
  and\ \citenamefont {Vogt}}]{Herzog:2017ohr}%
  \BibitemOpen
  \bibfield  {author} {\bibinfo {author} {\bibfnamefont {F.}~\bibnamefont
  {Herzog}}, \bibinfo {author} {\bibfnamefont {B.}~\bibnamefont {Ruijl}},
  \bibinfo {author} {\bibfnamefont {T.}~\bibnamefont {Ueda}}, \bibinfo {author}
  {\bibfnamefont {J.~A.~M.}\ \bibnamefont {Vermaseren}},\ and\ \bibinfo
  {author} {\bibfnamefont {A.}~\bibnamefont {Vogt}},\ }\bibfield  {title}
  {\bibinfo {title} {{The five-loop beta function of Yang-Mills theory with
  fermions}},\ }\href {https://doi.org/10.1007/JHEP02(2017)090} {\bibfield
  {journal} {\bibinfo  {journal} {JHEP}\ }\textbf {\bibinfo {volume} {02}},\
  \bibinfo {pages} {090}},\ \Eprint {https://arxiv.org/abs/1701.01404}
  {arXiv:1701.01404 [hep-ph]} \BibitemShut {NoStop}%
\end{thebibliography}%

\end{document}